\documentstyle[aas2pp4]{article}

\begin{document}
\title{A SPECTROSCOPIC STUDY OF A LARGE SAMPLE OF WOLF-RAYET GALAXIES}

\author{Natalia G. Guseva, Yuri I. Izotov\footnote[1]{Visiting astronomer,
Kitt Peak National Observatory, National Optical Astronomical Observatories,
operated by the Association of Universities for Research in Astronomy, Inc.,
under contract with the National Science Foundation.}}
\affil{Main Astronomical Observatory, Ukrainian National Academy of Sciences,
Goloseevo, Kiev 03680, Ukraine \\ Electronic mail: guseva@mao.kiev.ua,
izotov@mao.kiev.ua}
\and
\author{Trinh X. Thuan\footnotemark[1]}
\affil{Astronomy Department, University of Virginia, Charlottesville, VA 22903
\\ Electronic mail: txt@virginia.edu}


\begin{abstract}

   We analyze long-slit spectral
observations of 39 Wolf-Rayet (WR) galaxies with heavy element mass
fraction ranging over 2 orders of magnitude, from $Z_\odot$/50 to 2$Z_\odot$.
Nearly all galaxies in our sample show broad WR emission in the blue region 
of the spectrum (the blue bump) consisting of an unresolved blend of
N III $\lambda$4640, C III $\lambda$4650, C IV $\lambda$4658 and He II 
$\lambda$4686 emission lines. Broad C IV $\lambda$5808 emission (the red bump) 
is detected in 30 galaxies. Additionally, weaker WR emission lines 
are identified, most often the N III $\lambda$4512
and Si III $\lambda$4565 lines, which have very rarely or never been seen and 
discussed before in WR galaxies. These emission features are 
characteristic of WN7 -- WN8 and WN9 -- WN11 stars respectively.

  We derive the numbers of early WC (WCE) and late WN (WNL) stars from 
the luminosities of the red and blue bumps, and the number of O stars from the 
luminosity of the H$\beta$ emission line. Additionally, we 
propose a new technique for deriving the numbers
of WNL stars from the N III $\lambda$4512 and Si III $\lambda$4565 emission 
lines. This technique is potentially more precise than the blue bump method
because it does not suffer from contamination of WCE and early WN (WNE) stars 
and nebular gaseous emission.
 
 It is found that the relative number of WR stars $N$(WR)/$N$(O + WR) 
decreases with decreasing metallicity, in agreement with predictions of
evolutionary synthesis models. The relative number ratios $N$(WC)/$N$(WN) and
the equivalent widths of the blue bump $EW$($\lambda$4650) and of the 
red bump $EW$($\lambda$5808) derived  from
observations are also in satisfactory agreement with theoretical predictions,
except for the most metal-deficient WR galaxies. A possible
source of disagreement is too low a line emission luminosity adopted for
a single WCE star in low-metallicity models.

We assemble a sample of 30 H II regions with detected He II $\lambda$4686 
nebular emission to analyze the possible connection of this emission with the hard UV radiation of the WR stars. The theoretical predictions 
satisfactorily reproduce the observed intensities and equivalent widths of
the He II $\lambda$4686 nebular emission line. 
However, galaxies with nebular He II $\lambda$4686 emission do not always 
show WR emission. Therefore, in addition to the ionization of He$^+$ in the H II region by WR stars, other mechanisms for the origin of He II 
$\lambda$4686 such as radiative shocks probably need to be invoked.

\end{abstract}


\section {INTRODUCTION}

   Galaxies with Wolf-Rayet (WR) star features in their spectra have long been
known, beginning with the discovery of such features in the spectrum of the
blue compact galaxy He 2-10 by Allen, Wright \& Goss (1976). 
Osterbrock \& Cohen (1982) and Conti (1991) introduced the concept of WR 
galaxies, defining them to be those galaxies which
show broad stellar emission lines in their spectra. 
These WR galaxies are thought to be undergoing present or very
recent star formation which produces massive stars evolving to the WR stage.
WR galaxies are therefore ideal objects for studying the early phases of
starbursts, determining burst properties and constraining parameters of
the high mass end of the initial mass function.

More than 130 WR galaxies are now known (Conti 1999; Schaerer, Contini \&
Pindao 1999b). Morphologically, they constitute a
very inhomogeneous class, including such diverse objects as galaxies
with active galactic nuclei (AGN), ultraluminous far-infrared
galaxies, spiral galaxies, starbursts and blue compact dwarf (BCD) galaxies.
Individual WR stars cannot be usually observed as single stars in distant
galaxies. They have been observed only in our Galaxy (
Massey \& Conti 1980; Torres \& Conti 1984; Hamman,
Koesterke \& Wessolowski 1995; Koesterke \& Hamann 1995; 
Esteban \& Peimbert 1995; Crowther, Smith \& Hillier 1995b; Crowther et al. 
1995c; Crowther, Smith \& Willis 1995d; Crowther \& Bohannan 1997), 
in the Magellanic Clouds (Moffat et al. 1987; Conti \& Massey 1989; Bohannan \&
Walborn 1989;
Smith, Shara \& Moffat 1990a; Russell \& Dopita 1990; 
Crowther, Hillier \& Smith 1995a; Crowther \& Smith 1997;
Crowther \& Dessart 1998) and in some members of the Local
Group: M31 and M33 (Conti \& Massey 1981; D'Odorico \& Rosa 1981; 
Massey \& Conti 1983; Massey, Conti \& Armandroff 1987; 
Schild, Smith \& Willis 1990;
Willis, Schild \& Smith 1992; Smith, Crowther \& Willis 1995;
Crowther et al. 1997; Massey \& Johnson 1998), 
IC 10 (Massey, Armandroff \& Conti 1992), NGC 6822, IC 1613, NGC 300,
and NGC 55 (Massey \& Johnson 1998).

 Much effort has been expended recently to construct a reliable 
quantitative scheme for classifying WR stars, based on observations of these
stars in our Galaxy and in the nearest galaxies, mainly the LMC. 
As a result, a quantitative classification of WN, WC and WO stars has been 
elaborated (Smith, Shara \& Moffat 1990ab, 1996; Smith \& Maeder 1991; 
Kingsburgh, Barlow \& Storey 1995; Crowther, De Marco \& Barlow 1998).
Additionally, evolutionary synthesis models for young starbursts 
have been developed with the 
use of the latest stellar evolution models, theoretical stellar spectra and
compilation of observed emission line strengths from WR stars (Kr\"uger et al. 1992; Cervi\~no \& Mas-Hesse 1994; Meynet 1995; Schaerer \& Vacca 1998,
hereafter SV98). 
 
  WR stars in more distant objects are detected indirectly by observing 
integrated galaxy spectra. 
 Strong star formation activity in 
a galaxy results in a large number of massive stars, the 
most massive of which will
evolve through the WR phase. Thus, at a definite stage of the starburst 
evolution, many WR stars make their appearance during a short time
interval. 
Despite their small number relative to that of massive stars, 
especially in low-metallicity galaxies,
WR stars are numerous enough for their integrated emission to be detected.
In this paper, we shall follow Osterbrock \& Cohen (1982) and Conti (1991)
in defining a WR galaxy as one whose integrated spectrum of the whole or 
a part   
shows detectable WR broad features emitted by unresolved star clusters. 
While this 
definition is dependent on the quality of the spectrum and the 
location and size of the aperture, it has the advantage of distinguishing 
a distant WR galaxy from nearby galaxies, 
 where WR stars can be studied individually.
 
The ratio of the number of WR stars to that of all massive stars 
is a very strong function of metallicity.
Theoretical evolutionary models (Mas-Hesse \& Kunth 1991; Maeder 1991;
Kr\"uger et al. 1992; Cervi\~no \& Mas-Hesse 1994;
Maeder \& Meynet 1994; Meynet 1995; SV98) predict that for a fixed metallicity, 
this ratio varies strongly with time elapsed since the beginning of the 
starburst. Its maximum value decreases from 1 to 0.02 when the
heavy element mass fraction $Z$ decreases from $Z_\odot$ to $Z_\odot$/50.
The duration of the WR stage in the starburst also 
decreases with decreasing metallicity.
Hence the number of galaxies with extremely low metallicities containing a
WR stellar population is expected to be very small. 

A systematic search for WR features in emission-line galaxies was first carried 
out by Kunth \& Joubert (1985). They examined 45 extragalactic H II regions and 
found 17 WR galaxies, based on the detection in their spectra 
of a broad emission excess in the 
$\lambda$4600--4700 wavelength region. Conti (1991) compiled from 
the literature a first catalog containing 37 WR galaxies. 
Vacca \& Conti (1992) performed a search for WR stars in a sample of 14 
emission-line galaxies and developed a quantitative scheme for 
estimating WR populations. All these searches resulted in the
detection of WR features in the blue region of the spectrum at $\lambda$4650
(hereafter called the ``blue bump''). It is interesting to note that,
in the first WR galaxy observed, He 2-10, Allen et al.
(1976) detected not only the blue bump, but
also a broad N III feature at $\lambda$4511--4535. In subsequent studies of
WR galaxies this feature was not generally seen or discussed. An exception 
is the most metal-deficient WR galaxy known, I Zw 18, where Izotov et al. 
(1997) also identified the N III $\lambda$4511--4535 feature.

   Evolutionary synthesis models predict a significant number of
WC4 stars which radiate strong broad C III/C IV $\lambda$4650 and
C IV $\lambda$5808 features (Schaerer \& Vacca 1996; Schaerer et al. 1997).
The observed ratio $N$(WC)/$N$(WN) is $\sim$ 1
in the solar neighborhood (Conti \& Vacca 1990), and $\sim$ 0.1--0.2 
within a 10$''$ radius of the 30 Doradus nebula (Moffat et al. 1987).
In starbursting galaxies with 1/5 $\leq$ $Z$/$Z_\odot$ $\leq$ 1 $\sim$ 30\% of
WR stars are of WC subtype (Schaerer \& Vacca 1996; Schaerer et al. 1997). 
The first detections of the C IV $\lambda$5808 emission line (hereafter called
the ``red bump'') in integrated galaxy spectra were reported
by Kunth \& Schild (1986) and Dinerstein \& Shields (1986).
However, the red bump was not seen in other early observations
(Gonzalez-Riestra, Rego \& Zamorano 1987; Vacca \& Conti 1992).
Later higher signal-to-noise ratio observations allowed to detect the 
red WC bump (CIV $\lambda$ 5808) in an increasing number of
galaxies (Izotov, Thuan \& Lipovetsky 1994, 1997, hereafter ITL94 and ITL97; 
Izotov et al. 1996; Thuan, Izotov \& Lipovetsky 1996; Izotov \& Thuan 1998b,
hereafter IT98; Schaerer et al. 1997; Schaerer et al. 1999a; 
Huang et al. 1999). 
WR stars of both WN and WC subtypes have even
been detected in I Zw 18 with $Z$ $\sim$ $Z_\odot$/50 
(Izotov et al. 1997; Legrand et al. 1997; De Mello et al. 1998).
Hence, we have now at our disposal WR galaxies with metallicities 
ranging over two orders
of magnitude, from $Z_\odot$/50 to $\sim$ 2$Z_\odot$. This allows us to
study the properties of massive stellar populations in metal-poor environments 
which are not available in the Local Group. 

   In 1993, a program was begun to obtain high signal-to-noise
spectra for a large sample of a low-metallicity blue compact dwarf 
galaxies in order
to measure the primordial helium abundance (ITL94, ITL97, IT98, Thuan, Izotov \& Lipovetsky 1995). A large fraction of these spectra ( $\sim$ 50\% )
showed broad emission characteristic of WN and WC stars. While some of these
objects are known to contain WR stars from previous studies,
the majority are newly discovered WR galaxies. We have assembled here,
from our large data base, a sample of 39 WR galaxies with the aim of studying 
their spectroscopic properties.
In particular, we wish to determine the $N$(WR)/$N$(O+WR) and $N$(WC)/$N$(WN) 
ratios for all these galaxies, and use the data to 
constrain models of massive star evolution. Section 2 describes the
observations and data reduction. 
In section 3 we discuss heavy element abundances for a subsample of WR galaxies, with previously unpublished observations. In Section 4 we
describe the WR features detected in our spectra, some of which have not been
discussed before in the literature.
In section 5 we present a new method for deriving
the numbers of WR and O stars. In section 6 we compare our results with other
observational data as well as with theoretical models, and discuss the effects
of metallicity and starburst age on the WR stellar population. 
In section 7 we discuss the origin of the nebular He II $\lambda$4686 line
emission. Section 8 contains a summary of our main results and conclusions.

\section {OBSERVATIONS AND DATA REDUCTION}

   Spectrophotometric observations of 38 galaxies with detected or
suspected broad WR emission features
were obtained  with the Ritchey-Chr\'etien spectrograph at the 
Kitt Peak National Observatory (KPNO) 4m telescope, and with the GoldCam 
spectrograph at the 2.1m KPNO telescope. The 4m observations 
were carried out during the period 1993--1994 while the 2.1m observations
were obtained in 1996. The majority of the galaxies in our sample were selected from the First Byurakan Survey
(Markarian et al. 1989), and from a complete sample of $\sim$ 250 BCDs
(Izotov et al. 1993) discovered in the Second Byurakan Objective Prism
Survey (SBS, Markarian, Lipovetsky \& Stepanian 1983). Four 
low-metallicity galaxies are from the Michigan Survey 
(Salzer, MacAlpine \& Boroson 1989). We have also included the BCD II Zw 40, 
known to possess WR features (Conti 1991). In addition to these
38 galaxies observed at KPNO, we have included the BCD I Zw 18
because of its extremely low metallicity. The spectrum of I Zw 18 was
obtained in 1997 with the Multiple Mirror Telescope (Izotov et al. 1997). 
Hence the total sample contains 39 galaxies.
Because of the high signal-to-noise ratio required to determine an accurate 
helium abundance, our spectra allow to detect WR features not only in the blue
region, but also in the rarely observed C IV $\lambda$5808 region. Several other lower intensity WR emission lines are also seen.

  Table 1 lists the studied galaxies with their IAU names, 
coordinates $\alpha$, $\delta$ at the 1950.0 epoch, apparent magnitudes 
$m_{pg}$ from Markarian et al. (1989) and Zwicky et al. (1961--1968), absolute 
magnitudes $M_{pg}$ ( a Hubble constant $H_0$ = 75 km s$^{-1}$ Mpc$^{-1}$ is 
adopted ), observed redshifts {\sl z}, oxygen abundances 12 + log(O/H) 
and other designations. 

The journal of observations is given in Table 2.
All observations were performed in the same way.
The slit was centered on the brightest part of each galaxy.
 Total exposure times
varied between 10 and 180 minutes. Each exposure was broken up into 2 -- 6
subexposures to allow for more effective cosmic ray removal, 
 except for SBS 0948+532 for which only one exposure was obtained.  
For the majority of galaxies, a 2\arcsec\ wide slit was used set at the  
position angle given in Table 2. The exception is I Zw 18 which was observed
with a 1\farcs5 wide slit. The spectral resolution was 6-7 \AA\ 
in all cases.
Except for three objects observed at airmass $\sim$ 1.6 and position angle 
90 degrees, the vast majority of galaxies were observed at small airmasses, 
so that no correction for atmospheric dispersion was performed.
Several
spectrophotometric standard stars were observed each night for flux calibration. Spectra of He-Ne-Ar 
comparison lamps were obtained before and after observation of each galaxy 
to provide wavelength calibration. Data reduction was performed with the
IRAF{\footnote[2]{IRAF: the Image Reduction
and Analysis Facility is distributed by the National Optical Astronomy
Observatories, which is operated by the Association of Universities for
Research in Astronomy, In. (AURA) under cooperative agreement with the
National Science Foundation (NSF).}} software package and included bias 
subtraction, cosmic ray removal, flat-field correction, flux and wavelength
calibrations, correction for atmospheric extinction and subtraction of the
night sky background. The residuals of the night sky lines after 
subtraction are $\leq$ 1\%. 

\section{HEAVY ELEMENT ABUNDANCES}

   Line intensities and heavy element abundances for the majority of
the WR galaxies have been
given previously in a series of papers on the 
primordial helium abundance (ITL94, ITL97, IT98) and on 
heavy element abundances (Thuan et al. 1995, Izotov \& Thuan 1999).
To derive heavy element abundances for the twelve remaining galaxies we follow 
the procedure described in ITL94 and ITL97. 

The extracted one-dimensional spectra of the brightest regions in these twelve 
galaxies are shown in Figure 1. All emission line fluxes 
were corrected for interstellar extinction with the use of
the extinction coefficient $C$(H$\beta$) derived from the 
Balmer decrement. For this purpose we adopted the extinction curve of 
Whitford (1958) as fitted by ITL94. 
The theoretical ratios of hydrogen Balmer emission lines are taken from
Brocklehurst (1971) at the electron temperature derived from the observed
ratio [O III]($\lambda$4959 + $\lambda$5007) / $\lambda$4363 when the
auroral emission line $\lambda$4363 is detected. 
This is the case for 30 out of the 39 WR galaxies in our sample.
The electron temperature
for the nine galaxies with non-detected [O III]$\lambda$4363
was derived from the empirical calibration of total oxygen emission line flux
[O II]$\lambda$3727 + [O III]($\lambda$4959 + $\lambda$5007) vs. electron
temperature (Pagel et al. 1979). The observed 
$F$($\lambda$) and corrected $I$($\lambda$) emission line fluxes relative
to the H$\beta$ emission line fluxes for the 12 galaxies listed in Table 2 
are shown in Table 3.
Also given in Table 3 are the extinction coefficient $C$(H$\beta$), the observed absolute flux of the H$\beta$ emission line and its equivalent width
$EW$(H$\beta$) and the equivalent width of the hydrogen absorption lines 
$EW$(abs). 
The latter is set to be the same for all hydrogen lines.
We note that the intensity of the He I $\lambda$5876 emission line
in II Zw 40 is reduced, because of absorption by Galactic neutral 
sodium, just as in the case of I Zw 18 (Izotov \& Thuan 1998a). 
The He I $\lambda$5876 emission line in Mrk 178 is contaminated by the strong 
nearby WR C IV $\lambda$5808 emission line (Figure 1),
while this line in the spectrum of Mrk 1329
is on a bad CCD column, making its intensity unreliable.

The ionic and total heavy element abundances of the four galaxies 
in Table 2 with detected
[O III] $\lambda$4363 emission line are shown in Table 4. The oxygen abundance
12 + log (O/H) derived in these galaxies ranges from 7.82 to 8.23 and
is in fair agreement with previous determinations. For Mrk 178,
Gonzalez-Riestra et al. (1987) derived oxygen  12 + log(O/H) = 7.72 
as compared to our value of 7.82. For II Zw 40, Walsh \& Roy (1993), 
Masegosa, Moles \& Campos-Aguilar (1994) and Martin (1997) derived respectively 
an oxygen abundance of 8.25, 8.18 and 8.12 compared to our value of 8.09. 
As for Mrk 1236, Vacca \& Conti (1992) derived 12 + log(O/H) = 8.09 
as compared to our value of 8.07. The element abundance ratios in 
these four galaxies (Table 4) are in good agreement with typical ratios 
in low-metallicity blue compact dwarf galaxies (Izotov \& Thuan 1999).

We adopt the calibration between oxygen abundance
and [N II]($\lambda$6548 + $\lambda$6584) to H$\alpha$ $\lambda$6563 flux
ratio from van Zee et al. (1998) to derive oxygen abundance in the nine 
galaxies with nondetected [O III] $\lambda$4363
\begin{equation}
12 + \log ({\rm O/H}) = 1.02 \log ({\rm [N II]/H}\alpha) + 9.36.
\end{equation}
This calibration is insensitive to uncertainties in interstellar extinction
and agrees with that of Edmunds \& Pagel (1984), to within 0.1 dex in the
derived oxygen abundance. 
The oxygen abundances for the galaxies with nondetected [O III]
$\lambda$4363 
are shown in Table 1. It can be seen that they are considerably larger than 
those of galaxies with detected [O III]$\lambda$4363,
exceeding in some cases the solar oxygen abundance. 
The oxygen abundance of 9.03 derived us for Mrk 710 is in good agreement with
the values of 9.09 obtained by Vacca \& Conti (1992) and that of
9.00 obtained by Schaerer et al. (1999a).

Examination of Table 1 shows that the WR galaxies in our sample span 
an oxygen abundance range extending over two orders of magnitude ($Z_\odot$/50
to 2$Z_\odot$). This unique sample allows us to study 
the properties of WR stellar populations in a wide range of heavy 
element abundances.

\section{WR EMISSION FEATURES}

   We now discuss the properties of WR galaxy spectra. In low-resolution
spectra of WR galaxies the blue bump at $\lambda$4650 is most often seen.
This unresolved bump is a blend of the N V 
$\lambda$4605, 4620, N III $\lambda$4634, 4640, C 
III/C IV $\lambda$4650, 4658 and He II $\lambda$4686 broad WR lines. 
These are emitted mainly by late WN (WNL) and early WC (WCE) stars,
although some contribution of early WN (WNE) stars might be present 
(SV98). Superposed on the blue bump are much narrower [Fe III] $\lambda$4658, He II $\lambda$4686, He I + [Ar IV] $\lambda$4711 and [Ar IV] 
$\lambda$4740 nebular emission lines. The detectability of 
the weaker red bump, emitted mainly by WCE stars, is lower.
Hence, the observed characteristics of WR galaxies are restricted to a 
narrow range of WR star subtypes, most often WNL and WCE stars.
 
Can other WR subtypes be seen
in integrated spectra of WR galaxies? Each WR star, depending on the 
metallicity and mass of the progenitor, may evolve through different WR 
subtypes from WNL to WCE. WNE stars emit
N V $\lambda$4605, 4620, C IV $\lambda$4658, He II $\lambda$4686 and C IV 
$\lambda$5808 (e.g. Smith et al. 1996)
and hence cannot be distinguished from other WR stars in 
low-resolution spectra. Additionally, WNE stars are not as luminous as
WNL stars in the optical range. The WR star lifetime in the WNE 
stage is short compared to that in 
the WNL stage for nearly all metallicities and progenitor star masses (Maeder \& Meynet 1994). Therefore, in the majority of cases
the contribution of WNE stars to the blue bump is expected to be small.
However, evolution models of Maeder \& Meynet (1994) predict that at high 
metallicities and low WR progenitor star masses, the WNE lifetime is greater
than the WNL one. Hence, at late stages of instantaneous bursts in metal-rich
WR galaxies the contribution of WNE stars can be significant. 

The C III $\lambda$4650, $\lambda$5696 emission lines are
seen mainly in late-type WC stars (WC7 -- WC9). No normal WC star of type later 
than WC9 has been detected. However, 
some central stars of planetary nebulae have characteristics of WR stars
and are classified as [WC10] -- [WC12] (e.g. Leuenhagen \& Hamann 1994, 
1998; Leuenhagen, Hamann \& Jeffery 1996). In the spectra of these stars, C II 
$\lambda$4267 emission and some other C II lines are observed.  
It is thought that late WC stars are to be seen only in regions 
with metallicity greater than solar (Smith 1991; Smith \& Maeder 1991;
Phillips \& Conti 1992). Since the majority of starburst WR galaxies
have lower metallicities than solar, they are not expected to contain
WCL stars. 

As for the late WN (WN9 -- WN11) stars, they produce N II, N III 
and Si III lines (e.g. Smith et al. 1995; Crowther et al. 1997; Crowther \& 
Smith 1997) and are bright, therefore they should be detected in the
spectra of WR galaxies. However, high signal-to-noise ratio spectra are
required because these WNL features are weak. Probably, the first detection
of very late WN and WC stars was done by Kunth \& Schild (1986) in
the galaxy Tol 9. They saw N II $\lambda$5679, $\lambda$5747-67
and C III $\lambda$5696 and attributed those lines to WN10 and WC7 stars 
respectively. The presence of WC7 stars in Tol 9 is surprising because its
metallicity was lower than solar ($Z_\odot$/15). Later, Phillips \& Conti (1992) and Schaerer et al. (1999a) detected the C III $\lambda$5696 emission line in 
the high-metallicity galaxies NGC 1365 ( $\sim$ 3$Z_\odot$ ) 
and Mrk 710 ( 1.2$Z_\odot$ ) respectively. 

   In Figure 2 we show spectra of our sample of 39 WR galaxies. All spectra have been corrected to rest wavelengths and are restricted to the wavelength
range $\lambda$$\lambda$4150 -- 6100\AA\ which contains the WR features
of interest. The spectra are of varying quality depending on the brightness of 
the galaxy, exposure time and the size of the telescope used (2.1, 4m
or MMT). Identifications
for certain or suspected WR emission lines are given for each spectrum.
Many of the emission lines represent blends of several lines.
For simplicity, we mark the most often detected N III 
$\lambda$4511--4534 blend as N III $\lambda$4512, the Si III
$\lambda$4552--4576 blend as Si III $\lambda$4565, N V $\lambda$4605, 4620
as N V $\lambda$4619 and N III $\lambda$4634, 4640 as N III
$\lambda$4640. From those identifications, it is clear that several
new lines have been detected in the spectra of some of our WR galaxies which 
have rarely or never been seen before in the spectra of WR galaxies. 
Those new line identifications constitute one of the most important results
of our work. We discuss in more detail the emission WR features observed in each galaxy in the following.

{\sl 0112--011 $\equiv$ UM 311.} --- Masegosa, Moles \& del Olmo (1991)
first noticed the presence of a WR population in this galaxy. 
IT98 detected both blue and red bumps. The unresolved blue bump is
particularly strong in this galaxy. Additionally, N III $\lambda$4512, Si III $\lambda$4565 and He I/N II $\lambda$5047 are present, which are
characteristic of WN7 -- WN8  and WN9 -- WN10 stars respectively
(e.g. Massey \& Conti 1980, 1983; Crowther \& Smith 1997).
In the red part of the spectrum, the C IV $\lambda$5808 emission line
is clearly detected which indicates the presence of WCE stars.

{\sl 0207--105 $\equiv$ Mrk 1026.} --- Only two possible WR features N II
$\lambda$4620, $\lambda$5720--40 and one certain C IV $\lambda$5808 line
are detected in this galaxy. Most likely, these features are due to WNL and WCE 
stars. However, the spectrum is too noisy to allow definite conclusions.

{\sl 0211+038 $\equiv$ Mrk 589.} --- In the blue region, several features are 
present which we identify as N III $\lambda$4512, Si III $\lambda$4565, 
N V $\lambda$4619, 
N III $\lambda$4640, He II $\lambda$4686 and Si II $\lambda$5056 
emission lines. The C IV $\lambda$5808 line is present in the red part of 
spectrum. These lines suggest the presence of WCE, WNE and WNL stars.

{\sl 0218+003 $\equiv$ UM 420.} --- Noisy spectrum. However, broad 
C IV $\lambda$4658, He II $\lambda$4686 and C IV $\lambda$5808 indicate
the possible presence of WCE stars.

{\sl 0252--102 $\equiv$ Mrk 1063.} --- Broad bump consisting of 
N V $\lambda$4619,   
N III $\lambda$4640, He II $\lambda$4686. Additionally, N III $\lambda$4512, 
Si III $\lambda$4565 and Si II $\lambda$5046 are observed in the 
blue region. The C IV $\lambda$5808 line is observed in the red region. 

{\sl 0459--043 $\equiv$ NGC 1741 $\equiv$ Mrk 1089.} ---
Kunth \& Schild (1986) detected N III $\lambda$4640 and 
He II $\lambda$4686 lines. These features are also present in
our spectrum. Additionally, we note the presence of N III $\lambda$4512 and
Si III $\lambda$4565 emission lines. In the red part of the spectrum
N II $\lambda$5720-40 emission is detected. 
The WR population is dominated by WNL stars. 

{\sl 0553+033 $\equiv$ II Zw 40.} --- Kunth \& Sargent (1981) detected a
broadened He II $\lambda$4686 emission line. No broad WR emission was seen
by Walsh \& Roy (1993). Our observations reveal the
presence of Si III $\lambda$4565, N III $\lambda$4640/C IV $\lambda$4658,
He I/N II $\lambda$5047, N II $\lambda$5720-40 and C IV $\lambda$5808
emission lines. The WR lines in the blue bump are contaminated by the strong
nebular lines [Fe III] $\lambda$4658, He II $\lambda$4686, He I + [Ar IV]
$\lambda$4711 and [Ar IV] $\lambda$4740.

{\sl 0635+756 $\equiv$ Mrk 5.} --- Noisy spectrum. The N III $\lambda$4640 and 
He II $\lambda$4686 emission lines are detected, implying
the presence of WNL stars. 

{\sl 0720+335 $\equiv$ Mrk 1199.} --- 
The N III $\lambda$4512, Si III $\lambda$4565, N V $\lambda$4619, 
C IV $\lambda$4658, He II $\lambda$4686, Si II $\lambda$5056, 
N II $\lambda$5720-40 and C IV $\lambda$5808 emission lines
are detected. This implies that WNL, WNE and WCE populations are present.

{\sl 0723+692 $\equiv$ NGC 2363 $\equiv$ Mrk 71.} --- Roy et al. (1992) have
discussed the origin of the broad emission in nebular lines. The broad
blue bump was present in their spectrum but not noticed. 
Gonzalez-Delgado et al.
(1994) have detected both blue and red bumps. ITL97 confirmed
the finding by Gonzalez-Delgado et al. (1994). 
The Si III $\lambda$4565, N III $\lambda$4640,
C IV $\lambda$4658, He II $\lambda$4686, He I/N II $\lambda$5047 
and C IV $\lambda$5808 emission lines are present in
our high signal-to-noise ratio spectrum, indicating the presence of WNL and
WCE stars. The blue bump is strongly contaminated by nebular emission.

{\sl 0842+162 $\equiv$ Mrk 702.} --- Broad WR bumps are detected in 
the blue and red regions. The N III $\lambda$4512, Si III $\lambda$4565,
N V $\lambda$4619, N III $\lambda$4640, He II $\lambda$4686 and 
C IV $\lambda$5808 emission lines are seen. The WNL and WNE stars dominate in 
this galaxy.

{\sl 0926+606 $\equiv$ SBS 0926+606.} --- ITL97 first noticed the presence of WR stars in this galaxy. Reasonably strong Si III $\lambda$4565 and He I/N II 
$\lambda$5047 emission lines are detected indicating the 
presence of WN9 -- WN10 stars. There is possibly a N II $\lambda$5720--40 
emission feature. The N III $\lambda$4640 emission line is relatively weak, 
while the N III $\lambda$4512 line is not seen. The blue bump is strongly
contaminated by nebular emission.

{\sl 0930+554 $\equiv$ I Zw 18 $\equiv$ Mrk 116.} --- Izotov et al. (1997) and
Legrand et al. (1997) detected both the blue and red bumps in this BCD, the most
metal-deficient WR galaxy known. Izotov et al. (1997) also saw N III
$\lambda$4512 and Si III $\lambda$4565 emission lines. However, they
misidentified the latter line as Mg I $\lambda$4571.
Possibly, N II $\lambda$5720--40 and $\lambda$5928--52 lines are detected, 
indicating the presence of WNL stars. In contrast, Legrand et al. (1997) 
who observed with a different position angle of the slit found
evidence only for WCE stars.

{\sl 0946+558 $\equiv$ Mrk 22.} --- The WR stellar population in this galaxy
was detected by ITL94. The main features are N III $\lambda$4512, 
N III $\lambda$4640, broadened He II $\lambda$4686 and C IV $\lambda$5808.
The WR population is dominated by WNL and WCE stars.

{\sl 0947+008 $\equiv$ Mrk 1236.} --- 
A broad He II $\lambda$4686 emission line
in this galaxy was seen by Kunth \& Schild (1986) and confirmed by 
Vacca \& Conti (1992). However, those authors did not show the spectrum
of the galaxy. The most striking features in our spectrum are the
C II $\lambda$4267 and C III $\lambda$5696 emission lines which we
attribute to late WC stars. We also identify the feature seen at 4620\AA\ as a 
possible C II $\lambda$4620 emission line.
The presence of these features in a low-metallicity BCD such as Mrk 1236
( its oxygen abundance is 12 + log (O/H) = 8.06 or $\sim$ $Z_\odot$/7 )
is unexpected. New observations are required to confirm our
finding. Other lines are the C IV $\lambda$4658 / He II 
$\lambda$4686 blend and the C IV $\lambda$5808 emission
line, indicating the presence of WCE stars. No strong 
nitrogen lines are present, although the spectrum is noisy.

{\sl 0948+532 $\equiv$ SBS 0948+532.} --- Broad WR emission was
detected by ITL94. The Si III $\lambda$4565, N III $\lambda$4640 and
He II $\lambda$4686 lines are seen, indicating the presence
of WNL stars. Probably, N II $\lambda$5720--40 and C IV $\lambda$5808 
are also present, however the spectrum is noisy. The blue bump is strongly
contaminated by nebular emission.

{\sl 0952+095 $\equiv$ NGC 3049 $\equiv$ Mrk 710.} --- Kunth \& Schild (1986)
detected broad N III $\lambda$4640, He II $\lambda$4686 and C IV
$\lambda$5808 emission lines. A broad WR bump in the blue region was also
seen by Vacca \& Conti (1992). Schaerer et al. (1999a) discussed
the presence of N III $\lambda$4640, He II $\lambda$4686, C III $\lambda$5696 
and C IV $\lambda$5808 broad features indicating the presence of early and
late WC and late WN stars. We confirm their findings. Additionally, we
find N III $\lambda$4512, Si III $\lambda$4565 and N V $\lambda$4619 
features implying the presence of WNE and WNL stars. 

{\sl 1030+583 $\equiv$ Mrk 1434.} --- A broad He II $\lambda$4686 
emission line was detected by ITL97. The 
Si III $\lambda$4565 and C IV $\lambda$5808 emission lines are seen in this
low-metallicity galaxy ( 12 + log (O/H) = 7.79 or $Z$ $\sim$ $Z_\odot$/14 ).

{\sl 1036--069 $\equiv$ Mrk 1259.} --- Ohyama, Taniguchi \& Terlevich (1997)
identified broad N V $\lambda$4619, N III $\lambda$4640 and He II 
$\lambda$4686 emission lines. We confirm the presence of these lines
in our higher signal-to-noise ratio spectrum. Additionally, we identify
N III $\lambda$4512, N III $\lambda$4905 and Si II $\lambda$5056 features.
We find also the C IV $\lambda$5808 emission line in the red part of the 
spectrum.

{\sl 1053+064 $\equiv$ Mrk 1271.} --- IT98 detected a broad blue bump.
Here we identify the N III $\lambda$4512, Si III $\lambda$4565, N II
$\lambda$4620
and He II $\lambda$4686 emission lines. Possibly, C IV $\lambda$5808 is
also present, although higher signal-to-noise ratio observations are required
to be certain. The blue bump is strongly contaminated by nebular emission.

{\sl 1054+365 $\equiv$ CG 798.} --- ITL97 noticed a broad He II $\lambda$4686
emission line. Here we identify additionally Si III $\lambda$4565 
and possibly C III $\lambda$5696 emission lines.

{\sl 1130+495 $\equiv$ Mrk 178.} --- Gonzalez-Riestra, Rego \& Zamorano (1988)
detected a broad He II $\lambda$4686 emission line. In our high signal-to-noise
ratio spectrum the WR features are very pronounced. We identify
them with the N III $\lambda$4640, C IV $\lambda$4658 and He II $\lambda$4686
lines in the blue region, and with the C IV $\lambda$5808 emission line in
the red region. 

{\sl 1134+202 $\equiv$ Mrk 182.} --- Noisy spectrum. However, a broad WR 
emission is present in the blue part of the spectrum.

{\sl 1135+581 $\equiv$ Mrk 1450.} --- ITL94 noted broad WR features in the
blue and red regions. We identify the N III $\lambda$4512, Si III $\lambda$4565, N III $\lambda$4640, He II $\lambda$4686, N III $\lambda$4905,
He I/N II $\lambda$5047 and C IV $\lambda$5808 lines. The blue bump is strongly
contaminated by nebular emission.

{\sl 1139+006 $\equiv$ Mrk 1304 $\equiv$ UM 448.} --- Broad He II $\lambda$4686
was detected by Masegosa et al. (1991). Our observations show that
the intensities of WR features are atypical. Although
broadened, the He II $\lambda$4686 line is weak, while the N III $\lambda$4512,
Si III $\lambda$4565, N II $\lambda$4620, He I/N II $\lambda$5047, 
N II $\lambda$5720--40, C IV $\lambda$5808 lines are strong.

{\sl 1140--080 $\equiv$ Mrk 1305.} --- Noisy spectrum. N V $\lambda$4619,
C IV $\lambda$4658, He II $\lambda$4686 and C IV $\lambda$5808 features
are present. Higher signal-to-noise ratio observations are necessary to confirm
the detection of the WR features.

{\sl 1147+153 $\equiv$ Mrk 750.} --- Kunth \& Joubert (1985) found
broad He II $\lambda$4686 while Conti (1991) noted the
existence of N III $\lambda$4640. IT98 confirmed the presence of WR features.
The signal-to-noise ratio of the spectrum is relatively
low. C II $\lambda$4267, Si III $\lambda$4565, C II $\lambda$4620 are 
possibly present. The C IV $\lambda$4658 emission line is 
blended with the He II $\lambda$4686 line. Additionally, the C IV $\lambda$5808 line is detected.

{\sl 1150--021 $\equiv$ Mrk 1307 $\equiv$ UM 462.} --- The presence of WR stars 
in this galaxy is suspected. The He II $\lambda$4686 emission line is broadened. The Si III $\lambda$4565 and He I/N II $\lambda$5047 lines are detected. 

{\sl 1152+579 $\equiv$ Mrk 193.} --- Only Si III $\lambda$4565 and N II
$\lambda$5720--40 are fairly well detected. 
The presence of the N II $\lambda$4620, N III $\lambda$4640 
and He I/N II $\lambda$5047 lines is suspected. The blue bump is strongly
contaminated by nebular emission.

{\sl 1211+540 $\equiv$ SBS 1211+540.} --- Only the broadened He II $\lambda$4686
line is detected fairly well. The signal-to-noise ratio of the spectrum is 
too low to definitely detect other WR features in this very low-metallicity 
galaxy ( 12 + log (O/H) = 7.64 or $Z$ $\sim$ $Z_\odot$/19 ).

{\sl 1222+614 $\equiv$ SBS 1222+614.} --- ITL97 noted the presence
of both blue and red bumps. We identify N III $\lambda$4512,
Si III $\lambda$4565, He II $\lambda$4686, N II $\lambda$5720--40, 
C IV $\lambda$5808 emission lines. The N III $\lambda$4640 and C IV 
$\lambda$4658 lines are blended.

{\sl 1223+487 $\equiv$ Mrk 209.} --- ITL97 noted the presence of a blue
bump in this galaxy. We identify a broadened He II $\lambda$4686 emission line,
and weak N III $\lambda$4512, He I/N II $\lambda$5047 and C IV $\lambda$5808 
emission lines. The Si III $\lambda$4565 line is strong and the N II 
$\lambda$5720--40 emission line is suspected. The blue bump is strongly
contaminated by nebular emission.

{\sl 1234+072 $\equiv$ Mrk 1329.} --- Noisy spectrum. Strong broad 
He II $\lambda$4686 is present. We also identify N III 
$\lambda$4512, He II $\lambda$4541, Si III $\lambda$4565, N III $\lambda$4640, 
He I/N II $\lambda$5047, He II $\lambda$5411 and weak C IV $\lambda$5808 
emission lines.

{\sl 1249+493 $\equiv$ SBS 1249+493.} --- Noisy spectrum. Two broad features
(C IV $\lambda$4658 and He II $\lambda$4686) are probably present in this
low-metallicity galaxy ( 12 + log (O/H) = 7.72 or $Z$ $\sim$ $Z_\odot$/15 ). 
Further observations are required to confirm the presence of WR stars.

{\sl 1256+351 $\equiv$ NGC 4861 $\equiv$ Mrk 59.} --- Both blue and red
bumps have been detected by Dinerstein \& Shields (1986) and ITL97. 
We identify in our high quality spectrum 
N III $\lambda$4512, Si III $\lambda$4565, N V $\lambda$4619,
He II $\lambda$4686, He I/N II $\lambda$5047 and C IV $\lambda$5808 broad 
emission lines. The N III $\lambda$4640 and C IV $\lambda$4658 lines are 
blended. The blue bump is strongly contaminated by nebular emission.

{\sl 1319+579A $\equiv$ SBS 1319+579A.} --- Noisy spectrum. 
WR features were noted by ITL97. We detect the Si III
$\lambda$4565, C IV $\lambda$4658, He II $\lambda$4686,
He I/N II $\lambda$5047, N II $\lambda$5720-40 and C IV $\lambda$5808 emission 
lines which suggest the presence of WNL and WCE stars. The blue bump is strongly
contaminated by nebular emission.

{\sl 1437+370 $\equiv$ Mrk 475.} --- Conti (1991) noted the presence of a broad
He II $\lambda$4686 emission line and possibly of the N III $\lambda$4640
emission line. ITL94 detected strong blue and red bumps.
We identify Si III $\lambda$4565, He II $\lambda$4686, He I/N II $\lambda$5047 
and C IV $\lambda$5808 emission-line
features. Possibly, the N III $\lambda$4640 line is present as well.
The blue bump is strongly contaminated by nebular emission.

{\sl 1533+574B $\equiv$ SBS 1533+574B.} --- Noisy spectrum. We detect the
Si III $\lambda$4565, N III $\lambda$4640, He II $\lambda$4686, 
He I/N II $\lambda$5047, N II $\lambda$5720--40 and C IV $\lambda$5808 emission 
lines. The blue bump is strongly contaminated by nebular emission.

{\sl 2329+286 $\equiv$ Mrk 930.} --- Noisy spectrum. We identify the N III 
$\lambda$4512, Si III $\lambda$4565, He II $\lambda$4686, He I/N II 
$\lambda$5047, N II $\lambda$5720--40 and C IV $\lambda$5808 emission lines.

\section {DERIVATION OF THE NUMBER OF WR AND O STARS}

  The following quantitative analysis of massive stellar populations in
WR galaxies is based on only 33 objects. We have excluded from the original 
sample of 39 WR galaxies ( Table 1) the 6 galaxies 0207--105, 1134+202, 
1140--080, 1152+579, 1249+493, 1533+574B because their spectra were noisy 
and their WR features were either weak or uncertain.

\subsection{Flux measurements of H$\beta$ and WR emission lines} 

   In this section we describe our procedure for determining  
the number of O and WR stars from the spectroscopic data. 
In general, the number of WR stars is derived from the luminosity of a
WR line, while the number of O stars is deduced from the H$\beta$ luminosity
after subtracting the contribution of WR stars from it.
Because the H$\beta$ emission is extended and the slit usually does not 
cover the whole region of ionized gas emission, care should be exercised
to correct for lost light. Therefore, one-dimensional spectra with one
pixel step along the spatial axis of the slit have been extracted from the
two-dimensional spectrum. The H$\beta$ fluxes were then measured for
each one-dimensional spectrum where this line was detected.
As for the equivalent width of H$\beta$, it was measured in the one-dimensional
spectrum extracted in the largest possible aperture with diameter equal to the
whole spatial extent of the slit.
The WR fluxes are derived from one-dimensional spectra with aperture covering 
the regions where WR emission is present. 
The measured H$\beta$ flux cannot be used directly for the determination 
of the number of ionizing O stars because of the
aperture effect discussed by Conti (1991) and Vacca \& Conti (1992):
while the region with the brightest H$\beta$ emission where the stellar cluster
is located is usually covered by the slit, a significant fraction of the 
H$\beta$ emission from the extended H II region may be outside of it. 
H$\beta$ emission can spread far away from the center of the O and WR star
cluster. 

To take into account this effect the following procedure has been adopted. 
We assume that
the brightness distribution of each object has a circular symmetry,
with the center of the cell with the maximum H$\beta$ flux ( all H$\beta$ fluxes
are extinction-corrected ) chosen to be the center of symmetry. 
However, if the flux in adjacent cells exceeds 0.7 
the flux in the brightest cell, the 
center was shifted by half a cell in the direction of that cell. Since the
distribution of fluxes on opposite sides from the 
center would generally not be equal, we consider each object as a set of
homogeneous half-rings. This yields a correction factor $C^i_{cor}$
for each cell $i$ along the slit of the form:
\begin{equation}
C^i_{cor}= \left\{ \begin{array}{ll}
\frac{\pi(r^2_{max}-r^2_{min})}{2ab}, & 
\mbox{if $\frac{r_{min}+r_{max}}{2} \ge \frac{b}{2}$, } \\
1,                                    &
\mbox{otherwise. }
\end{array}
\right. 
\end{equation}
Here $a$ and $b$ are the cell sizes along and across the slit expressed in 
arcsec, $r_{max}$ and $r_{min}$ are the distances  from the center of symmetry
to the near and far edges of the cell. $C^i_{cor}$ is thus just the ratio of the area of a half-ring to that of a cell. 
This procedure is used for all cells along the slit. The H$\beta$ fluxes in
each half-ring are then summed together to give the extinction and
aperture-corrected H$\beta$ flux and the mean correction factor $C_{cor}$ is 
derived as the ratio of the extinction and aperture-corrected
H$\beta$ flux to the H$\beta$ flux corrected only for extinction. 
The corrected H$\beta$ fluxes are shown in Table 5 together with the correction 
factors $C_{cor}$. Here $F_{cor}$(H$\beta$), is the
absolute flux, corrected for interstellar extinction 
and aperture, while $F$(H$\beta$) is the absolute flux measured along
the slit and corrected only for interstellar extinction. We also show in Table 5 other relevant parameters for the H II regions:
the coefficient for interstellar extinction $C$(H$\beta$) derived from the
observed Balmer decrement in the spectrum of the brightest
region, the equivalent width of the H$\beta$ emission line, 
the age of the star formation burst derived from $EW$(H$\beta$) and the
parameter $\eta_0$ to be discussed below.
 $EW$(H$\beta$) is also subject to aperture effects and its aperture
correction factor may be as large as $C_{cor}$ for the H$\beta$ flux. 
Therefore, H$\beta$ equivalent widths of star-forming regions not corrected for aperture effects
are lower limits and ages of star formation bursts derived from 
noncorrected $EW$(H$\beta$) should be  considered as an upper limits. 
We cannot however
derive correction factors for $EW$(H$\beta$) from the spectra alone, 
as the underlying galaxy can also contribute to the continuum and decrease
$EW$(H$\beta$) as compared to the case where the continuum is produced by 
the star-forming region only.

 The above aperture correction procedure has not been applied to the 
WN and WC bump fluxes
because that emission comes from very compact regions with
angular sizes usually not greater than the 
slitwidth. Those fluxes have been measured instead in the integrated 
spectrum obtained by summing in the spatial direction all one-pixel-sized  
one-dimensional spectra where WR emission features are detected.
In only five galaxies 0635+756, 0952+095, 1030+583, 1053+064 and 1211+540, is 
the angular size of WR regions along the slit slightly larger than 
the slit width. For those galaxies, the measured fluxes of WR features should be considered as lower limits.
The correction factors are however very close to unity.
  
  A careful subtraction of the continuum is essential for
deriving accurate WR bump fluxes and hence numbers of WR stars.
To define the continuum, we carefully select several points in spectral 
regions free of nebular and stellar lines. The continuum is then fitted by
cubic splines. The quality of the continuum fit is visually checked and if
deemed satisfactory is subtracted from the spectrum. 
 
On top of the blue bump are superposed [Fe III] $\lambda$4658, He II 
$\lambda$4686, [Ar IV] + He I $\lambda$4712 and [Ar IV] $\lambda$4740 narrow 
nebular emission lines. These nebular lines have been subtracted from the
bump using the IRAF SPLOT software package. We then measure the fluxes of the 
broad components within a specified wavelength range which is varying 
because of the changing appearance of broad components in galaxies with
different burst ages. Similar measurements have been done for other weaker 
WR lines. The results of the measurements are shown in Table 6. Here
$F$($\lambda$) and $EW$($\lambda$) are the total fluxes corrected for 
interstellar extinction and equivalent widths of different WR lines.

\subsection{The number of WR stars}

 We estimate the number of WR stars from the luminosity of the blue
($\lambda$4650) and red ($\lambda$5808) bumps. 
In principle, if the luminosity of one WR star in a specific
broad line or in the whole bump is known, we can derive the number of WR stars:
 \begin{equation}
N_{\rm WR}=\frac{L_{\rm WR}}{L^0_{\rm WR}}
\end{equation}
where $L_{\rm WR}$ is the absolute luminosity of the WR bump 
(or of one WR line) corrected for interstellar
extinction, and $L^0_{\rm WR}$ is the 
luminosity of the WR bump (or of one WR line) of a single WR star. 
Usually, WC4 stars are considered as
representatives of WCE stars, while WN7 stars are considered as
representatives of WNL stars. In the following, to derive the numbers of
WR stars we shall use parameters relevant to WC4 and WN7 stars. However, one should keep in mind that WNE stars can contribute to the blue bump 
at late WR stages in high-metallicity galaxies.

The luminosity of the red bump includes only the C IV 
$\lambda$5808 line and gives directly the number of WCE stars. The situation 
with the blue bump is more complicated. To determine the number of WNL stars, 
the contribution of WCE stars to the blue bump should be removed. 
To do this we use the luminosity of the red bump and introduce the coefficient
\begin{equation}
k=\frac{L_{\rm WC4}(\lambda4658)}{L_{\rm WC4}(\lambda5808)}.
\end{equation}
Then the luminosity of WC4 stars which should be subtracted from the total 
luminosity of the blue bump is equal to $k$$L_{\rm WC4}$($\lambda$5808). 
The value of the coefficient $k$ not well known. Smith (1991) gives
$k$ = 1.52, while SV98 obtain $k$ = 
1.71$\pm$0.53, consistent with the Smith value within the errors.
The value of $k$ is uncertain because the relative fluxes of 
individual lines vary strongly even within the same WR subtype (see
Tables 1 and 2 in SV98). 

Luminosities of single WR stars are also known
poorly. Smith (1991) adopts a luminosity of 3.2$\times$10$^{36}$ ergs s$^{-1}$ 
for a single WNL star in the blue bump and for a single WC4 star in the red bump from observations of WR stars in the Large Magellanic Cloud, 
while Vacca \& Conti (1992) favor a luminosity of
2.5$\times$10$^{36}$ ergs s$^{-1}$ for a single WC4 star in the red bump.
SV98 have compiled the luminosities of
WR stars in different emission lines. They find the luminosities of single
WNL and WC4 stars in the red bump to be 9.9$\times$$10^{34}$ and
3.0$\times$$10^{36}$ ergs s$^{-1}$ respectively. 
Furthermore, the luminosity of a single WNL 
star in the blue bump is metallicity-dependent because of
the varying contribution of the N III $\lambda$4640 emission line, which is
smaller at lower metallicity. The luminosity of a single WNL star in the blue 
bump is equal to (2.0 -- 2.6)$\times$10$^{36}$ ergs s$^{-1}$ in the range of 
heavy element mass fraction from $Z$ = 0.007 (LMC) to $Z$ = 0.02 (Milky Way), 
while the contribution of a single WC4 star to the blue bump is 
5.1$\times$$10^{36}$ ergs s$^{-1}$. 

To derive the number of WCE and WNL stars we have adopted the calibration by 
SV98. For this purpose, only a few lines are needed. We use the flux of the C IV $\lambda$5808 line to derive the number of WCE stars. 
In the blue region, to derive the number of WNL stars we measure the flux of the whole bump including the unresolved N III $\lambda$4640, C III / C IV 
$\lambda$4650 / $\lambda$4658 and He II $\lambda$4686 emission lines, and 
subtract from it the 
flux contributed by WCE stars. We adopt the luminosity of a single WNL star
in the blue bump to be 2.0$\times$10$^{36}$ ergs s$^{-1}$ for $Z$ $<$ $Z_\odot$ 
and 2.6$\times$10$^{36}$ ergs s$^{-1}$ for
$Z$ $\geq$ $Z_\odot$ (SV98). For a single WCE star,
a luminosity $L$(C IV $\lambda$5808) of 3.0$\times$10$^{36}$ ergs s$^{-1}$
and a coefficient $k$ = 1.71 are adopted. 
The numbers of WNL and WCE stars so derived are given in Table 7.

    This technique does not always give reasonable results because of 
observational uncertainties and uncertainties in the adopted 
values of $k$ and the luminosities of single WR stars.
The number of WCE stars in the two galaxies 1211+540 and 1223+487 is less
than unity. The spectra of the galaxies 0218+003 and 2329+286 are noisy. This
leads to too low or even negative numbers of WNL stars. In those cases,
another technique to be discussed in the following is used.
 
\subsection{The number of WNL and WCL stars from the weak WR lines}

   The detection of many weak features in the spectra of our galaxies
allows us to derive the number of WR stars by another independent method and
even to study the distribution of stars with different subtypes.
Our spectra show mainly features of WNL stars, most often
N III $\lambda$4512 which is characteristic of WN7 -- WN8 stars.
The Si III $\lambda$4565 line present 
in WN9 -- WN11 stars is also often seen. 
Alternatively, this line and some other permitted lines such
as N II $\lambda$4620, Si II $\lambda$5056 could be nebular in origin
and excited by absorption of UV starlight or strong UV 
He I line emission (e.g., Grandi 1976). However in our objects, the
intensities of these lines relative to H$\beta$ (Tables 5 -- 6) 
are in general several times larger than those predicted by Grandi. 
Thus, despite a possible contribution of nebular emission, 
the Si III $\lambda$4565 line is most likely related to WR stars.
This line and the N III $\lambda$4512 line are not contaminated by 
the emission of other WR subtypes
and can in principle yield quite reliable determinations of the
numbers of WN stars. In contrast, the determination of the number of the
WNL stars from the blue bump at $\lambda$4650 is subject to more uncertainties 
caused by the contamination from WCE and WNE stars, strong nebular lines and 
uncertainties in the poorly known intensity ratio 
C IV $\lambda$4658/$\lambda$5808 (SV98).

    However, the intensities of weak N III $\lambda$4512 and Si III 
$\lambda$4565 in single stars are not well known. We estimate from the
spectrum of a WN8 star by Massey \& Conti (1980) that the strength of
N III $\lambda$4512 is $\sim$ 3 times weaker than the strength of 
N III $\lambda$4640 + He II $\lambda$4686 lines. Similar values
are obtained by Crowther et al. (1995a) from the spectra of 
WN9 stars for both the N III $\lambda$4512 and Si III $\lambda$4565 lines.
Therefore, we adopt for the luminosity of WNL stars in these lines the value of 
6.6$\times$10$^{35}$ ergs s$^{-1}$, which is 1/3 that for the 
N III $\lambda$4640 + He II $\lambda$4686 lines derived by SV98.

   The results are shown in Table 7. The numbers of WNL
stars derived from different lines are in fair agreement 
( they are generally within a factor of 2 ) when uncertainties in input data
are taken into account. In particular, for the BCD I Zw 18 the numbers of WNL 
stars derived from the blue bump, and from the N III $\lambda$4512 and Si III 
$\lambda$4565 lines agree well. Further observations with higher signal-to-noise ratio and better calibration of the N III $\lambda$4512 and Si III $\lambda$4565 line luminosities are desirable to test this method which promises to yield
more reliable determinations of the number of WNL stars.

    The emission line C III $\lambda$5696 which is characteristic of WCL
stars is detected in three galaxies. We estimate the number of WCL stars
adopting the luminosity of a single WC7 star  in the C III $\lambda$5696 
emission line to be $L^0_{\rm WC7}$ = 8.1$\times$10$^{35}$
ergs s$^{-1}$ (SV98). Then, for the C III $\lambda$5696 
extinction-corrected fluxes of 1.91$\times$10$^{-15}$,
4.39$\times$10$^{-15}$ and 0.50$\times$10$^{-15}$ ergs s$^{-1}$ cm$^{-2}$
in the galaxies 0947+008 (Mrk 1236), 0952+095 (Mrk 710) and
1054+365 (CG 798) respectively, the derived numbers of WCL stars are
178, 254 and 5. These values are comparable to the numbers of WCE and
WNL stars. Although Schaerer et al. (1999a) have not derived the number of
WCL stars in Mrk 710 from their data, their measured flux of C III 
$\lambda$5696 (relative to H$\beta$) is in satisfactory agreement with ours.
While the presence of WCL stars in this high-metallicity galaxy 
( $Z$ $\sim$ 1.3 $Z_\odot$ ) is 
expected (Smith \& Maeder 1991), the presence of late WC stars in the lower 
metallicity galaxies Mrk 1236 and CG 798 
( both with $Z$ $\sim$ $Z_\odot$/8 ) is more surprising. 
New observations are required to confirm the presence of WCL stars in
these galaxies.

\subsection{The number of O stars}

  The number of O stars can be derived from the number of ionizing photons 
$Q_0^{cor}$ which is related to the total luminosity of the H$\beta$ emission
line $L_{cor}$(H$\beta$) by
\begin{equation}
L_{cor}({\rm H}\beta) = 4.76\times10^{-13}Q^{cor}_0.
\end{equation}
We take as representative an O7V star and set the
number of Lyman continuum photons emitted by such a star to be
$Q_0^{{\rm O7V}}$ = 1$\times$10$^{49}$ s$^{-1}$ (Leitherer 1990).
The total number of O stars is then derived from the number of O7V stars by
correcting for other O stars subtypes. In order to estimate the number 
of OV stars in WR galaxies, Vacca \& Conti (1992) and Vacca (1994) have 
considered the parameter $\eta_0$, defined to be the ratio of the number of O7V stars to the number of all OV 
stars. The quantity $\eta_0$ depends on the parameters of the initial mass 
function for massive stars and is, in general, a function of time
because of massive star evolution (Schaerer 1996). SV98 
have calculated $\eta_0$ as a function of time elapsed
from the beginning of an instantaneous burst, and we use their calculations to
derive $\eta_0$$(t)$ for each of our galaxies at age $t$ as
determined from the equivalent width $EW$(H$\beta$)
(SV98). We adopt an IMF with a Salpeter slope
$\alpha$ = 2.35 and low and upper mass limits of 0.8 $M_\odot$ and 120
$M_\odot$. Derived ages and $\eta_0$$(t)$ are shown in Table 5.

  It is also necessary to subtract the contribution of WR stars to the total
number of ionizing photons. Following Schaerer et al. (1999a) who assumed that 
the average Lyman continuum flux per WR star is comparable to $Q_0^{\rm O7V}$,
we adopt the average Lyman continuum photon flux per WR star to be 
$Q^{\rm WR}_0$ = $Q^{\rm O7V}_0$ = 1.0$\times$$10^{49}$ s$^{-1}$.
Then the number of O stars $N$(O) is given by:

\begin{equation}
N({\rm O})=\frac{Q^{cor}_0-N_{\rm WR}Q^{\rm WR}_0}{\eta_0(t)Q^{\rm O7V}_0}.
\label{eq:NO}
\end{equation}
The number of O stars derived from Eq. (\ref{eq:NO}) is given
in Table 7 together with the relative numbers of massive stars
$N$(WR)/$N$(O+WR) and $N$(WC)/$N$(WN), where $N$(WR) = $N$(WN) + $N$(WC)
and assuming $N$(WC) = $N$(WCE) and $N$(WN) = $N$(WNL).
The number of WNL stars is derived from the $\lambda$4650 bump 
except for two galaxies, 0218+003 and 2329+286.
The blue bump gives a too small $N$(WNL) in the case of 0218+003, and a
negative $N$(WNL) in the case of 2329+286.
We use instead the flux of the Si III $\lambda$4565 emission line to derive
$N$(WNL) in both cases.

\section{COMPARISON WITH EVOLUTIONARY SYNTHESIS MODELS}

 Figure 3 shows the luminosity of the blue bump at $\lambda$4650 in our sample
galaxies  vs. metallicity (as denoted by 12+log(O/H)). It is seen that
there is a general trend for the luminosity of the blue bump 
to decrease with decreasing metallicity. However there is a very large 
spread of points at a fixed metallicity. At 12 + log (O/H) $\sim$ 7.9, the 
spread spans some three orders of magnitude. This large spread 
is thought to reflect different 
stages during the burst of star formation. 
To check for possible selection effects, we have plotted in different symbols
the bright ($M_B$ $<$ --18) non-dwarf and dwarf ($M_B$ $\geq$ --18) galaxies. 
It is seen that all galaxies with large $L$($\lambda$4650) 
($>$ 10$^{39}$ ergs s$^{-1}$) are bright. 
Can the lack of galaxies with faint $L$($\lambda$4650) in the 
8.2 $\leq$ 12 + log (O/H) $\leq$ 9.1 abundance range simply be due to the 
selection effect that, in bright galaxies, 
WR bumps with small $L$($\lambda$4650) 
(those with $<$ 10$^{39}$ ergs s$^{-1}$) cannot be detected?
We cannot completely exclude this possibility as we observed only 
nuclear starbursts in the bright highest-metallicity galaxies. 
There exists the possibility that WR stars can be found with high S/N
observations in faint non-nuclear metal-rich H II regions. 
However, we believe that the most likely reason for   
the non-detection of galaxies with faint $L$($\lambda$4650) in 
the 8.2 $\leq$ 12 + log (O/H) $\leq$ 9.1 abundance range, is not because their 
WR bumps cannot be detected, but simply because there are no dwarf galaxies
in this high metallicity range.

Thus we believe the general decrease of $L$($\lambda$4650) with O/H to be real.   
  This decrease is expected from massive stellar evolution models 
  (Maeder 1991; Maeder \&
Meynet 1994; Meynet 1995; SV98) and it is consistent with 
earlier observations of WR galaxies (Vacca \& Conti 1992; Masegosa et al. 1991;
Kunth \& Schild 1986; Kunth \& Joubert 1985) where no galaxy with oxygen
abundance 12 + log (O/H) $\leq$ 7.9 
was ever seen to contain WR stars. 
A possible exception was the galaxy Zw 0855+06, although there was some
controversy about its oxygen abundance: Vacca \& Conti (1992) derived
12 + log (O/H) = 7.72 while Kunth \& Joubert (1985) obtained a value of 8.40. 
However, the increase in detector sensitivity and the 
use of large telescopes for the detection of weak low-contrast WR features 
in galaxies have changed the situation. The discovery of WR stars in 
I Zw 18 ( $Z_\odot$/50, Izotov et al. 1997;
Legrand et al. 1997) and SBS 0335--052 ($Z_\odot$/40, Izotov et al. 1999a) 
implies that the luminosity of the WR bump cannot keep on decreasing with
decreasing metallicity. Instead, the data in Fig. 3 
for those galaxies suggest that the 
luminosity of the blue bump appears to become roughly constant 
for 12 + log (O/H) $\le$ 7.9, although this conclusion 
is based only on a few data points. The data on WR populations in 
low-metallicity galaxies is 
very scarce for two reasons. First, such low-metallicity objects are 
extremely rare and second, the time spent by massive stars 
in the WR stage is very short at low metallicities, which make them hard 
to detect.  

    The discovery of WR stellar populations in the two most metal-deficient
galaxies known has increased substantially the metallicity range for WR 
galaxies. Our WR galaxy sample spans two orders of magnitude in metallicity,
from $Z_\odot$/50 (I Zw 18) to
$\sim$ 2$Z_\odot$ (0720+335 $\equiv$ Mrk 1199), allowing to compare
observed and predicted relative numbers of WR stars and check the validity 
of theoretical models in a wide range of metal abundances.
   In the following we compare the properties of the observed WR population in
our galaxies with predictions of evolutionary synthesis models by SV98 
for the range of heavy element mass fraction $Z$ = 0.001 -- 0.02
and by Schaerer (1998, private communication) for $Z$ = 0.0004. 

\subsection{Metallicity effects on the relative numbers of WR stars}

 Figure 4 shows the number ratio $N$(WR) / $N$(O + WR) for our sample
galaxies (filled circles) vs. metallicity. 
For comparison we also show the data from
Schaerer et al. (1999a) (open circles), Kunth \& Joubert (1985) (asterisks) 
and Vacca \& Conti (1992) (diamonds). While the relative numbers of 
WR stars are derived by Schaerer et al. (1999a) in the same way as ours, they 
are derived differently by Kunth \& Joubert (1985) and Vacca \& Conti (1992)
and cannot be compared directly with our data. Therefore, we have recalculated 
the relative numbers of WR stars
by the same method, using the fluxes and equivalent widths of 
nebular and WR emission lines published by those authors. 
However, the red bumps were not detected in their observations and therefore we 
have no information about the relative contribution
of WCE stars to the blue bump. Furthermore, for other authors' data, we cannot
correct the flux in the H$\beta$ emission line for aperture effects
for lack of information. 
Inspection of Table 5 shows that the mean correction factor for aperture effects is $\sim$ 2 -- 3. Therefore, H$\beta$ fluxes are underestimated and the 
relative numbers of WR to (O + WR) stars derived from the data by Kunth \& Joubert (1985), 
Vacca \& Conti (1992) and Schaerer et al. (1999a) are slightly
overestimated as is evidenced by a small upward shift of their data compared
to ours. Note that the detectability limit of WR populations 
in our galaxies is better compared to other observations because 
of the generally higher signal-to-noise ratio of our spectra.

   For each of the four galaxies in our sample with metallicity greater than 
solar ( 0211+038, 0720+335, 0952+095, 1036--069 )
we show two values of $N$(WR) / $N$(O + WR) connected by dashed 
lines. Lower limits are calculated assuming the correction factor $\eta_0$ to be 1. However, the equivalent widths of the H$\beta$ emission line in these 
galaxies are $\leq$ 30\AA\ and hence the ages of 
the star bursts are $\geq$ 5.3 Myr (Table 5). For these ages, 
high-metallicity O stars are not expected, the definition of $\eta_0$ becomes
meaningless and the ionizing flux is provided by stars
of type B or later and WR stars (SV98). Therefore, upper 
limits are set to $N$(WR)/$N$(O+WR) = 1 for the four metal-rich galaxies.
At the low-metallicity end, the $N$(WR)/$N$(O+WR) ratio for I Zw 18 
(filled circle) is probably underestimated by a factor of $\sim$ 2 because the 
equivalent width 
of the H$\beta$ emission line in the BCD's NW component, equal to 67\AA, and the parameter $\eta_0$ = 0.2 give a relatively large age of 5.9 Myr. 
A burst age of 4--5 Myr,
seems more reasonable as it is only slightly larger than the age 
estimated from the H$\beta$ equivalent width of the SE component of I Zw 18
( 128\AA\ ) where WR stars are not seen. To illustrate how $N$(WR)/$N$(O + WR)
changes as a function of adopted age, we show respectively 
by an open rectangle and an open 
triangle the $N$(WR)/$N$(O+WR) ratios expected in I Zw 18 for 
instantaneous burst ages of 4 Myr ($EW$(H$\beta$) $\sim$ 100\AA) and 3 
Myr ($EW$(H$\beta$) $\sim$ 200\AA), in addition to the ratio for a burst age
of 5.9 Myr (filled circle).
For the other galaxies, the $N$(WR)/$N$(O+WR) ratio may increase slightly
when their H$\beta$ equivalent width is corrected for aperture effects. This
is because a correction would increase $EW$(H$\beta$), which would increase
$\eta_0$ and decrease the number of O stars.

Despite these uncertainties, it is clear from Fig. 4 that the fraction of WR 
stars relative to other massive stars increases with increasing 
metallicity, as first noted by Kunth \& Schild (1986). This increase 
is in agreement with predictions of evolutionary
synthesis models (Mas-Hesse \& Kunth 1991; Kr\"uger et al. 1992; 
Cervi\~no \& Mas-Hesse 1994; Meynet 1995; SV98). 

The solid line in Fig. 4 shows the maximum theoretical values of 
$N$(WR)/$N$(O+WR) ratios as a function of metallicity as predicted by 
SV98 and Schaerer (1998, private communication). 
We have used the instantaneous burst models with a Salpeter IMF 
slope $\alpha$=2.35 and $M_{up}$=120$M_{\odot}$, except for the metallicity 
$Z$ = $Z_\odot$/50 for which $M_{up}$ = 150 $M_\odot$ is adopted.
Nearly all our galaxies lie below the solid line, in agreement with
the theoretical predictions. The spread of points can be understood as caused by
the strong dependence on time of the WR star relative numbers during the short
WR stage of the starburst. Evolutionary synthesis models predict that 
the number of WR stars rises to a maximum value and then falls down to zero on 
the short time scale of 1 Myr for $Z$ = $Z_\odot$/50 and of 5 Myr for $Z$ = 
$Z_\odot$. The maximum value depends on the IMF and the 
heavy element mass fraction $Z$, since at a higher metallicity the
minimum mass limit for a star which undergoes the WR phase is lower. In
Fig. 4 the evolutionary track of a galaxy undergoing a star formation burst 
would describe a loop. An object rushes up and reaches the maximum 
$N$(WR)/$N$(O+WR) value (solid line), 
and after that drops down, drifting slightly to higher $Z$, because massive 
stars would enrich the surrounding environment with heavy elements via stellar 
winds and supernova explosions (Maeder \& Meynet 1994; 
Esteban \& Peimbert 1995; Pilyugin 1994). 
In this scenario we would expect different objects 
at different stages of WR evolution to scatter in the region under the solid 
line, as is observed. 
The lack of galaxies with high $Z$ and small 
relative number of WR stars can be understood as a 
selection effect.  
We have observed only nuclear starbursts in
highest-metallicity galaxies as nuclear regions constitute
an especially favorable environment for the formation of numerous
massive stars.
It is likely that this region will be
filled in when higher signal-to-noise ratio observations of non-nuclear 
metal-rich H II regions are carried out.
Another factor which may account for the observed spread of points is 
the finite duration of the starburst. The assumption of an instantaneous burst
where all stars are formed at the same time, represents a limiting case.
High spatial resolution studies of
blue compact dwarf galaxies (e.g. Thuan, Izotov \& Lipovetsky 1997; Papaderos
et al. 1998) show that star forming regions consist of several clusters with
an age spread of several Myr. Thus real star formation should be intermediate
between the instantaneous burst and continuous star formation cases. 
The dot-dashed line in Fig. 4 shows the theoretical prediction for the other limiting case, a
continuous star formation model. The data scatters nicely between the two
limiting cases, implying that the models are basically correct. 

\subsection{The $N$(WC)/$N$(WN) ratio}

 Figure 5 shows the dependence of the WC-to-WN star number ratio on 
metallicity. It is seen that galaxies with relatively large numbers of WC stars (the so-called WC galaxies) possess a very narrow range of metallicites, 
with 12 + log (O/H) going only from 7.8 to 8.2.
These large $N$(WC)/$N$(WN) ratios can be explained in terms of the
bursting nature of the star formation. The thick solid line in Figure 5 shows
the maximum values of $N$(WC)/$N$(WN) as a function of oxygen abundance
as predicted by theoretical models (SV98; Schaerer 1998, 
private communication). These instantaneous burst models use stellar evolution 
models with enhanced mass-loss (Maeder \& Meynet 1994), and are characterized by the same parameters as described before. The theoretical curve delineates quite 
well the left boundary of the region where the data points in the range of 
oxygen abundance 12 + log(O/H) = 7.6 -- 8.3 scatter. 

However, the agreement for the extremely metal-deficient 
galaxy I Zw 18 is not so good. 
The observed $N$(WC)/$N$(WN) in this galaxy is $\sim$ 0.28 while models predict 
$N$(WC)/$N$(WN) $\approx$ 0.09. This discrepancy cannot be due to
our adopted values of line luminosities of single WR stars since we use
the same values as SV98. In contrast to the relative number of WR stars, the $N$(WC)/$N$(WN) ratio does not suffer from aperture effects. 
The numbers of WNL stars in I Zw 18 of 45 and 36 derived from the weaker lines N III $\lambda$4512 and Si III $\lambda$4565 respectively (Table 7) are very close to the value of 44 derived from the blue bump. These values are derived
with a distance of 20 Mpc to I Zw 18 (Izotov et al. 1999b). 
On the other hand, if the same distance is adopted, Legrand et al. (1997)
detect only 3 -- 6 WCE stars and no WN star, so that their $N$(WC)/$N$(WN) ratio is uncertain. De Mello et al. (1998) obtain $N$(WN) $\sim$ 30 from
analysis of {\sl Hubble Space Telescope} images, which results in an even larger
$N$(WC)/$N$(WN) ratio for I Zw 18. Hence, we conclude that models fail to predict the correct $N$(WC)/$N$(WN) ratio at very low metallicity. However,
because of the small numbers of WR stars in I Zw 18 ( $\sim$ 45 WNL and 
$\sim$ 12 WCE stars) and the small slit width of 1\farcs5 used in the spectroscopic 
observations, we cannot exclude the possibility that we are observing in I Zw 18
a region with a locally
enhanced number of WCE stars. Therefore, high signal-to-noise ratio (SNR)
two-dimensional spectroscopic mapping is necessary to check this possibility.
The existing two-dimensional spectroscopic data on I Zw 18 by V\'ilchez \& 
Iglesias-P\'aramo (1998) have too low a SNR for WR features to be seen.

  The $N$(WC)/$N$(WN) ratio derived for the galaxies in our sample is very
different from that expected in the case of continuous star formation as
derived empirically by observations of Local Group galaxies.
We show by diamonds in Fig. 5 the $N$(WC)/$N$(WN) ratios
in the Local Group galaxies M33, M31, NGC 6822, SMC, LMC, IC 10 and the 
Milky Way based on observations of individual WR stars (Massey \& Johnson 1998). The linear fit to these data excluding the 2 deviant points of
IC 10 and the Milky Way given by Massey \& Johnson (1998) is shown by the dashed line. This fit can be considered as an empirical determination of the
dependence of the $N$(WC)/$N$(WN) ratio on metallicity in the case of
continuous star formation. It predicts $N$(WC)/$N$(WN) to approach 0 for 
galaxies with oxygen abundance 12 + log(O/H) $\leq$ 8.1 and agrees well with the predictions of massive stellar evolution models with enhanced mass-loss for 
continuous star formation with a constant rate. 

   The galaxy IC 10 was excluded by Massey \& Johnson (1998) from the 
determination of the correlation because of its very large $N$(WC)/$N$(WN) ratio of $\sim$ 2 ( although that ratio may be as small as 1.4 if 
uncertainties are taken into account ). Those authors suggest that such a high 
value is caused by the bursting nature of star formation in IC 10.
Indeed, this galaxy, together with our galaxies with oxygen abundances
in the range from 7.8 to 8.2, can be understood in terms of models with an
instantaneous starburst as shown by the solid line in Fig. 5.

   The five galaxies in our sample at the high end of oxygen abundance 
12 + log(O/H) $\geq$ 8.6 lie systematically 
below the fit of Massey \& Johnson (1998) while 
no galaxy with high metallicity is found above it. The same result has been 
obtained by Schaerer et al. (1999a) for two metal-rich galaxies He 2-10 
($Z$ $\sim$ $Z_\odot$/2) and Mrk 710 ($Z$ $\sim$ $Z_\odot$). These low values
may be partly due to the assumption that WCE stars dominate 
among WC stars. If we assume instead that WC7 stars are the main contributors
to the C IV $\lambda$5808 emission line in high-metallicity galaxies then
the relative number of WC-to-WN stars is increased by a factor of $\sim$ 4
because of the lower luminosity of WC7 stars, just accounting for the 
discrepancy between the values given by the fit for continuous star formation 
(dashed line) and the observed values. However, in this case an appreciable 
C III $\lambda$5696 emission line is expected in the spectra
of high-metallicity galaxies because the luminosity ratio 
$L$(C III $\lambda$5696)/$L$(C IV $\lambda$5808) for a single WC7 star is $\sim$
0.5 (SV98). Thus far, the C III $\lambda$5696 line has
been seen only in Mrk 710 (Schaerer et al. 1999a; this paper).
Additionally, WNE stars can significantly contribute to
the luminosity of blue bump in these galaxies, decreasing even more 
the $N$(WC)/$N$(WN) ratio
because the luminosity of a single WNE star in this bump is $\sim$ 3 times
lower than that of a WNL star (SV98).

But, probably, the main reason for the low $N$(WC) / $N$(WN) ratios in high-metallicity 
galaxies is the bursting nature of star formation in these galaxies, with 
different durations of the successive WNL, WNE and WCE stages. All
five galaxies have low equivalent widths of H$\beta$ (Table 5) and hence 
they are in the late stage of the WR episode with starburst ages $\geq$ 5.3 Myr,
when mainly WN stars are present while the number of WC stars drops to 0
(SV98).

In summary, analysis of the $N$(WC)/$N$(WN) ratio shows good general 
agreement between the observational data and the model predictions, except 
in the case of the very low metallicity galaxy I Zw 18. The detection of WR 
stars in other extremely metal-deficient galaxies is necessary to clarify 
the situation at the low-metallicity end.

\subsection{Age dependence}

 Due to the bursting nature of star formation in WR galaxies, the relative
number of WR stars is a sensitive function of time elapsed since the beginning
of the star formation episode. The equivalent width of the 
H$\beta$ emission line $EW$(H$\beta$) can be used with a certain degree of confidence as an indicator of the age of
star formation region. The hydrogen ionizing flux of a star cluster 
gradually decreases as the most massive stars disappear. As a result the 
equivalent width of H$\beta$ decreases with time. The theoretical behavior
of $EW$(H$\beta$) with time for an instantaneous burst is given by 
SV98 for a set of metallicities ranging from $Z_\odot$/20 to 2$Z_\odot$.

   Figure 6 shows the dependence of the $N$(WR) / $N$(O+WR) ratio as a function
of the age of an instantaneous burst occuring at time
$t$ = 0. Because this ratio is a strong function of the heavy element mass
fraction $Z$, we divided objects from our sample (filled circles) 
and from Kunth \& Joubert (1985) (asterisks), Vacca \& Conti (1992) (diamonds) 
and Schaerer et al. (1999a) (open circles) into four groups: 
the highest-metallicity objects with 12 + log (O/H) $>$ 8.63
are shown in Figure 6a, those with  8.43 $<$ 12 + log (O/H) $\leq$ 8.63
are shown in Figure 6b, those with  7.93 $<$ 12 + log (O/H) $\leq$ 8.43
are shown in Figure 6c and the lowest-metallicity objects with 
12 + log (O/H) $\leq$ 7.93 are shown
in Figure 6d. We compare the observed distributions with predictions of
evolutionary synthesis models by SV98 shown by solid lines
and labeled by their heavy element mass fraction.
For metal-rich galaxies we again show two values 
of $N$(WR)/$N$(O+WR): 1 and the ratio expected for the parameter $\eta_0$ = 1. 
They are connected by dashed lines in Figure 6a. At the low-metallicity end
(Fig. 6d), three values of $N$(WR)/$N$(O+WR) are shown for I Zw 18 
corresponding as before to three choices of starburst age: 3, 4 and 5.9 Myr. 
They are connected by a dashed line. Taking into account the uncertainties in 
deriving the $N$(WR)/$N$(O+WR) ratio
(observational errors, uncertainties in the age of starburst, breakdown of the 
instantaneous burst approximation and other model assumptions), we find a 
good general agreement between the relative numbers of WR stars predicted by 
the theoretical models and those inferred from observations, especially for 
objects in
the range of heavy element mass fraction $Z$ = 0.002 -- 0.010 (Figures 6b, 6c). 
The agreement is not so good for the most metal-deficient galaxies (Figure 6d), 
including I Zw 18. For the latter, we expect that part of the 
disagreement may come from
uncertainties in the age determination with the use of the H$\beta$ 
equivalent width. It is known that the ionized gas distribution in the
NW component of I Zw 18 is very complex with WR stars located in a hole with 
very low $EW$(H$\beta$) (Izotov et al. 1997). 
The presence of a large H$\alpha$ halo in 
I Zw 18 suggests that some of the H II regions 
in it are density-bounded (Dufour \& Hester
1990; Martin 1996). In this case part of the ionizing photons will
escape the H II region, resulting in a lower equivalent width of the H$\beta$ 
emission line. The most deviant point in Figure 6d belongs to the galaxy 
Mrk 178. This discrepancy may be caused by small statistics in the number of WR 
stars which are only a few ( $\sim$ 2 -- 3) in this galaxy, and by 
uncertainties in the burst age determination. 

The equivalent widths of the blue and red bumps as a function of H$\beta$ 
equivalent width are shown respectively in Figures 7 and 8. 
We again divide our galaxy sample into 4 subsamples according to
heavy element mass fraction, in the same manner as in Figure 6.
Solid lines show the theoretical predictions by SV98
for stellar populations with
IMF slope $\alpha$ = 2.35, while dashed lines in Fig. 7d and 8d are those 
with IMF slope $\alpha$ = 1.0. We expect the main contribution to the
red bump equivalent width to come from the C IV $\lambda$5808 line. However, 
because of the low resolution and large width of this line, some contribution of
N II $\lambda$5720--40, although small, might be present. By contrast,
different subtypes of WR stars contribute to the blue bump. Therefore, while the
theoretical curves in Figure 8 represent only $EW$($\lambda$5808),
for the blue bump the theoretical curves shown in Figure 7
are the sum of the N III $\lambda$4640, C III/C IV $\lambda$4658 and He II 
$\lambda$4686
WR emission line equivalent widths. Note that the points with high values
( $EW$($\lambda$4650) $>$ 5\AA\ and $EW$($\lambda$5808) $>$ 3\AA\ ) in Figures 7d and 8d 
are highly uncertain, because they belong to galaxies with very few ($\leq$ 1 -- 3) WR stars (1130+495, 1223+487, 1437+370 in Tables 6 and 7),
or galaxies with noisy spectra (0218+003). We do not consider
these galaxies in our analysis. Again, we find general good agreement between 
observations and theory for $EW$($\lambda$4650) and $EW$($\lambda$5808) 
of objects with heavy element fraction $Z$ $>$ 0.002 (Fig. 7a--7c and 8a--8c). 
The relatively low values of the observed $EW$($\lambda$5808) in Fig. 8a and 8b
can be understood in the following manner. 
High-metallicity objects in our sample are likely to be in a 
late WR stage and hence they are observed after the WC bump equivalent width 
reaches its maximum value. This supports our previous conclusion in section 6.2 that the low $N$(WC)/$N$(WN) ratios in high-metallicity
galaxies are lower than those predicted by the continuous star formation 
empirical relationship 
(Fig. 5) because of the bursting nature of star formation in these galaxies. 
In both Figures 7 and 8, there appears to be a slight systematic shift of 
the data points to the left of the theoretical curves by log $EW$(H$\beta$) 
$\sim$ 0.4. This implies that the
models predict systematically larger $EW$(H$\beta$) by a factor of $\sim$ 2.5,
i.e. smaller burst ages as compared to those implied by the observations.
However, we expect that correction for aperture effects will increase
$EW$(H$\beta$) making the agreement better.

    Agreement is not so good however for the galaxies with lowest metallicities.
Theoretical predictions with the Salpeter 
IMF slope $\alpha$ = 2.35 are below our observed data points for both 
$EW$($\lambda$4650) and $EW$($\lambda$5808). Models with a very shallow 
IMF slope $\alpha$ = 1.0 also fail to explain the observed points. 
We note that, in contrast, de Mello et al. (1998) found satisfactory agreement
between the predicted $EW$($\lambda$4650) and $EW$($\lambda$5808) and the 
values observed by Legrand et al. (1997). However, those authors' observations
did not include the region of maximum WR emission. This region was found by
Izotov et al. (1997) to have several times larger 
$EW$($\lambda$4650) and $EW$($\lambda$5808), 
leading to the discrepancy between models and observations
discussed here. It is likely,
that the disagreement comes from the different properties of WCE stars 
at low metallicities as compared to those in the Galaxy and
Local Group galaxies. We suspect that single
low-metallicity WCE star luminosities in the C IV $\lambda$4658 and 
C IV $\lambda$5808 lines are $\sim$ 2 -- 4 times larger than those adopted by
SV98. 
The use of Smith's (1991) WR star emission line luminosities would increase
$EW$($\lambda$4650) and $EW$($\lambda$5808) ( SV98, de Mello
et al. 1998 ), but by an amount not large enough to account for the observed
values in I Zw 18. Observational data for single WR stars are not 
available at such low metallicities. However, some arguments in favor of a larger WCE star 
line luminosity at low metallicity come from the calculations of Maeder \&
Meynet (1994). They have shown that, for a given fixed mass of the progenitor 
star, both the total luminosity of a WCE star and its surface C/He 
abundance ratio are larger at very low metallicities. The luminosity of a WNL 
star also increases with decreasing metallicity, but by not such a
large amount. Hence, we expect that the net effect might be an increase of both
the blue and red bump model equivalent widths, in better agreement with 
observations. A higher line luminosity of low-metallicity WCE stars would also 
result in a better agreement between the observed and predicted $N$(WC)/$N$(WN) 
ratio for I Zw 18 (Fig. 5), since the observed value would be reduced. 

   There is another factor which may increase the equivalent widths of the
blue and red bumps. Theory predicts that WR stars in massive close binary
systems in the late ( $\ga$ 5 Myr ) phases of an instantaneous burst of star
formation may have larger equivalent widths ( SV98 ).
However the properties of WR stars in binaries are still poorly known.

\section{THE ORIGIN OF THE NEBULAR HE II $\lambda$4686 EMISSION}

   The strong nebular He II $\lambda$4686 emission line has been detected
in many low-metallicity blue compact galaxies ( in $\sim$ 50\% of the sample )
observed by ITL94, ITL97,
Thuan et al. (1995), IT98, Izotov et al. (1996, 1997). Its intensity, exceeding 
$\sim$ 3\% that of H$\beta$ in some objects, is several orders of 
magnitude larger 
than theoretical values predicted by models of photoionized H II regions
(e.g. Stasi\'nska 1990). It was suggested by
Bergeron (1977), that the He II emission in dwarf emission-line galaxies could
arise in the atmospheres of Of stars. Garnett et al. (1991) 
obtained observations of nebulae in nearby dwarf galaxies with strong narrow
He II $\lambda$4686 emission lines and examined several possible 
excitation mechanisms, concluding that the radiation field associated with 
star-forming regions can be harder than previously suspected. More recently
Schaerer \& de Koter (1997) calculated non-LTE atmosphere models
taking into account line blanketing and stellar winds and found that the flux
in the He II continuum is increased by 2 to 3 orders of magnitudes compared to
predictions from plane-parallel non-LTE model atmospheres and by 3 to 6 orders
of magnitudes compared to predictions from plane parallel LTE model atmospheres.
However, for young starbursts dominated by O stars ($t$ $\leq$ 3 Myr), typical
values of $I$(He II)/$I$(H$\beta$) are between 5$\times$10$^{-4}$ and 
2$\times$10$^{-3}$, still far below the observed intensities.
Schaerer (1996) synthesized the nebular and Wolf-Rayet He II $\lambda$4686 
emission in young starbursts. For heavy element mass fractions $Z_\odot$/5 
$\leq$ $Z$ $\leq$$Z_\odot$, he predicted a strong nebular 
He II emission due to a significant fraction of WC stars in the early WR phases 
of the burst, and concluded that the predictions (typically 
$I$(He II)/$I$(H$\beta$) 
$\sim$ 0.01 -- 0.025) agree well with the observations. SV98
proposed that hot WN stars may also play a role. Another mechanism, 
suggested by Garnett et al. (1991), is that radiative shocks in giant H II 
regions can produce relatively strong He II emission under 
certain conditions. The strength of
the He II emission is sensitive mostly to the velocity of the shock, reaching a
maximum for $V$$_{shock}$ $\sim$ 120 km s$^{-1}$, and dropping rapidly at higher
velocities. 

   To study the possible mechanisms responsible for the emission of the nebular
He II $\lambda$4686 line, we gathered a sample of galaxies from the
present study and previous papers ( ITL94, ITL97, IT98, Thuan et al. 1995) 
showing nebular He II $\lambda$4686. 
The galaxies are listed in Table 8. 
No nebular He II $\lambda$4686 emission was detected in galaxies with 
12 + log(O/H) $>$ 8.13.  
Also shown are the intensities and 
equivalent widths of the nebular He II $\lambda$4686 line together with the 
equivalent width of the H$\beta$ emission line and the oxygen abundance.
WR 
stellar emission was detected only in 18 out of the 30 H II regions in Table 8,
 a 60\% detection rate. A non-detection of WR features means that a 
galaxy contains no or too few WR stars (several to a few tens depending on the 
distance of the galaxy) to produce detectable WR emission.
The number of WR stars responsible for a broad feature
with peak intensity comparable to the continuum rms is approximatively 
given by  
$d^2/(S/N)$, where $d$ is the distance to the galaxy in Mpc. 
Typically, our spectra with
no WR lines detected have $S/N$ $\sim$ 30. Then, at a distance of
$\sim$ 10 Mpc, a broad blue bump produced by $\ga$ 3 WR stars can be detected.
We believe that the non-detection of WR emission in some galaxies 
is not the result of a selection effect as the spectra of all galaxies 
possess approximately  
the same signal-to-noise ratio: it has to be large enough so we could measure 
accurately the intensities of the weak 
He I lines necessary for the primordial helium problem.   
Moreover, they have approximatively the same relative
intensity and equivalent width of the nebular He II $\lambda$4686 emission
line. 
Because the 
nebular He II emission region is generally smaller than the H$\beta$ 
emission region, we have not corrected the He II $\lambda$4686 line intensity 
for aperture effect. 
We present in Table 8 He II $\lambda$4686 line intensities relative 
 H$\beta$ for two cases: in column 4, H$\beta$ is corrected 
only for extinction, while in column 5 H$\beta$ is corrected for both
extinction and aperture effect. 
The aperture correction factors for H$\beta$ in 
the WR galaxies are given in Table 5. For the remaining objects we assume 
the correction factor to be 2.5, a typical value for the objects in Table 5.

   Figure 9 displays the dependence of the equivalent width and intensity of the
nebular He II $\lambda$4686 line as a function of the H$\beta$ equivalent width.
Theoretical predictions are also shown for stellar populations with $Z$ = 0.0004, 0.001, 0.004 by dashed, solid and dotted lines, respectively
(SV98, Schaerer 1998, private communication).
The observed equivalent widths of the nebular He II $\lambda$4686 emission
line in WR galaxies (filled circles) and in the galaxies with nondetected
WR features (open circles) are in good agreement with values 
predicted by models with $Z$ = 0.0004 and 0.001 (Figure 9b). As before, the 
observed values of $EW$(H$\beta$) are
systematically shifted to the left of the theoretical curves by 
log $EW$(H$\beta$) $\sim$ 0.4, implying that predicted burst ages are 
systematically younger as compared to observations. However, as previously,
correction for aperture effect would increase $EW$(H$\beta$) and 
result in better agreement. The agreement
with data is not so good however for the model with $Z$ = 0.004 which predicts
$EW$(He II $\lambda$4686) that are systematically higher (Figure 9a). 

Figure 9b shows that the intensities of He II $\lambda$4686 
relative to H$\beta$, the latter being corrected only for 
interstellar extinction, are systematically larger than predicted values. 
Correction of H$\beta$ for aperture effect does result in a 
better agreement (Figure 9c). 
Again we see a systematic shift of the data points to 
the left of the evolutionary synthesis models which may be due in part
to the fact that the H$\beta$ equivalent width 
has not been corrected for aperture effect.
Overall, we find a satisfactory 
general agreement between models and observations, implying that the hard 
radiation of WR stars can account for many properties of the 
He II $\lambda$4686 nebular emission in WR galaxies.

    However, the galaxies with lowest metallicities tend to have slightly
larger nebular He II $\lambda$4686 line intensities compared to those with
higher metallicities (Table 8). No nebular He II $\lambda$4686 emission has 
been observed in galaxies with $Z$ $>$ 0.004.
This trend is just the opposite of what is expected from the evolutionary
synthesis models by SV98 which predict nebular
$I$(He II $\lambda$4686)/$I$(H$\beta$) as high as 0.05 -- 0.10 at solar and
higher metallicities. We conclude that high-metallicity stellar
models overpredict the number of photons with $\lambda$ $<$ 228 \AA\ responsible
for the ionization of He$^+$.

Additionally, the points representing 
galaxies with detected and nondetected WR features mingle indistinctly in
Figure 9. 
This implies that, despite the good agreement
between the observed and predicted characteristics of the nebular He II
$\lambda$4686 emission line, WR stars are probably not the sole origin of He II
$\lambda$4686 emission in star-forming regions. Other mechanisms
such as for example radiative shocks, are probably also at work.
These mechanisms may play an important role in the late stages of 
star formation
bursts (indicated by a lower equivalent width of the H$\beta$ emission line) 
when the supernova activity increases.

\section {SUMMARY AND CONCLUSIONS}

    We present here the results of a spectroscopic study 
of a sample of 39 WR galaxies spanning two orders of magnitude in
heavy element abundances, from $Z_\odot$/50 to 2$Z_\odot$. Our main goal is to 
search for WN and WC stars in these galaxies, compare their numbers with 
predictions from evolutionary synthesis models and test massive stellar 
evolution models in a wide range of metallicities. 

    Our principal results are the following:

1. The broad WR emission in the blue region of the spectrum at $\lambda$4650
(the blue bump), an unresolved blend of N V $\lambda$4605, 4620,
N III $\lambda$4634, 4640, 
C III $\lambda$4650, C IV $\lambda$4658 and He II $\lambda$4686 emission lines,
is present in 37 galaxies and suspected in 2 more. The red bump mainly
produced by the emission of broad C IV $\lambda$5808 is detected in 
30 galaxies. The WR population in the majority of our galaxies is dominated by 
late WN and early WC stars. However, a nonnegligible population of early WN 
stars can be present in the highest-metallicity galaxies in our sample.

2. Weak WR emission lines are present in the spectra of many of our galaxies 
which are very rarely or never seen before. The N III $\lambda$4512 and Si III 
$\lambda$4565 lines are most often present and they are 
tracers of WN7--WN8 and WN9--WN11 stars respectively. These features have
been detected in particular in the most metal-deficient blue compact galaxy 
known, I Zw 18 (Izotov et al. 1997). The C III $\lambda$5696 emission line is detected in three galaxies suggesting the presence of late WC stars. We confirm 
the detection of C III $\lambda$5696 by Schaerer et al. (1999a) in the spectrum 
of the high-metallicity galaxy Mrk 710. This line is expected in 
high-metallicity environments. However, its presence 
in the spectra of Mrk 1236 and CG 798 with $\sim$ $Z_\odot$/8 is 
more surprising and needs confirmation by higher signal-to-noise ratio 
spectral observations.

3. A new technique is proposed for the determination of the WNL star numbers 
from the fluxes of the N III $\lambda$4512 and Si III $\lambda$4565
emission lines. The advantage of this technique is that these lines are only
seen in WNL stars, while the blue bump, usually used to derive
their number, is a mixture of WNE, WNL and WCE stellar emission
contaminated by nebular gaseous emission. The numbers of WNL stars
derived from the N III $\lambda$4512 and Si III $\lambda$4565 emission lines
are in satisfactory agreement with values derived from the blue bump.
This new technique allows potentially to study the distribution of WNL stars
within narrow subtypes. However, a more precise calibration of the
N III $\lambda$4512 and Si III $\lambda$4565 emission lines in WR stars 
is necessary.

4. Good general agreement is found between the relative numbers of WR stars
$N$(WR)/$N$(O+WR) inferred from observations and those predicted by
evolutionary synthesis models. The relative numbers of WR stars
decrease with decreasing metallicity in the whole metallicity range 
discussed in this paper ($Z_\odot$/50 -- 2$Z_\odot$), in agreement with 
predictions by massive stellar evolution models with enhanced stellar wind 
(Maeder \& Meynet 1994).

5. The relative numbers $N$(WC) / $N$(WN) of WR stars of different subtypes 
in the galaxies of our sample can be explained by the bursting nature
of star formation, and are in general good agreement with predictions of evolutionary
synthesis models by SV98. The relative numbers 
$N$(WR) / $N$(O+WR) and observed equivalent widths of the blue and
red bumps also compare favorably with predictions of evolutionary synthesis 
models by the same authors for metallicities larger than $\sim$ 1/10 solar.
However, the agreement is not so good for galaxies at the low-metallicity 
end, where $N$(WC) / $N$(WN), $EW$($\lambda$4650) and $EW$($\lambda$5808)
derived from observations are several times larger compared to model 
predictions. Part of the disagreement may come from the poor statistics
of WR stars in these low-metallicity WR galaxies. In the case of I Zw 18
($Z_\odot$/50) however, the difference between observations
and models may be explained by too low single WCE star line luminosities
adopted in the SV98 models, or by an additional contribution by WR stars in
binaries.

6. The nebular He II $\lambda$4686 emission is analyzed in 30 H II regions 
to study its origin, whether it can be due to the ionization of He$^+$ by the 
hard UV radiation of WR stars. A WR population is detected in only 18
of these H II regions. 
No H II region with detected nebular He II $\lambda$4686 
emission has oxygen abundance 12 + log(O/H) $>$ 8.13. Models
by SV98 reproduce satisfactorily the observed intensities
and equivalent widths of the nebular He II $\lambda$4686 emission line.
Hence, in WR galaxies, He II $\lambda$4686 emission can in general 
be accounted for by the hard UV radiation of WR stars.
However, their models predict the existence of nebular He II $\lambda$4686
emission at the early stages of a burst of star formation with $EW$(H$\beta$) = 
200 -- 300\AA\, while the data are systematically shifted to lower
$EW$(H$\beta$)$\leq$ 100\AA\ implying larger ages. For I Zw 18 in particular,
models predict the existence of nebular He II $\lambda$4686 emission with
$EW$(H$\beta$) $>$ 300\AA, while the observed $EW$(H$\beta$) is only 67\AA. 
While this disagreement can be explained in part by aperture effects in
the $EW$(H$\beta$) measurements, WR stars cannot be the sole cause of 
He II $\lambda$4686 emission. 
The properties of this emission are similar whether galaxies
contain WR stars or not (Fig. 9). Hence, to addition to WR star
ionization other mechanisms (e.g. radiative shocks) need to be invoked 
to account for He II $\lambda$4686 nebular emission, 
especially at the late stages of the star formation burst.

\acknowledgements
It is a pleasure to thank Daniel Schaerer for the use of his
evolutionary synthesis models and useful comments. 
Phil Massey and the referee Crystal Martin also contributed helpful comments.
This international collaboration was 
possible thanks to the partial financial support of INTAS grant No. 97-0033
for which N.G.G. and Y.I.I. are grateful. T.X.T. and Y.I.I.
acknowledge the partial financial support of NSF grant AST-9616863. Y.I.I. 
thanks the staff of the University of Virginia 
for their kind hospitality.


\clearpage



\begin{deluxetable}{lrrccccl}
\tablenum{1}
\tablecolumns{8}
\tablewidth{0pt}
\tablecaption{General Parameters of Galaxies}
\tablehead{
\colhead{Galaxy}&\colhead{$\alpha$(1950.0)}&\colhead{$\delta$(1950.0)}&\colhead{$m_{pg}$}&\colhead{$M_{pg}$}&\colhead{$z$}&\colhead{12 + log (O/H)}
&\multicolumn{1}{l}{Other name}  }
\startdata
 0112$-$011                  &1 13 00.5   &$-$01 07 22  &17.0  &--14.8  &0.00570  &8.31&UM 311              \nl
 0207$-$105                  &2 07 50.2   &$-$10 33 19  &15.0  &--18.6  &0.01338  &9.04\tablenotemark{a}&Mrk 1026, NGC 848  \nl
 0211+038                    &2 11 08.7   &+03 52 08    &14.3  &--19.0  &0.01148  &8.99\tablenotemark{a}&Mrk 589, UGC 1716  \nl
 0218+003                    &2 18 20.4   &+00 19 42    &16.5  &--20.3  &0.05844  &7.93&UM 420              \nl
 0252$-$102                  &2 52 08.0   &$-$10 13 46  &13.0  &--18.5  &0.00501  &8.46\tablenotemark{a}&Mrk 1063, NGC 1140  \nl
 0459$-$043                  &4 59 09.4   &$-$04 19 43  &15.0  &--18.7  &0.01343  &8.04&Mrk 1089, NGC 1741  \nl
 0553+033                    &5 53 04.9   &+03 23 07    &15.5  &--14.5  &0.00249  &8.09&II Zw 40, UGCA 116  \nl
 0635+756                    &6 35 24.4   &+75 40 14    &17.0  &--13.1  &0.00264  &8.04&Mrk 5, UGCA 130     \nl
 0720+335                    &7 20 28.5   &+33 32 24    &13.7  &--20.0  &0.01353  &9.13\tablenotemark{a}&Mrk 1199, UGC 3829  \nl
 0723+692                    &7 23 23.7   &+69 17 33    &11.6  &--16.1  &0.00036  &7.85&Mrk 71, NGC 2363    \nl
 0842+162                    &8 42 45.3   &+16 16 46    &15.7  &--20.9  &0.05303  &8.60\tablenotemark{a}&Mrk 702, PG 0842+162 \nl
 0926+606                    &9 26 20.0   &+60 40 02    &17.5  &--16.2  &0.01371  &7.91&                    \nl
 0930+554                    &9 30 30.3   &+55 27 46    &17.6  &--13.9\tablenotemark{b}  &0.00274  &7.16&I Zw 18, Mrk 116, UGCA 166    \nl
 0946+558                    &9 46 03.1   &+55 48 46    &15.6  &--16.0  &0.00525  &8.00&Mrk 22, UGCA 184    \nl
 0947+008                    &9 47 19.9   &+00 51 00    &13.5  &--18.5  &0.00626  &8.07&Mrk 1236            \nl
 0948+532                    &9 48 10.2   &+53 13 41    &18.0  &--18.3  &0.04629  &8.00&                    \nl
 0952+095                    &9 52 10.2   &+09 30 32    &13.5  &--18.0  &0.00494  &9.03\tablenotemark{a}&Mrk 710, NGC 3049, UGC 5325 \nl
 1030+583                    &10 30 56.3  &+58 19 20    &16.5  &--15.9  &0.00757  &7.79&Mrk 1434            \nl
 1036$-$069                  &10 36 03.0  &$-$06 54 37  &13.5  &--18.8  &0.00718  &8.95&Mrk 1259, IC 630    \nl
 1053+064                    &10 53 33.3  &+06 26 24    &14.8  &--15.9  &0.00339  &7.99&Mrk 1271            \nl
 1054+365                    &10 54 59.8  &+36 31 30    &16.0  &--13.5  &0.00198  &7.97&CG 798              \nl
 1130+495                    &11 30 45.2  &+49 30 43    &13.9  &--13.5  &0.00074  &7.82&Mrk 178, UGC 6541   \nl
 1134+202                    &11 34 17.7  &+20 12 14    &17.0  &--17.6  &0.02083  &8.74\tablenotemark{a}&Mrk 182             \nl
 1135+581                    &11 35 51.3  &+58 09 04    &15.5  &--15.2  &0.00340  &7.98&Mrk 1450, PG 1136+581 \nl
 1139+006                    &11 39 38.5  &+00 36 42    &13.7  &--20.6  &0.01830  &7.99&UM 448, Mrk 1304, UGC 6665 \nl
 1140$-$080                  &11 40 24.6  &$-$08 03 18  &15.5  &--17.6  &0.01033  &8.72\tablenotemark{a}&Mrk 1305, IC 723    \nl
 1147+153                    &11 47 28.1  &+15 18 05    &15.4  &--14.5  &0.00243  &8.11&Mrk 750             \nl
 1150$-$021                  &11 50 03.8  &$-$02 11 28  &14.1  &--16.6  &0.00342  &7.95&UM 462, Mrk 1307, UGC 6850 \nl
 1152+579                    &11 52 51.9  &+57 56 34    &16.5  &--17.9  &0.01930  &7.81&Mrk 193             \nl
 1211+540                    &12 11 33.9  &+54 01 58    &17.8  &--12.7  &0.00308  &7.64&                    \nl
 1222+614                    &12 22 44.5  &+61 25 46    &17.0  &--12.9  &0.00243  &7.95&                    \nl
 1223+487                    &12 23 50.6  &+48 46 07    &15.3  &--12.6  &0.00096  &7.77&Mrk 209, UGCA 281, I Zw 36 \nl
 1234+072                    &12 34 29.9  &+07 11 59    &14.6  &--17.0  &0.00533  &8.23&Mrk 1329, IC 3591, UGC 7790 \nl
 1249+493                    &12 49 35.6  &+49 19 43    &18.8  &--16.1  &0.02430  &7.72&                    \nl
 1256+351                    &12 56 38.2  &+35 06 50    &12.8  &--17.4  &0.00272  &7.99&Mrk 59, NGC 4861, UGC 8098 \nl
 1319+579A                   &13 19 25.2  &+57 57 09    &18.5  &--13.6  &0.00650  &8.11&                    \nl
 1437+370                    &14 37 03.0  &+37 01 07    &17.0  &--11.7  &0.00137  &7.93&Mrk 475, CG 493     \nl
 1533+574B                   &15 33 03.3  &+57 27 04    &15.5  &--17.8  &0.01143  &8.11&VII Zw 611          \nl
 2329+286                    &23 29 29.5  &+28 40 18    &15.0  &--19.3  &0.01847  &8.06&Mrk 930, PG 2329+286 \nl
\enddata
\tablenotetext{a}{the [O III] $\lambda$4363 emission line is not detected. The oxygen abundance is derived using 
the calibration by van Zee et al. (1998) (see text).}
\tablenotetext{b}{the distance of 20 Mpc is adopted (Izotov et al. 1999b).}
\end{deluxetable}

\clearpage

\begin{deluxetable}{lllcccr}
\tablenum{2}
\tablecolumns{7}
\tablewidth{0pt}
\tablecaption{Journal of Observations}
\tablehead{
\colhead{Galaxy}&\colhead{Telescope}&\colhead{Date}&\colhead{Number of}&\colhead{Exposure Time}&\colhead{Airmass}&\multicolumn{1}{r}{P.A.} \\
\colhead{}&\colhead{}&\colhead{}&\colhead{Exposures}&\colhead{(minutes)}&\colhead{}&\multicolumn{1}{r}{(deg)}  }
\startdata
0112$-$011   &2.1m  &12/12/96     &3   &60       &1.2  &90  \nl
0207$-$105\tablenotemark{1}  &2.1m  &14/12/96     &2   &10       &1.4  &90  \nl
0211+038\tablenotemark{1}    &2.1m  &16/12/96     &5   &60       &1.2  &90  \nl
0218+038     &2.1m  &16/12/96     &3   &60       &1.2  &90  \nl
0252$-$102\tablenotemark{1}  &2.1m  &15/12/96     &3   &15       &1.4  &90  \nl
0459$-$043   &2.1m  &17/12/96     &2   &40       &1.6  &90  \nl
0553+033\tablenotemark{1}    &2.1m  &14,17/12/96  &6   &120\,~   &1.2  &90  \nl
0635+756     &2.1m  &16/12/96     &3   &60       &1.4  &90  \nl
0720+335     &2.1m  &19/02/96     &3   &60       &1.1  &90  \nl
0723+692     &4m    &18/03/94     &6   &29       &1.3  &77  \nl
0842+162\tablenotemark{1}    &2.1m  &16/12/96     &3   &60       &1.3  &90  \nl
0926+606     &4m    &17/03/94     &3   &45       &1.2  &115 \nl
0930+554     &MMT   &29,30/04/97  &6   &180\,~   &1.2  &139 \nl
0946+558     &4m    &23/03/93     &2   &30       &1.2  &90  \nl
0947+008\tablenotemark{1}    &2.1m  &19/12/96     &2   &40       &1.2  &90  \nl
0948+532     &4m    &24/03/93     &1   &15       &1.2  &90  \nl
0952+095\tablenotemark{1}    &2.1m  &14,16/12/96  &4   &60       &1.4  &90  \nl
1030+583     &4m    &16/03/94     &2   &30       &1.1  &92  \nl
1036$-$069\tablenotemark{1}  &2.1m  &19/02/96     &4   &20       &1.4  &91  \nl
1053+064     &2.1m  &15/12/96     &3   &60       &1.6  &90  \nl
1054+365     &4m    &18/03/94     &3   &45       &1.0  &97  \nl
1130+495\tablenotemark{1}    &2.1m  &17/12/96     &3   &55       &1.1  &133 \nl
1134+202\tablenotemark{1}    &2.1m  &18/02/96     &2   &10       &1.0  &91  \nl
1135+581     &4m    &24/03/93     &3   &30       &1.2  &90  \nl
1139+006     &2.1m  &16/12/96     &2   &40       &1.6  &90  \nl
1140$-$080\tablenotemark{1}  &2.1m  &17/02/96     &2   &10       &1.3  &35  \nl
1147+153     &2.1m  &17/12/96     &2   &30       &1.1  &90  \nl
1150$-$021   &2.1m  &16/12/96     &3   &60       &1.3  &62  \nl
1152+579     &4m    &24/03/93     &3   &45       &1.1  &90  \nl
1211+540     &4m    &23/03/93     &3   &60       &1.1  &90  \nl
1222+614     &4m    &16/03/94     &3   &45       &1.2  &90  \nl
1223+487     &4m    &17/03/94     &4   &55       &1.1  &145 \nl
1234+072\tablenotemark{1}    &2.1m  &18/02/96     &2   &30       &1.4  &91  \nl
1249+493     &4m    &23/03/93     &3   &55       &1.1  &90  \nl
1256+351     &4m    &17/03/94     &8   &70       &1.0  &59  \nl
1319+579A    &4m    &18/03/94     &2   &45       &1.1  &44  \nl
1437+370     &4m    &24/03/93     &2   &35       &1.1  &90  \nl
1533+574B    &4m    &16/03/94     &3   &45       &1.1  &112 \nl
2329+286     &2.1m  &16/12/96     &3   &60       &1.0  &90  \nl
\enddata
\tablenotetext{1}{Galaxies with not yet published data. The data for other
galaxies are appeared in ITL94, ITL97, Izotov et al. 1997, IT98, Thuan et al.
1995.}
\end{deluxetable}

\clearpage
%
%
\plotfiddle{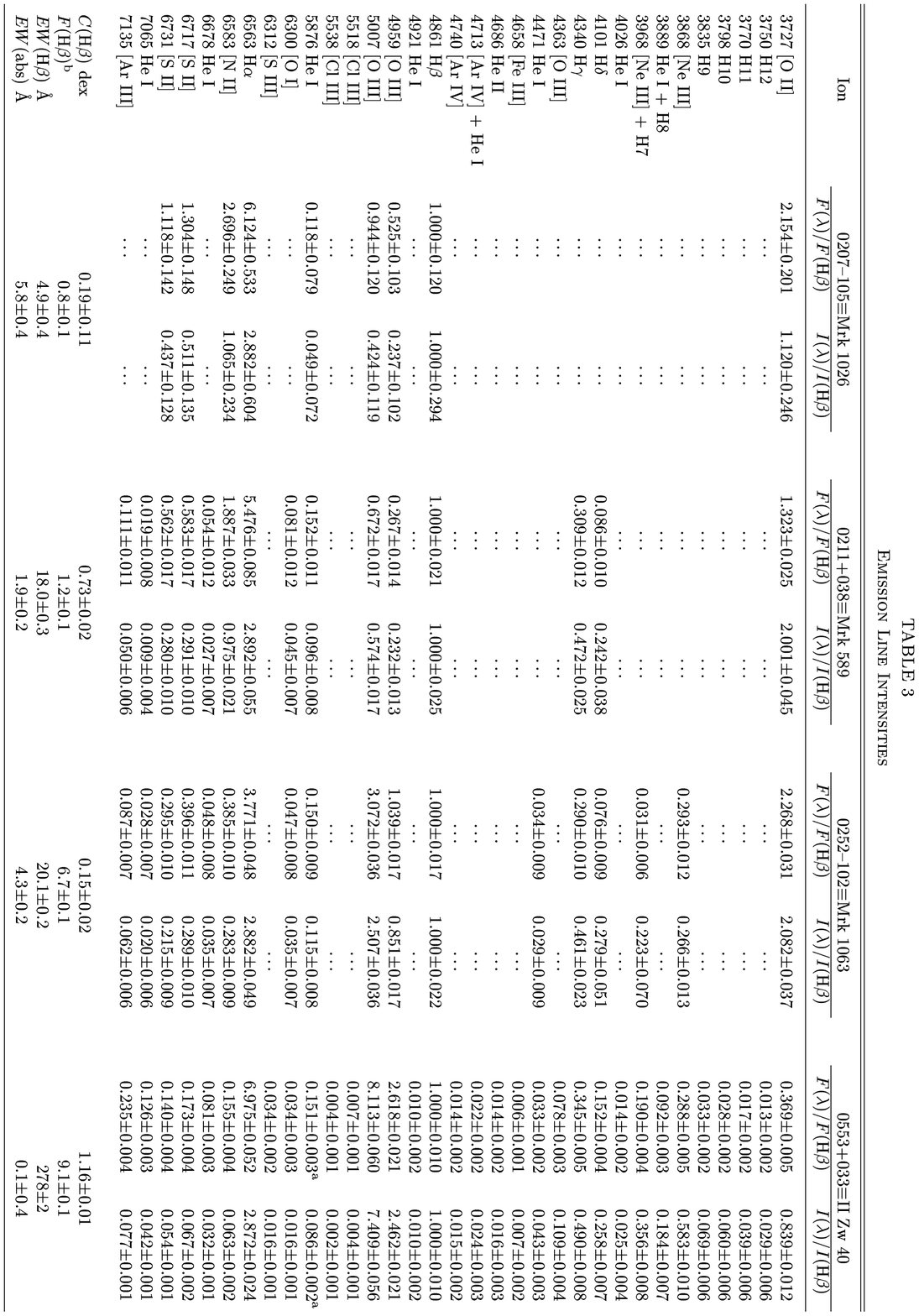}{0.cm}{180.}{80.}{80.}{400.}{80.}
 
\clearpage
%
%
\plotfiddle{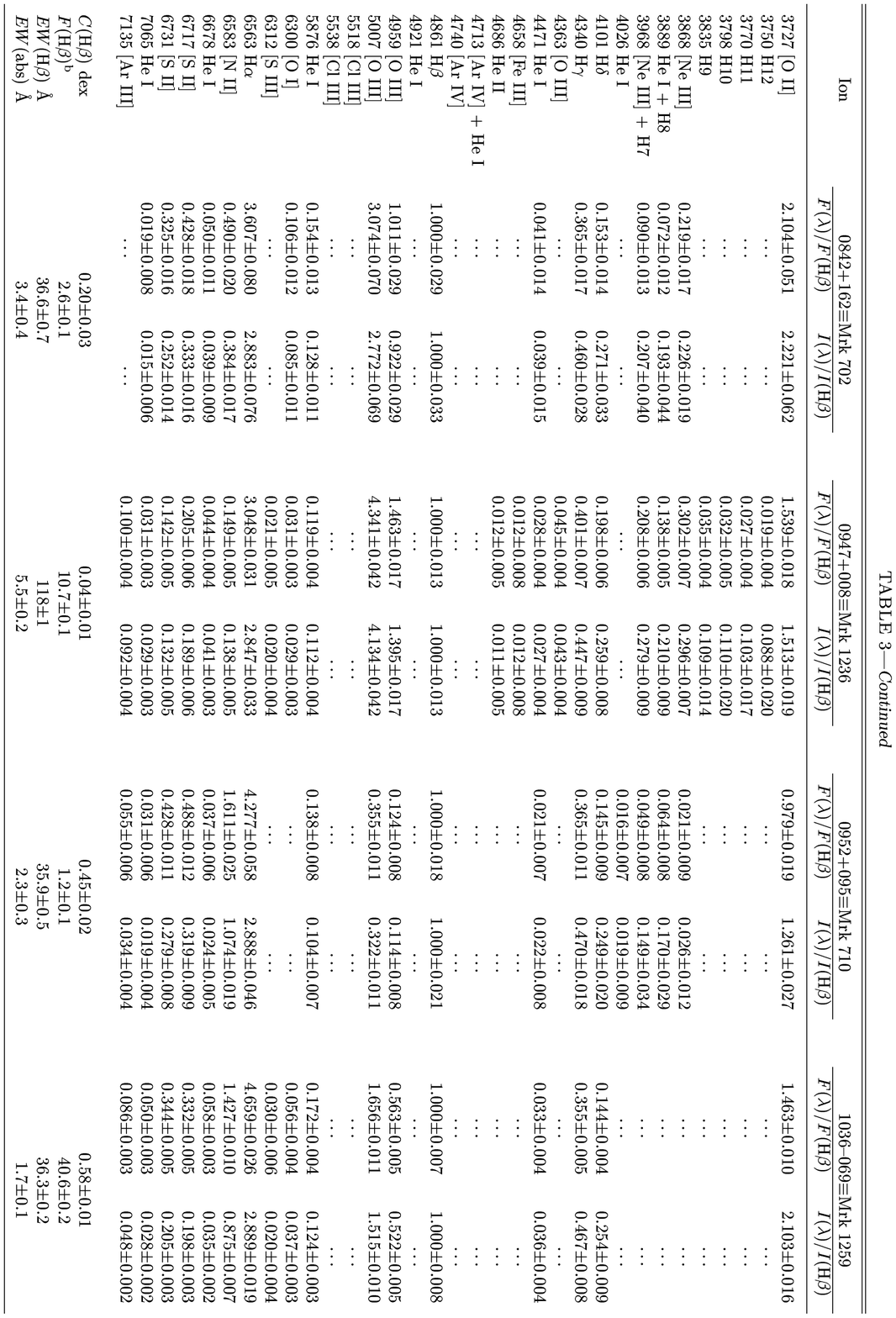}{0.cm}{180.}{80.}{80.}{400.}{80.}

\clearpage
%
%
\plotfiddle{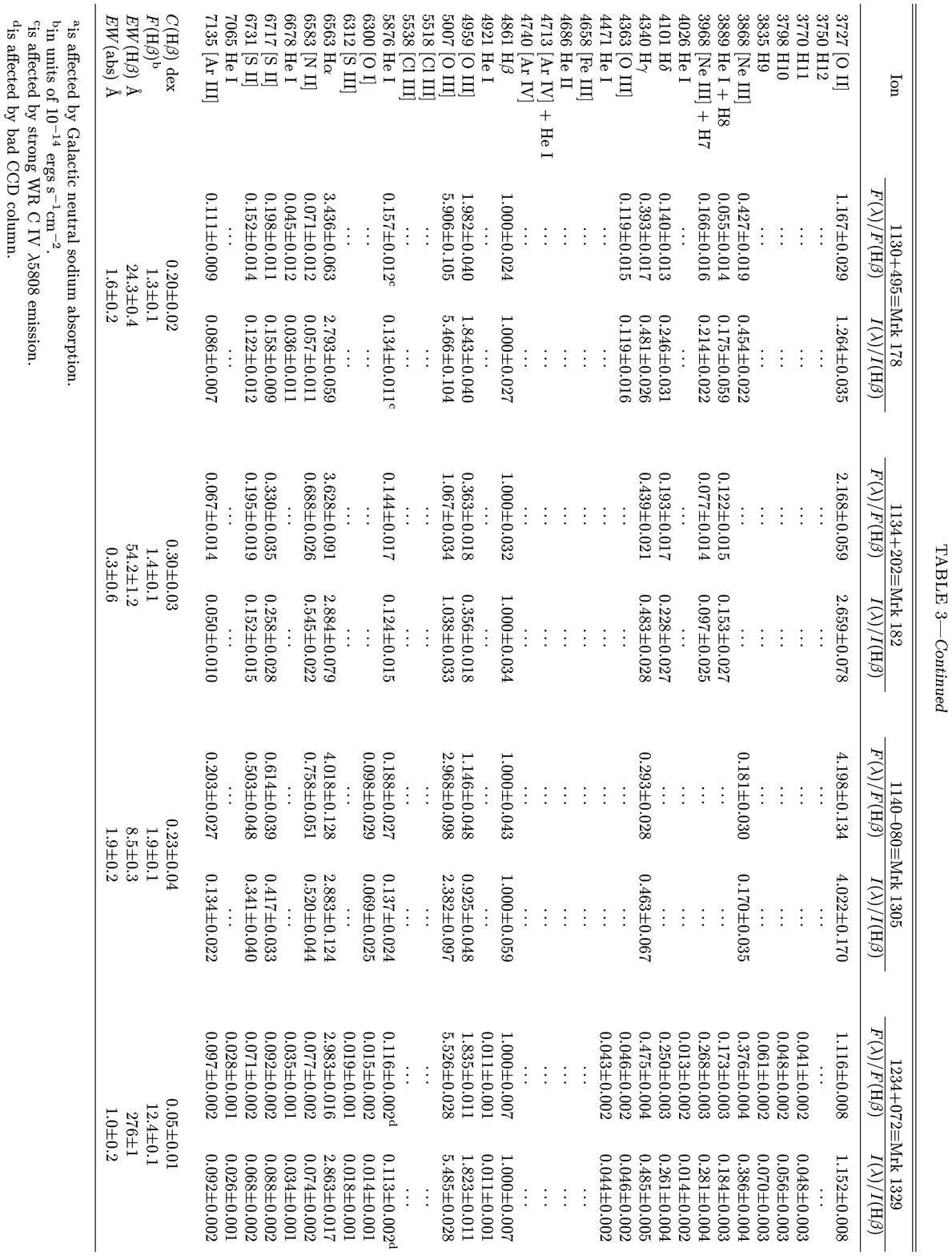}{0.cm}{180.}{80.}{80.}{400.}{80.}

\clearpage
%
%
 
\begin{deluxetable}{lcccc}
\tablenum{4}
\tablecolumns{5}
\tablewidth{400pt}
\tablecaption{Heavy Element Abundances}
\tablehead{
\colhead{Property}&\colhead{II Zw 40}&\colhead{Mrk 1236}&\colhead{Mrk 178}&\colhead{Mrk 1329}  }
\startdata
$T_e$(O III)(K)                     &13,400$\pm$190     &11,700$\pm$390     &15,800$\pm$1,000   &10,900$\pm$150     \nl
$T_e$(O II)(K)                      &12,900$\pm$180   &12,000$\pm$360     &14,100$\pm$\,~~860 &11,500$\pm$140     \nl
$T_e$(S III)(K)                     &13,200$\pm$160   &11,900$\pm$320     &14,800$\pm$\,~~840 &11,200$\pm$130     \nl
$N_e$(S II)(cm$^{-3}$)              &   190$\pm$50      &    10$\pm$10      &   100$\pm$160     &   130$\pm$40      \nl \nl
O$^+$/H$^+$($\times$10$^5$)         &0.85$\pm$0.02      &2.73$\pm$0.27      &1.33$\pm$0.23      &2.47$\pm$0.11      \nl
O$^{++}$/H$^+$($\times$10$^4$)      &1.24$\pm$0.03      &0.88$\pm$0.09      &0.52$\pm$0.08      &1.46$\pm$0.06      \nl
O$^{+++}$/H$^+$($\times$10$^5$)     &0.19$\pm$0.03      &0.10$\pm$0.04      &\nodata            &\nodata            \nl
O/H($\times$10$^4$)                 &1.22$\pm$0.04      &1.16$\pm$0.09      &0.66$\pm$0.09      &1.71$\pm$0.06      \nl
12 + log(O/H)                       &8.09$\pm$0.02      &8.07$\pm$0.03      &7.82$\pm$0.06      &8.23$\pm$0.02      \nl \nl
N$^{+}$/H$^+$($\times$10$^6$)       &0.63$\pm$0.03      &1.62$\pm$0.16      &0.48$\pm$0.08      &0.96$\pm$0.04      \nl
ICF(N)                              &10.3\,~~~~~~~~~~   &4.26~~~~~~~~       &4.95~~~~~~~~       &6.90~~~~~~~~       \nl 
log(N/O)                            &--1.27$\pm$0.03\,~~&--1.23$\pm$0.08\,~~&--1.45$\pm$0.13\,~~&--1.41$\pm$0.04\,~~\nl \nl
Ne$^{++}$/H$^+$($\times$10$^5$)     &2.01$\pm$0.09      &1.58$\pm$0.17      &0.95$\pm$0.16      &2.70$\pm$0.13      \nl
ICF(Ne)                             &1.13~~~~~~~~       &1.32~~~~~~~~       &1.25~~~~~~~~       &1.17~~~~~~~~       \nl 
log(Ne/O)                           &--0.73$\pm$0.03\,~~&--0.75$\pm$0.07\,~~&--0.74$\pm$0.11\,~~&--0.73$\pm$0.03\,~~\nl \nl
S$^+$/H$^+$($\times$10$^6$)         &0.16$\pm$0.01      &0.48$\pm$0.03      &0.31$\pm$0.03      &0.26$\pm$0.01      \nl
S$^{++}$/H$^+$($\times$10$^6$)      &1.21$\pm$0.10      &2.20$\pm$0.54      &\nodata            &2.51$\pm$0.21      \nl
ICF(S)                              &2.49~~~~~~~~       &1.47~~~~~~~~       &\nodata            &1.90~~~~~~~~       \nl 
log(S/O)                            &--1.55$\pm$0.03\,~~&--1.47$\pm$0.08\,~~&\nodata            &--1.51$\pm$0.03\,~~\nl \nl
Ar$^{++}$/H$^+$($\times$10$^6$)     &0.35$\pm$0.01      &0.51$\pm$0.03      &0.32$\pm$0.04      &0.58$\pm$0.02      \nl
Ar$^{+3}$/H$^+$($\times$10$^6$)     &0.29$\pm$0.04      &\nodata            &\nodata            &\nodata            \nl
ICF(Ar)                             &1.01~~~~~~~~       &1.76~~~~~~~~       &1.90~~~~~~~~       &2.26~~~~~~~~       \nl
log(Ar/O)                           &--2.27$\pm$0.03\,~~&--2.11$\pm$0.04\,~~&--2.04$\pm$0.07\,~~&--2.12$\pm$0.02\,~~\nl \nl
Fe$^{++}$/H$^+$($\times$10$^6$)     &0.19$\pm$0.05      &0.41$\pm$0.30      &\nodata            &\nodata            \nl
ICF(Fe)                             &12.8~~~~~~~~~~\,   &5.32~~~~~~~~       &\nodata            &\nodata            \nl 
log(Fe/O)                           &--1.70$\pm$0.03\,~~&--1.72$\pm$0.14\,~~&\nodata            &\nodata            \nl
[O/Fe]                              &0.28$\pm$0.03      &0.30$\pm$0.14      &\nodata            &\nodata            \nl
\enddata
\end{deluxetable}

\clearpage

\begin{deluxetable}{lcrcrrcc}
\tablenum{5}
\tablecolumns{9}
\tablewidth{0pt}
\tablecaption{Parameters of H II Regions}
\tablehead{
\colhead{Galaxy}
&\colhead{$C$(H$\beta$)}
&\multicolumn{1}{c}{$F$(H$\beta$)\tablenotemark{a,c}}
&\colhead{$C_{cor}$}
&\multicolumn{1}{c}{$F_{cor}$(H$\beta$)\tablenotemark{b, c}}
&\multicolumn{1}{c}{$EW$(H$\beta$)}
&\colhead{Age\tablenotemark{d}}
&\colhead{$\eta_0$} }
\startdata
0112$-$011    & 0.15&  15.8&2.06&  32.6& 252.2& 4.2& 0.40 \nl
0211+038      & 0.72&  38.7&2.24&  86.9&  17.9& 5.8& \nodata\tablenotemark{e} \nl
0218+003      & 0.31&   5.7&1.78&  10.2&  99.8& 4.5& 0.30 \nl
0252$-$102    & 0.16&  16.7&3.11&  52.0&  21.8& 6.0& 0.30 \nl
0459$-$043    & 0.28&  31.6&4.90& 155.0&  56.1& 4.8& 0.25 \nl
0553+033      & 1.20& 238.0&2.86& 680.0& 272.2& 4.0& 0.50 \nl
0635+756      & 0.42&  11.6&2.17&  25.2& 118.6& 4.2& 0.45 \nl
0720+335      & 0.55&  44.5&2.06&  91.7&  20.3& 5.5& \nodata\tablenotemark{e} \nl
0723+692      & 0.12& 114.0&5.05& 576.0& 264.4& 3.6& 0.85 \nl
0842+162      & 0.20&   4.5&2.64&  11.9&  35.0& 5.3& 1.00 \nl
0926+606      & 0.18&   6.5&3.60&  23.4& 108.6& 4.4& 0.35 \nl
0930+554      & 0.13&   3.9&2.18&   8.5&  67.0& 5.9& 0.20 \nl
0946+558      & 0.18&   6.3&2.24&  14.1&  99.5& 4.5& 0.40 \nl
0947+008      & 0.04&  16.1&3.80&  61.1& 124.3& 4.1& 0.50 \nl
0948+532      & 0.04&   4.1&2.00&   8.1& 169.8& 3.9& 0.65 \nl
0952+095      & 0.52&  21.9&2.30&  52.6&  33.0& 5.3& \nodata\tablenotemark{e} \nl
1030+583      & 0.03&   3.7&2.55&   9.4&  84.1& 4.7& 0.35 \nl
1036$-$069    & 0.60& 210.0&2.13& 448.0&  32.1& 5.3& \nodata\tablenotemark{e} \nl
1053+064      & 0.16&  13.6&2.62&  35.7&  78.0& 4.7& 0.30 \nl
1054+365      & 0.05&   9.1&2.70&  24.6&  64.3& 4.8& 0.25 \nl
1130+495      & 0.21&   4.0&5.02&  19.9&  20.1& 8.0& 0.10 \nl
1135+581      & 0.14&  29.9&2.25&  67.4& 116.5& 4.2& 0.50 \nl
1139+006      & 0.34&  56.5&4.05& 229.0&  48.9& 5.3& 0.20 \nl
1147+153      & 0.14&  25.2&2.16&  54.4& 130.1& 4.3& 0.35 \nl
1150$-$021    & 0.29&  37.7&6.02& 227.0&  87.8& 4.6& 0.30 \nl
1211+540      & 0.10&   3.0&2.04&   6.0& 115.1& 4.0& 0.50 \nl
1222+614      & 0.00&   9.3&3.37&  31.2&  85.2& 4.6& 0.30 \nl
1223+487      & 0.06&  17.2&3.10&  53.4& 206.2& 3.9& 0.70 \nl
1234+072      & 0.05&  17.2&2.00&  34.4& 166.5& 4.2& 0.45 \nl
1256+351      & 0.09&  22.1&3.80&  84.1& 144.1& 4.1& 0.50 \nl
1319+579A     & 0.01&   5.1&2.07&  10.6& 173.0& 4.1& 0.50 \nl
1437+370      & 0.14&   8.4&3.17&  26.6& 127.4& 4.1& 0.50 \nl
2329+286      & 0.26&  13.4&2.46&  33.0&  90.4& 4.7& 0.30 \nl
\enddata
\tablenotetext{a}{flux measured along the slit and corrected for extinction.}
\tablenotetext{b}{total flux after correction for extinction and aperture: 
$F_{cor}$(H$\beta$) = $C_{cor}$$F$(H$\beta$).}
\tablenotetext{c}{in units of 10$^{-14}$ ergs cm$^{-2}$s$^{-1}$.}
\tablenotetext{d}{in Myr.}
\tablenotetext{e}{parameter $\eta_0$ is undefined (see text).}
\end{deluxetable}

\clearpage

%
%
\plotfiddle{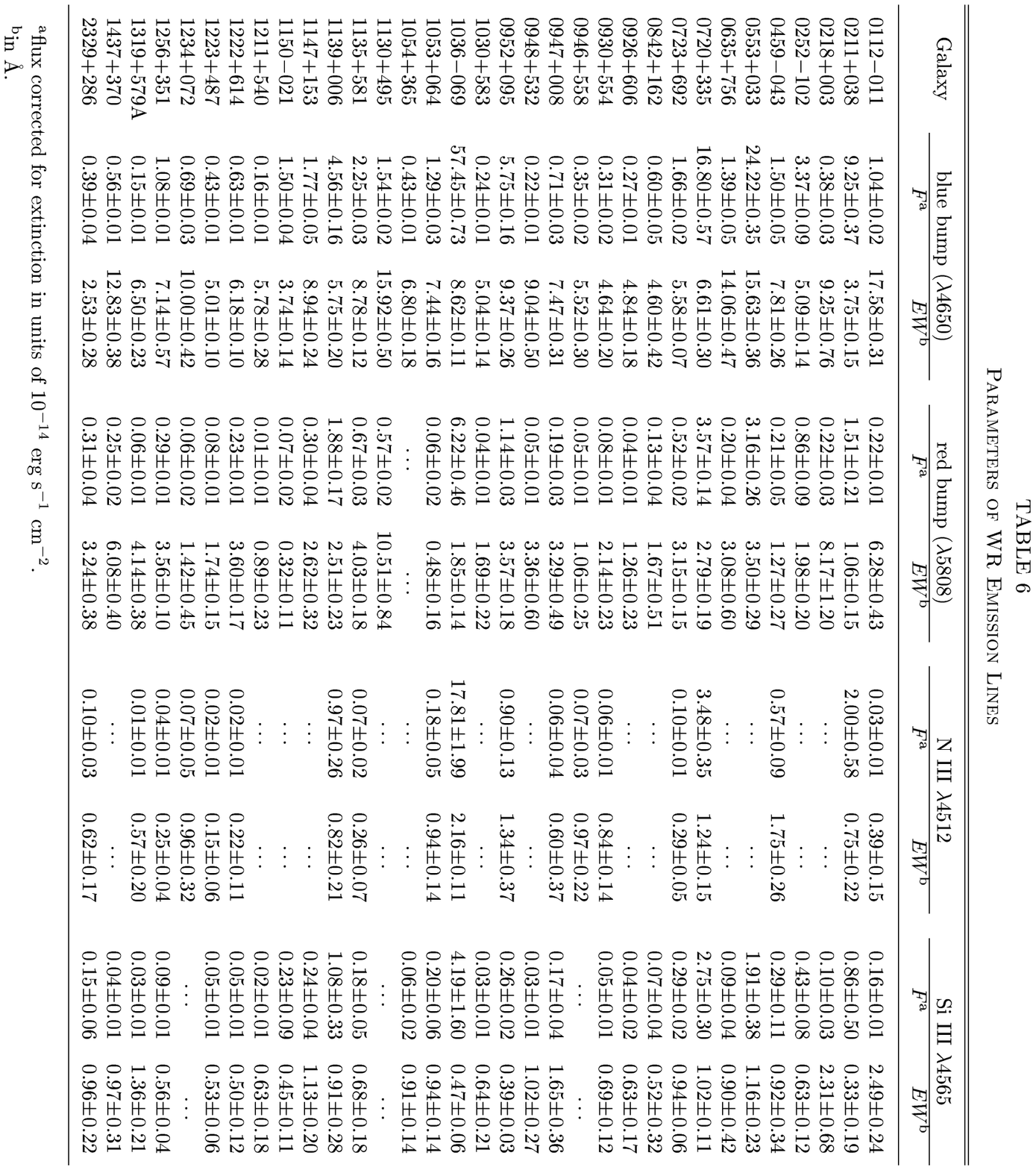}{0.cm}{180.}{90.}{90.}{400.}{150.}

\clearpage
 
\begin{deluxetable}{lrrrrrcc}
\tablenum{7}
\tablecolumns{8}
\tablewidth{0pt}
\tablecaption{Numbers of Massive Stars}
\tablehead{
\colhead{Galaxy}&\colhead{$N$(O)}&\multicolumn{3}{c}{$N$(WNL)}&\colhead{$N$(WCE)}&
\multicolumn{1}{c}{$N$(WR)/$N$(O+WR)\tablenotemark{a}}&\colhead{$N$(WC)/$N$(WN)\tablenotemark{a}} \\ \cline{3-5} 
\colhead{ }&\colhead{}&\colhead{$\lambda$4650}&\colhead{$\lambda$4512}&\colhead{$\lambda$4565}&\colhead{ }&
\multicolumn{1}{c}{}&\colhead{}
 }
\startdata
0112$-$011   &  10038.6&  209.6&   25.7&  147.4&   45.6& 0.025$\pm$0.001& 0.218$\pm$0.018 \nl
0211+038     &      0.0\tablenotemark{b}& 6508.1& 7634.1& 3281.0& 1277.6& 1.000\tablenotemark{b}& 0.196$\pm$0.032 \nl
0218+003     & 419093.2&  504.2&\nodata&10026.1& 4702.2& 0.034$\pm$0.007& 0.469$\pm$0.156 \nl
0252$-$102   &  15546.9&  461.0&\nodata&  309.4&  137.5& 0.037$\pm$0.003& 0.298$\pm$0.041 \nl
0459$-$043   & 441985.6& 1977.9& 2993.4& 1516.2&  246.5& 0.005$\pm$0.001& 0.125$\pm$0.028 \nl
0553+033     &  31497.3& 1125.0&\nodata&  343.8&  125.5& 0.038$\pm$0.001& 0.112$\pm$0.010 \nl
0635+756     &   1397.1&   70.5&\nodata&   19.2&    8.9& 0.054$\pm$0.004& 0.127$\pm$0.027 \nl
0720+335     &      0.0\tablenotemark{b}&14526.5&18452.0&14574.0& 4179.1& 1.000\tablenotemark{b}& 0.288$\pm$0.020 \nl
0723+692     &   2082.2&    5.7&    2.1&    6.5&    2.6& 0.004$\pm$0.001& 0.453$\pm$0.034 \nl
0842+162     & 122502.4&10002.0&\nodata& 5691.0& 2426.3& 0.092$\pm$0.019& 0.243$\pm$0.096 \nl
0926+606     &  49483.1&  368.0&\nodata&  218.9&   47.0& 0.008$\pm$0.001& 0.128$\pm$0.026 \nl
0930+554     &   3996.3\tablenotemark{c}&   43.6\tablenotemark{c}&   45.0\tablenotemark{c}&   36.0\tablenotemark{c}&   12.0\tablenotemark{c}& 0.014$\pm$0.001& 0.276$\pm$0.044 \nl
0946+558     &   3723.4&   68.2&   55.5&\nodata&    9.4& 0.020$\pm$0.002& 0.137$\pm$0.036 \nl
0947+008     &  18923.0&  145.0&   70.3&  187.4&   47.5& 0.010$\pm$0.001& 0.328$\pm$0.068 \nl
0948+532     & 102938.3& 2707.7&\nodata& 1617.6&  709.0& 0.032$\pm$0.004& 0.262$\pm$0.062 \nl
0952+095     &      0.0\tablenotemark{b}&  687.7&  633.7&  180.2&  178.0& 1.000\tablenotemark{b}& 0.259$\pm$0.013 \nl
1030+583     &   5890.1&   89.5&\nodata&   53.5&   16.3& 0.018$\pm$0.001& 0.182$\pm$0.027 \nl
1036$-$069   &      0.0\tablenotemark{b}&17891.3&26592.9& 6255.6& 2051.1& 1.000\tablenotemark{b}& 0.115$\pm$0.009 \nl
1053+064     &   5061.9&  130.9&   60.5&   67.3&    4.8& 0.026$\pm$0.001& 0.036$\pm$0.012 \nl
1054+365     &   1490.3&   16.3&\nodata&    7.2&\nodata& 0.011$\pm$0.001& \nodata         \nl
1130+495     &    389.4&    3.0&\nodata&\nodata&    2.0& 0.013$\pm$0.001& 0.668$\pm$0.049 \nl
1135+581     &   5938.1&  122.7&   24.1&   60.3&   49.8& 0.028$\pm$0.001& 0.406$\pm$0.028 \nl
1139+006     &1504191.1& 4350.3& 9422.0&10497.9& 4024.9& 0.006$\pm$0.001& 0.925$\pm$0.248 \nl
1147+153     &   3463.1&   71.8&\nodata&   40.8&   11.3& 0.023$\pm$0.001& 0.158$\pm$0.021 \nl
1150$-$021   &  35147.6&  155.8&\nodata&   79.0&    5.3& 0.005$\pm$0.001& 0.034$\pm$0.012 \nl
1211+540     &    435.6&   12.3&\nodata&    5.3&    0.9& 0.029$\pm$0.002& 0.071$\pm$0.020 \nl
1222+614     &   2402.3&   13.5&    4.0&    9.1&    8.6& 0.009$\pm$0.001& 0.638$\pm$0.065 \nl
1223+487     &    279.2&    2.5&    0.4&    1.4&    0.5& 0.011$\pm$0.001& 0.199$\pm$0.019 \nl
1234+072     &   8370.9&  163.1&   60.1&\nodata&   10.4& 0.020$\pm$0.001& 0.064$\pm$0.020 \nl
1256+351     &   4906.4&   41.8&    9.0&   19.5&   13.7& 0.011$\pm$0.001& 0.328$\pm$0.013 \nl
1319+579A    &   3540.7&   19.7&   16.9&   39.4&   15.7& 0.010$\pm$0.001& 0.798$\pm$0.182 \nl
1437+370     &    391.9&    2.3&\nodata&    2.4&    3.0& 0.013$\pm$0.001& 1.316$\pm$0.327 \nl
2329+286     & 143970.0&--445.9& 1032.6& 1531.2&  668.6& 0.015$\pm$0.004& 0.437$\pm$0.164 \nl
\enddata
\tablenotetext{a}{for all galaxies, the ratio is derived from $N$(WNL) obtained
with the blue bump $\lambda$4650, except for 0218+003 and 2329+286 where
$N$(WNL) obtained with Si III $\lambda$4565 is used.}
\tablenotetext{b}{O stars are not expected (Schaerer \& Vacca 1998) and the 
ratio $N$(WR)/$N$(O+WR) is set to 1.}
\tablenotetext{c}{the distance of 20 Mpc is adopted (Izotov et al. 1999b).}
\end{deluxetable}

\clearpage
 
\begin{deluxetable}{lrcccc}
\tablenum{8}
\tablecolumns{6}
\tablewidth{0pt}
\tablecaption{Parameters of the Nebular He II $\lambda$4686 Emission Line}
\tablehead{
\colhead{Galaxy}&\multicolumn{1}{c}{$EW$(H$\beta$)\tablenotemark{a}}&\colhead{12+log(O/H)}
&\colhead{$I$($\lambda$4686)/$I$(H$\beta$)\tablenotemark{b}}&\colhead{$I$($\lambda$4686)/$I_c$(H$\beta$)\tablenotemark{c}}&\colhead{$EW$($\lambda$4686)\tablenotemark{a}} 
}
\startdata
0335$-$052   &217.2& 7.30& 0.0262$\pm$0.0018& 0.0105$\pm$0.0007&  4.82$\pm$0.26 \nl
0553+033     &272.2& 8.09& 0.0156$\pm$0.0026& 0.0055$\pm$0.0009&  3.61$\pm$0.45 \nl
0723+692     &264.4& 7.85& 0.0093$\pm$0.0003& 0.0018$\pm$0.0001&  2.72$\pm$0.07 \nl
0749+568     &117.1& 7.85& 0.0177$\pm$0.0084& 0.0071$\pm$0.0034&  2.20$\pm$1.00 \nl
0917+527     & 85.9& 7.86& 0.0229$\pm$0.0030& 0.0092$\pm$0.0012&  1.64$\pm$0.07 \nl
0926+606     &108.6& 7.91& 0.0161$\pm$0.0024& 0.0045$\pm$0.0007&  1.64$\pm$0.05 \nl
0930+554     & 67.0& 7.16& 0.0340$\pm$0.0022& 0.0156$\pm$0.0010&  2.54$\pm$0.10 \nl
0947+008     &124.3& 8.07& 0.0110$\pm$0.0047& 0.0029$\pm$0.0012&  1.17$\pm$0.23 \nl
0948+532     &169.8& 8.00& 0.0150$\pm$0.0005& 0.0060$\pm$0.0002&  1.56$\pm$0.14 \nl
1030+583     & 84.1& 7.79& 0.0239$\pm$0.0021& 0.0094$\pm$0.0008&  1.91$\pm$0.07 \nl
1053+064     & 78.0& 7.99& 0.0055$\pm$0.0033& 0.0021$\pm$0.0013&  0.45$\pm$0.12 \nl
1102+450     & 56.0& 8.12& 0.0155$\pm$0.0052& 0.0062$\pm$0.0021&  0.78$\pm$0.13 \nl
1102+294     & 69.8& 7.83& 0.0249$\pm$0.0064& 0.0100$\pm$0.0026&  1.55$\pm$0.17 \nl
1116+583B    &107.3& 7.68& 0.0245$\pm$0.0095& 0.0098$\pm$0.0038&  2.53$\pm$0.31 \nl
1139+006     & 48.9& 7.99& 0.0100$\pm$0.0034& 0.0025$\pm$0.0008&  0.48$\pm$0.09 \nl
1147+153     &130.1& 8.11& 0.0089$\pm$0.0031& 0.0041$\pm$0.0014&  0.99$\pm$0.18 \nl
1150$-$021   & 87.8& 7.95& 0.0092$\pm$0.0024& 0.0015$\pm$0.0004&  0.83$\pm$0.10 \nl
1152+579     &189.3& 7.81& 0.0129$\pm$0.0003& 0.0052$\pm$0.0001&  2.12$\pm$0.09 \nl
1159+545     &249.3& 7.49& 0.0106$\pm$0.0006& 0.0042$\pm$0.0002&  2.29$\pm$0.20 \nl
1205+557     & 63.7& 7.75& 0.0181$\pm$0.0074& 0.0072$\pm$0.0030&  1.19$\pm$0.10 \nl
1222+614     & 85.2& 7.95& 0.0171$\pm$0.0017& 0.0051$\pm$0.0005&  1.30$\pm$0.05 \nl
1223+487     &206.2& 7.77& 0.0117$\pm$0.0005& 0.0038$\pm$0.0002&  3.25$\pm$0.17 \nl
1249+493     &124.1& 7.72& 0.0124$\pm$0.0010& 0.0050$\pm$0.0004&  1.32$\pm$0.22 \nl
1319+579A    &173.0& 8.11& 0.0077$\pm$0.0014& 0.0037$\pm$0.0007&  1.60$\pm$0.05 \nl
1319+579B    & 64.6& 8.09& 0.0138$\pm$0.0044& 0.0067$\pm$0.0021&  1.21$\pm$0.07 \nl
1319+579C    & 16.6& 7.92& 0.0158$\pm$0.0135& 0.0076$\pm$0.0065&  0.30$\pm$0.25 \nl
1415+437     &163.6& 7.59& 0.0251$\pm$0.0005& 0.0100$\pm$0.0002&  3.65$\pm$0.10 \nl
1420+544     &217.3& 7.75& 0.0124$\pm$0.0005& 0.0050$\pm$0.0002&  2.42$\pm$0.17 \nl
1441+294     & 56.2& 7.99& 0.0189$\pm$0.0126& 0.0076$\pm$0.0050&  1.37$\pm$0.17 \nl
1851+693     &368.4& 7.79& 0.0131$\pm$0.0012& 0.0052$\pm$0.0005&  4.90$\pm$0.14 \nl
\enddata
\tablenotetext{a}{in \AA.}
\tablenotetext{b}{corrected for interstellar extinction only.}
\tablenotetext{c}{corrected for both interstellar extinction and aperture.}
\end{deluxetable}

\clearpage
%
%
\begin{figure*}
\figurenum{1}
\epsscale{1.6}
\plotfiddle{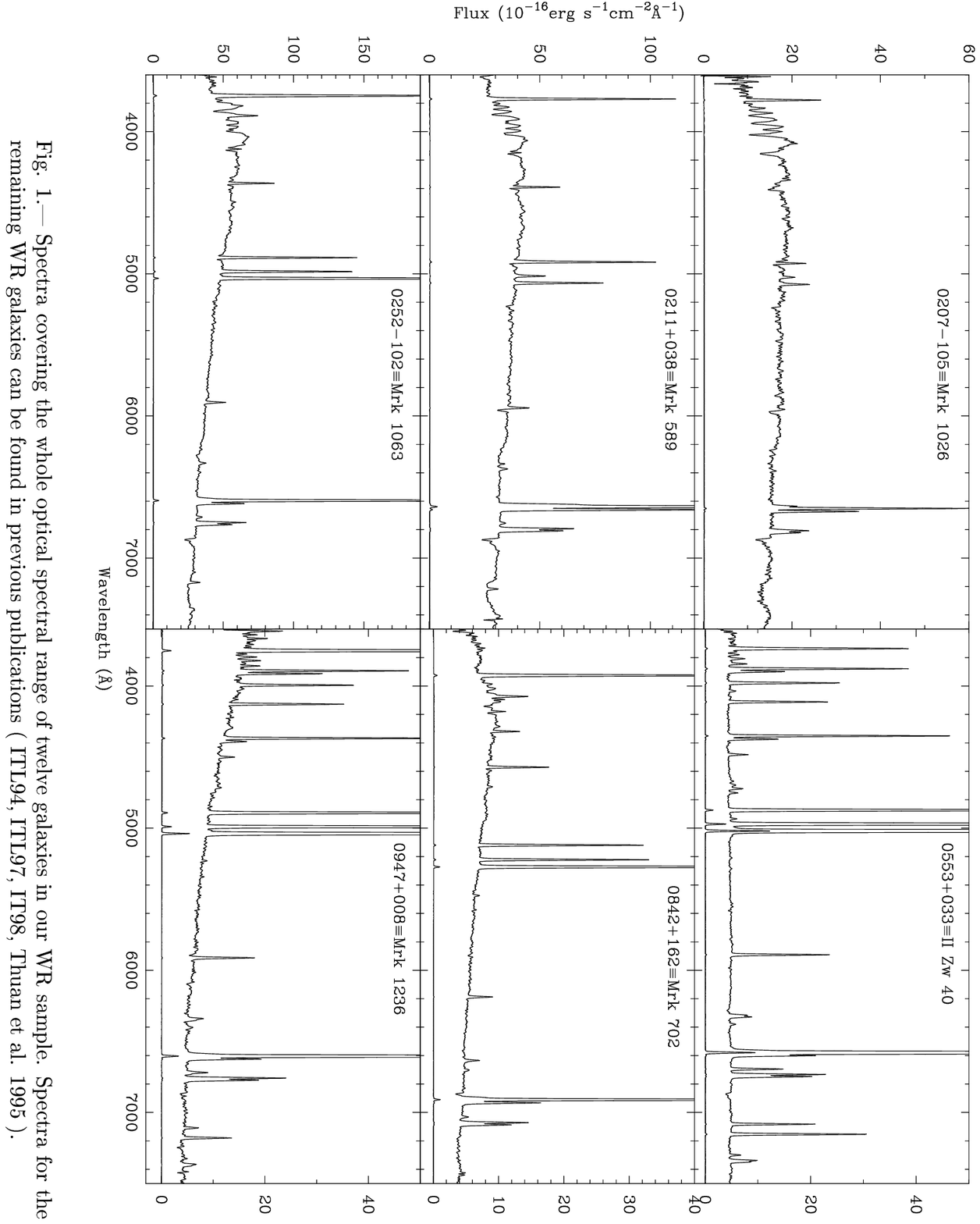}{0.cm}{180.}{100.}{100.}{250.}{500.}
\end{figure*}

\clearpage
%
%
\begin{figure*}
\figurenum{1}
\epsscale{1.6}
\plotfiddle{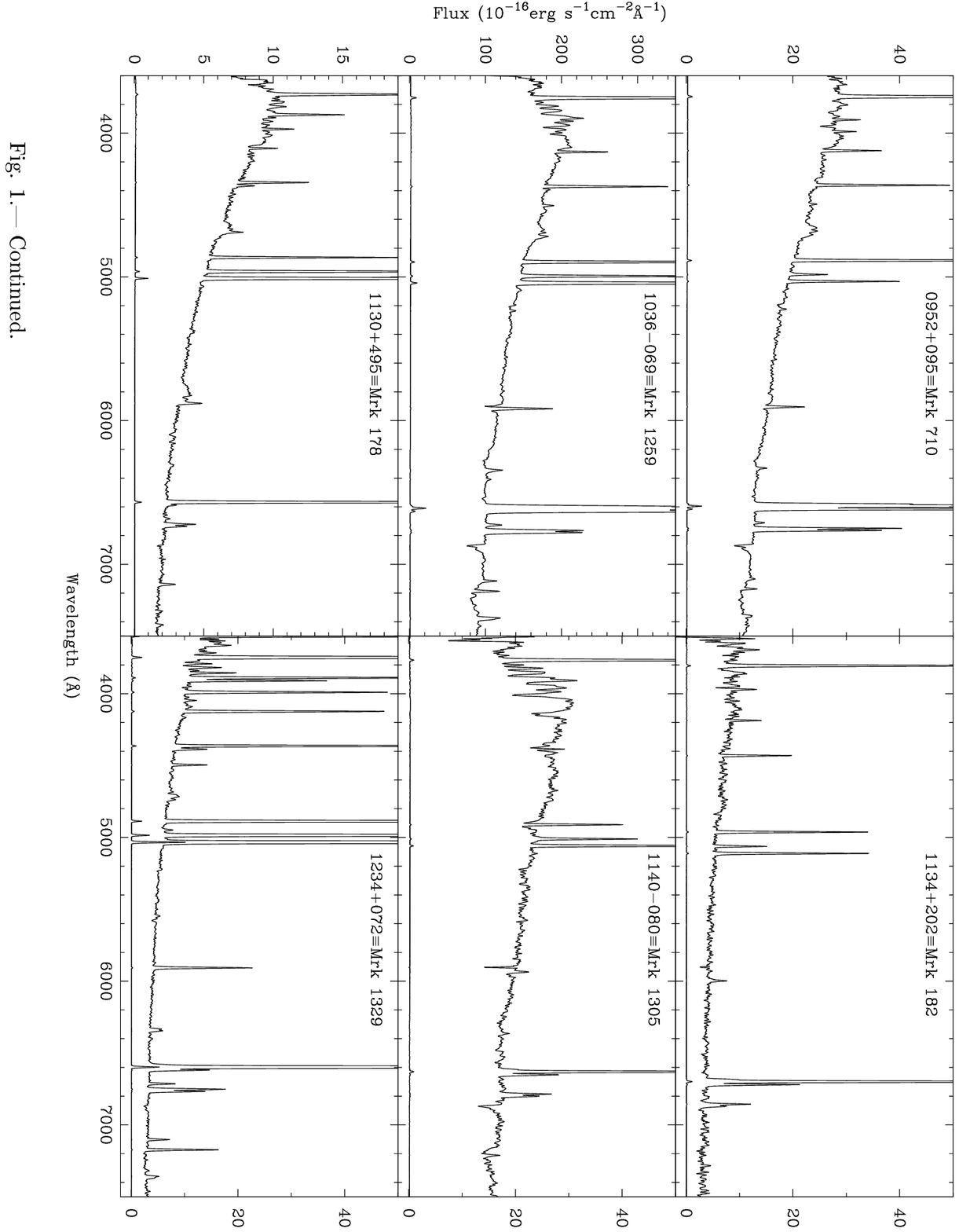}{0.cm}{180.}{100.}{100.}{250.}{500.}
\end{figure*}

\clearpage
%
%
\begin{figure*}
\figurenum{2}
\epsscale{1.6}
\plotfiddle{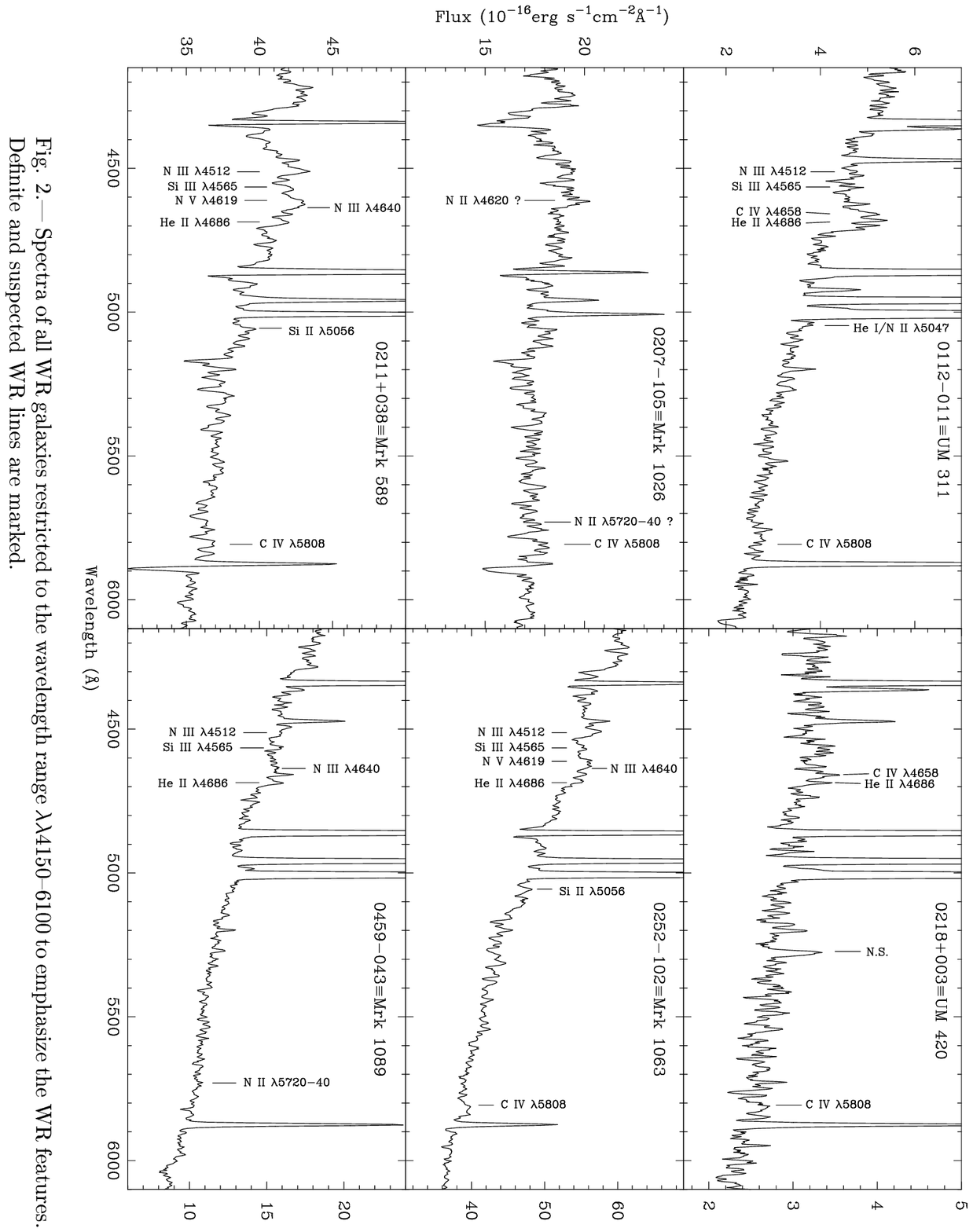}{0.cm}{180.}{100.}{100.}{250.}{500.}
\end{figure*}

\clearpage
%
%
\begin{figure*}
\figurenum{2}
\epsscale{1.6}
\plotfiddle{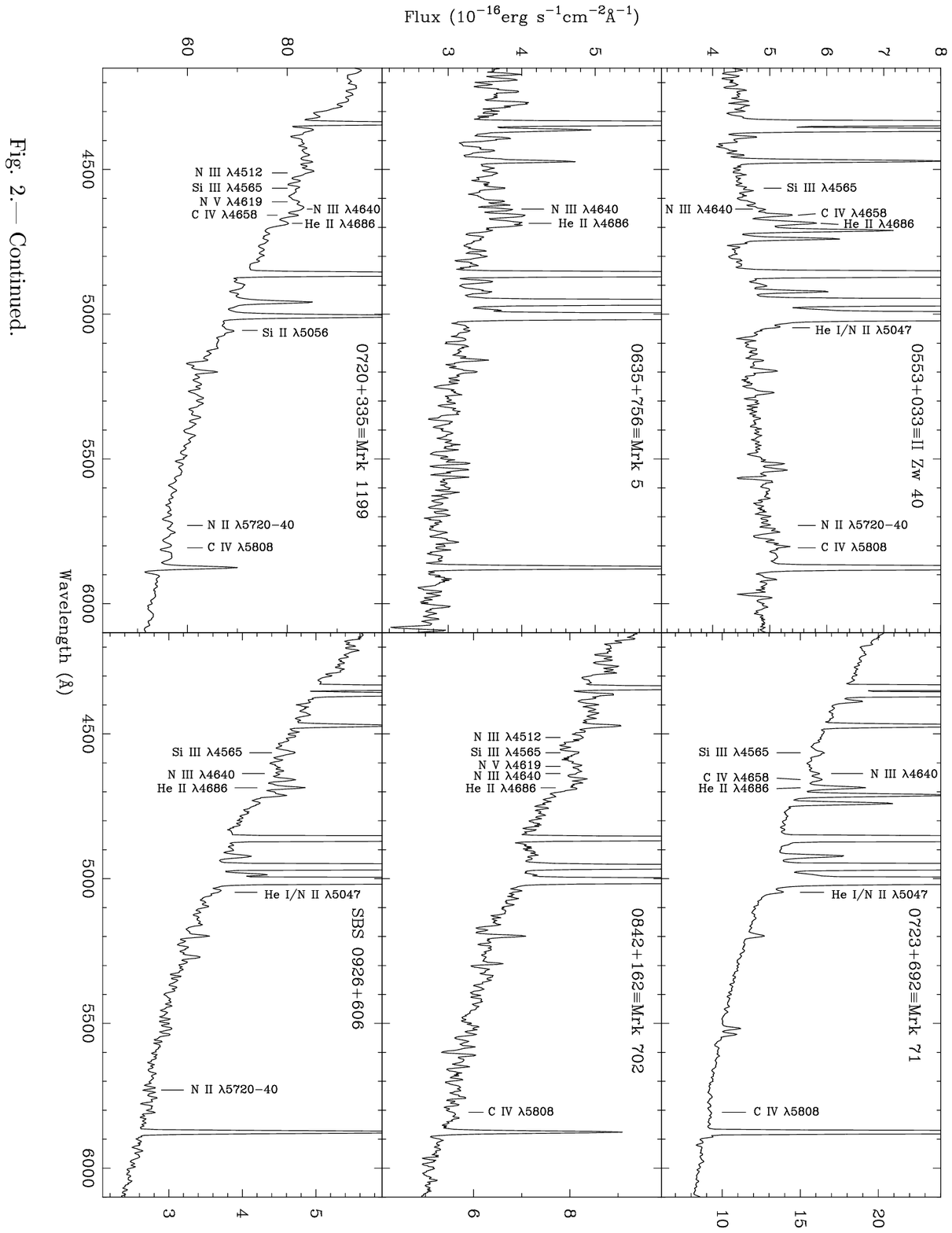}{0.cm}{180.}{100.}{100.}{250.}{500.}
\end{figure*}

\clearpage
%
%
\begin{figure*}
\figurenum{2}
\epsscale{1.6}
\plotfiddle{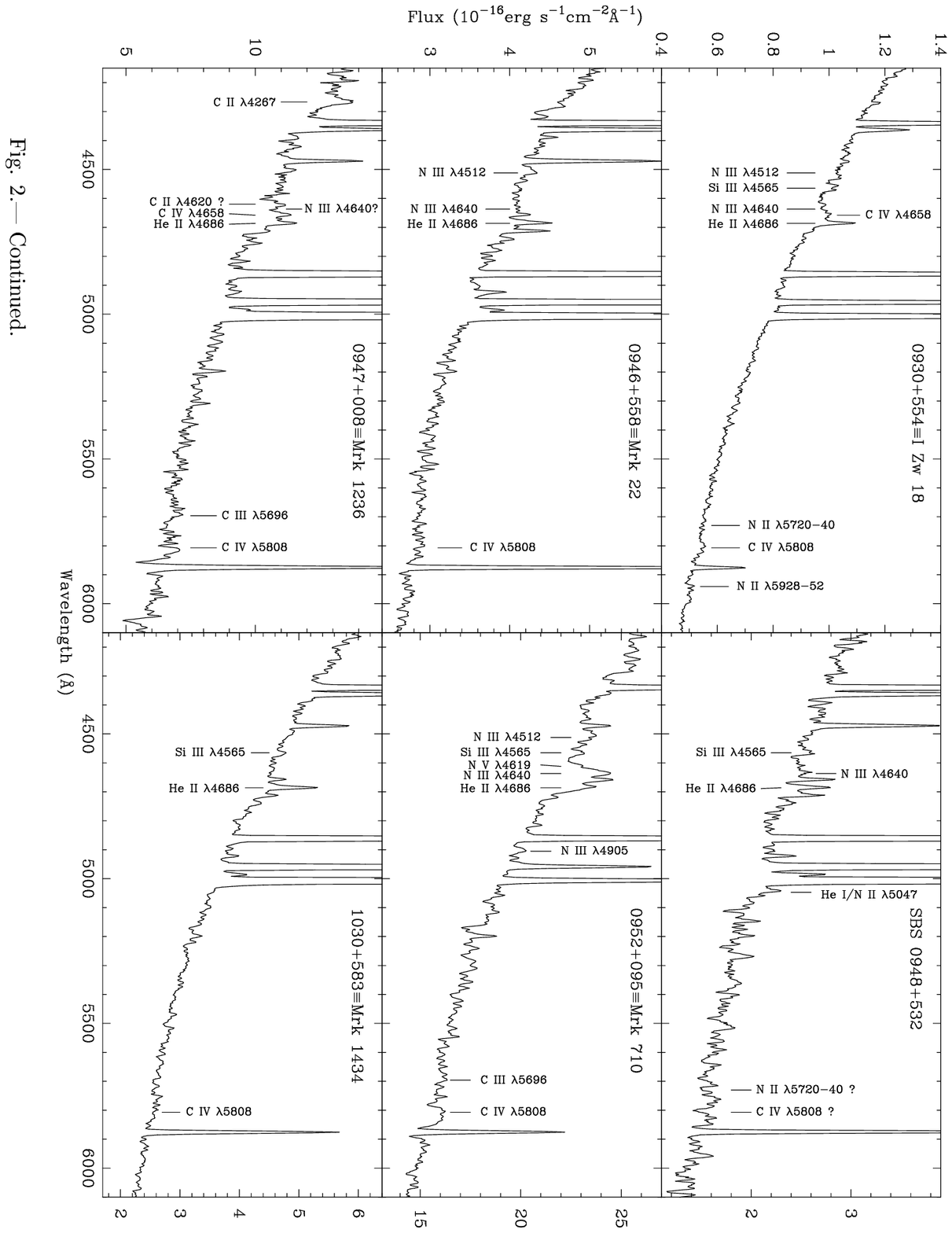}{0.cm}{180.}{100.}{100.}{250.}{500.}
\end{figure*}

\clearpage
%
%
\begin{figure*}
\figurenum{2}
\epsscale{1.6}
\plotfiddle{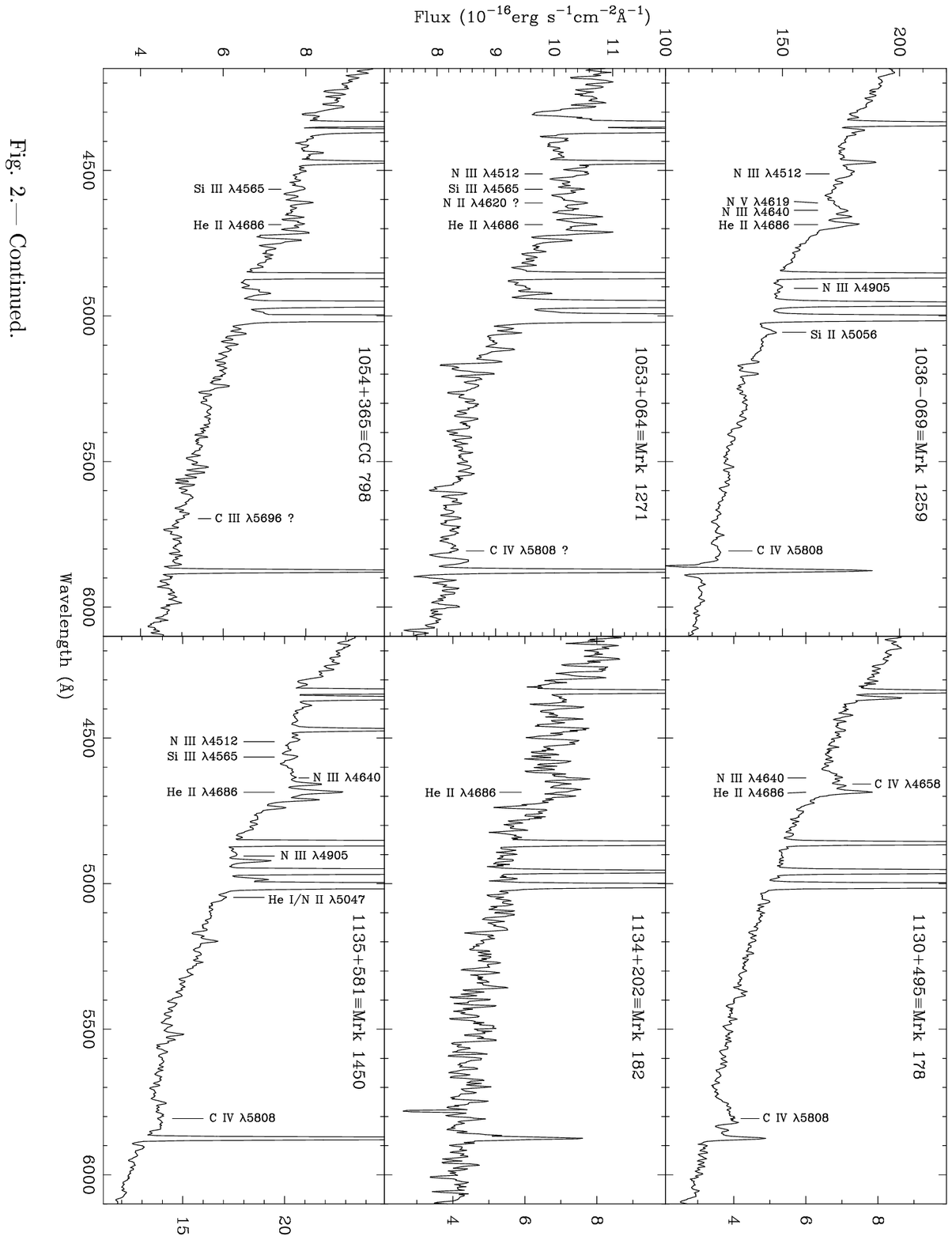}{0.cm}{180.}{100.}{100.}{250.}{500.}
\end{figure*}

\clearpage
%
%
\begin{figure*}
\figurenum{2}
\epsscale{1.6}
\plotfiddle{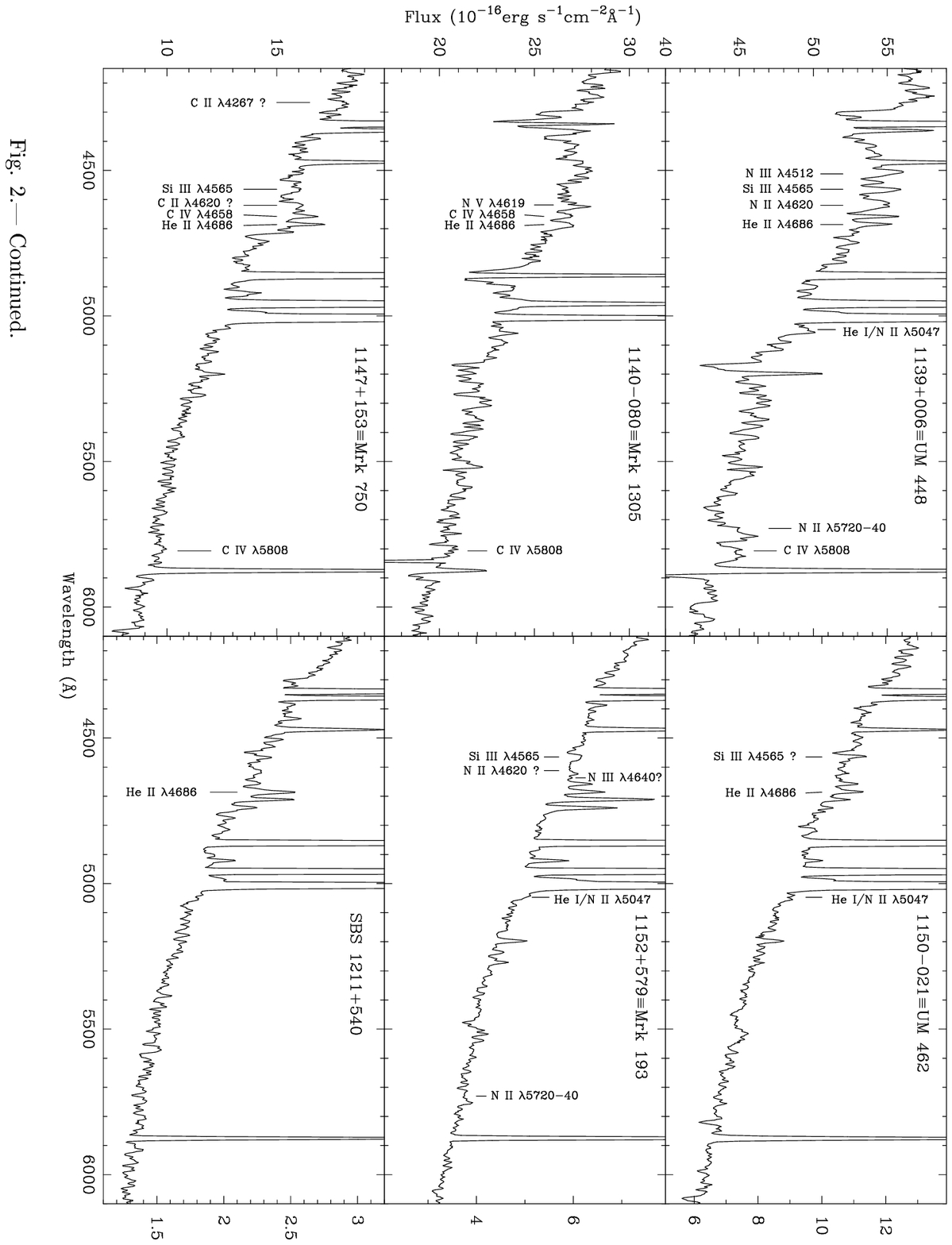}{0.cm}{180.}{100.}{100.}{250.}{500.}
\end{figure*}

\clearpage
%
%
\begin{figure*}
\figurenum{2}
\epsscale{1.6}
\plotfiddle{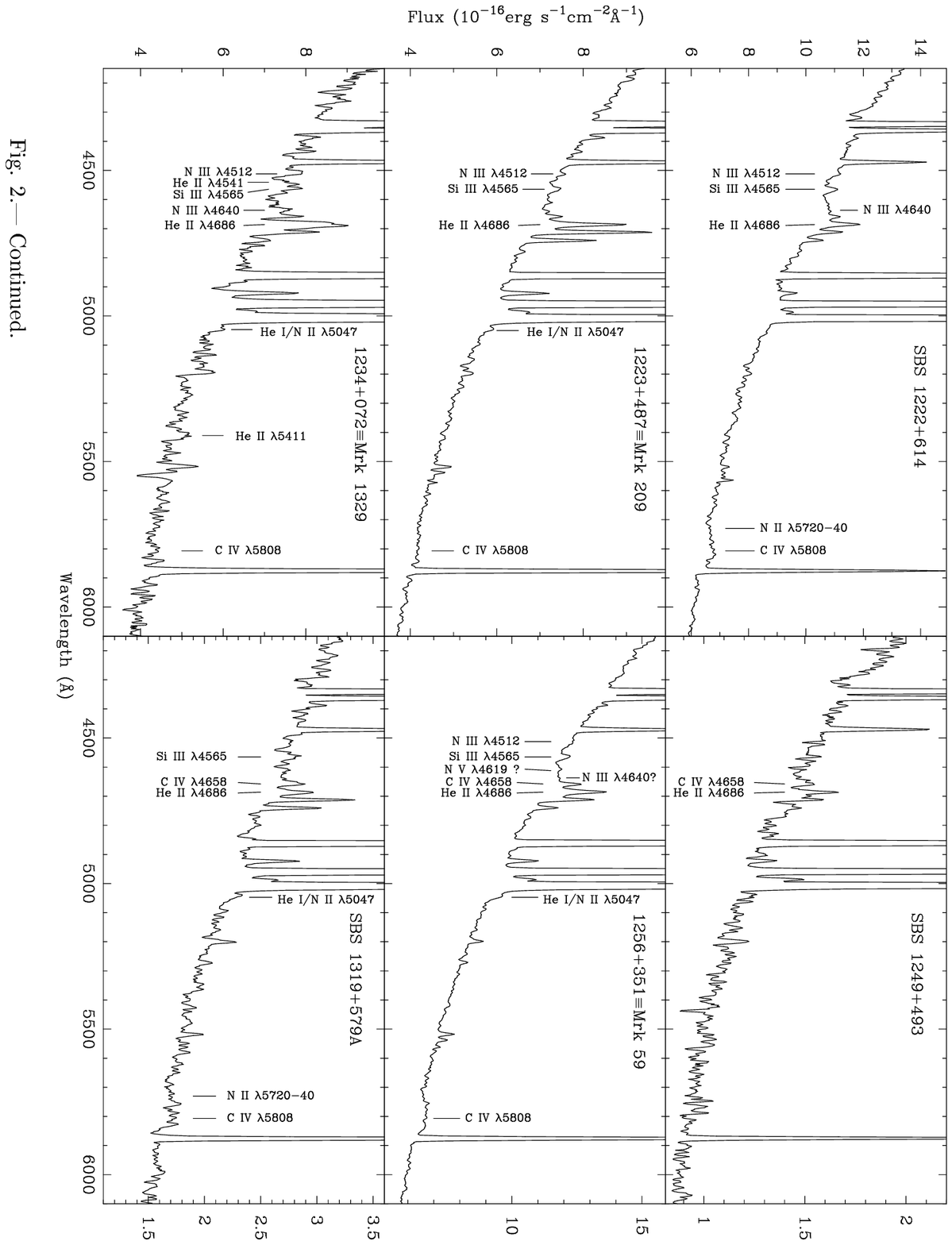}{0.cm}{180.}{100.}{100.}{250.}{500.}
\end{figure*}

\clearpage
%
%
\begin{figure*}
\figurenum{2}
\epsscale{1.6}
\plotfiddle{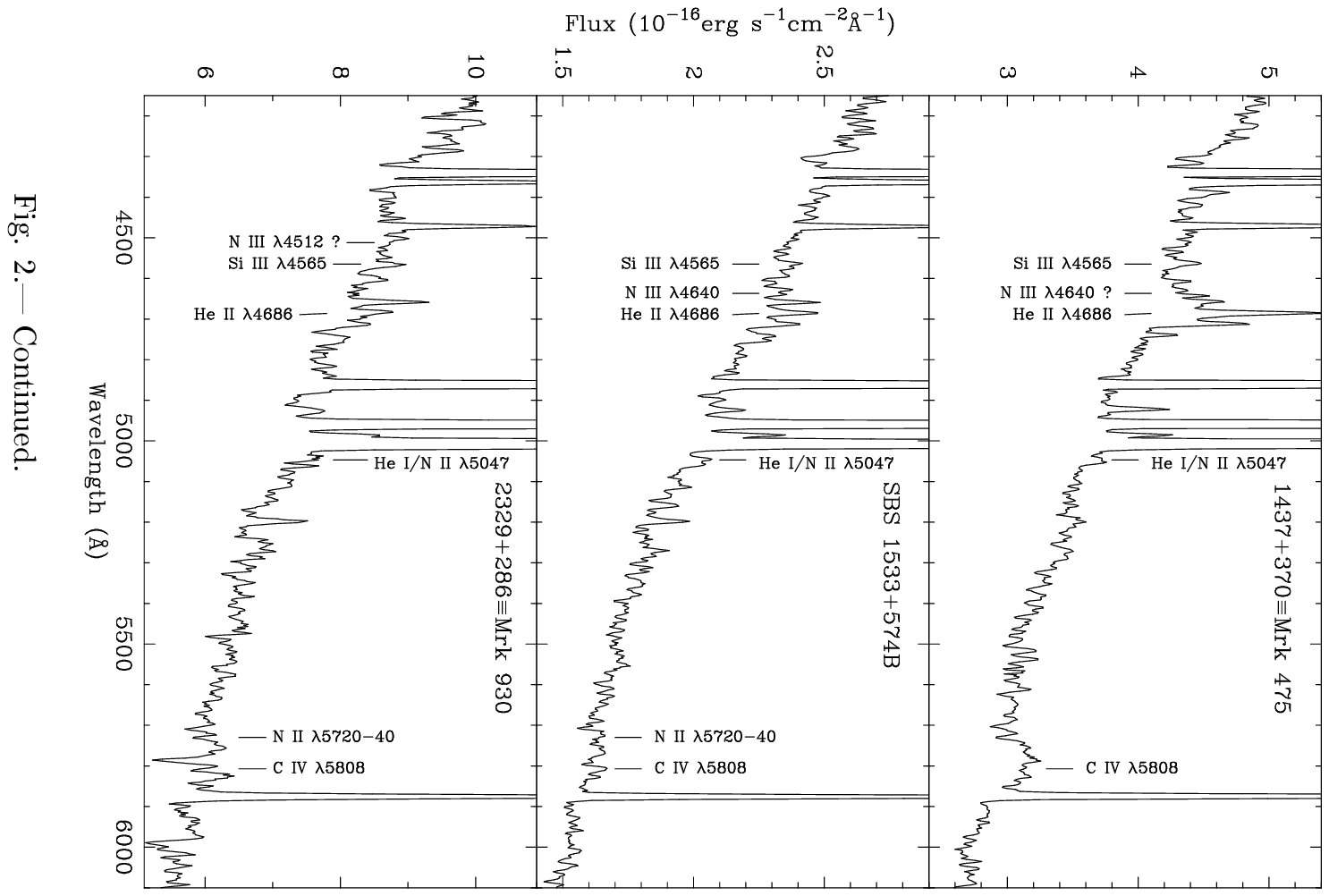}{0.cm}{180.}{100.}{100.}{250.}{500.}
\end{figure*}

\clearpage
%
%
\begin{figure*}
\figurenum{3}
\epsscale{1.6}
\plotfiddle{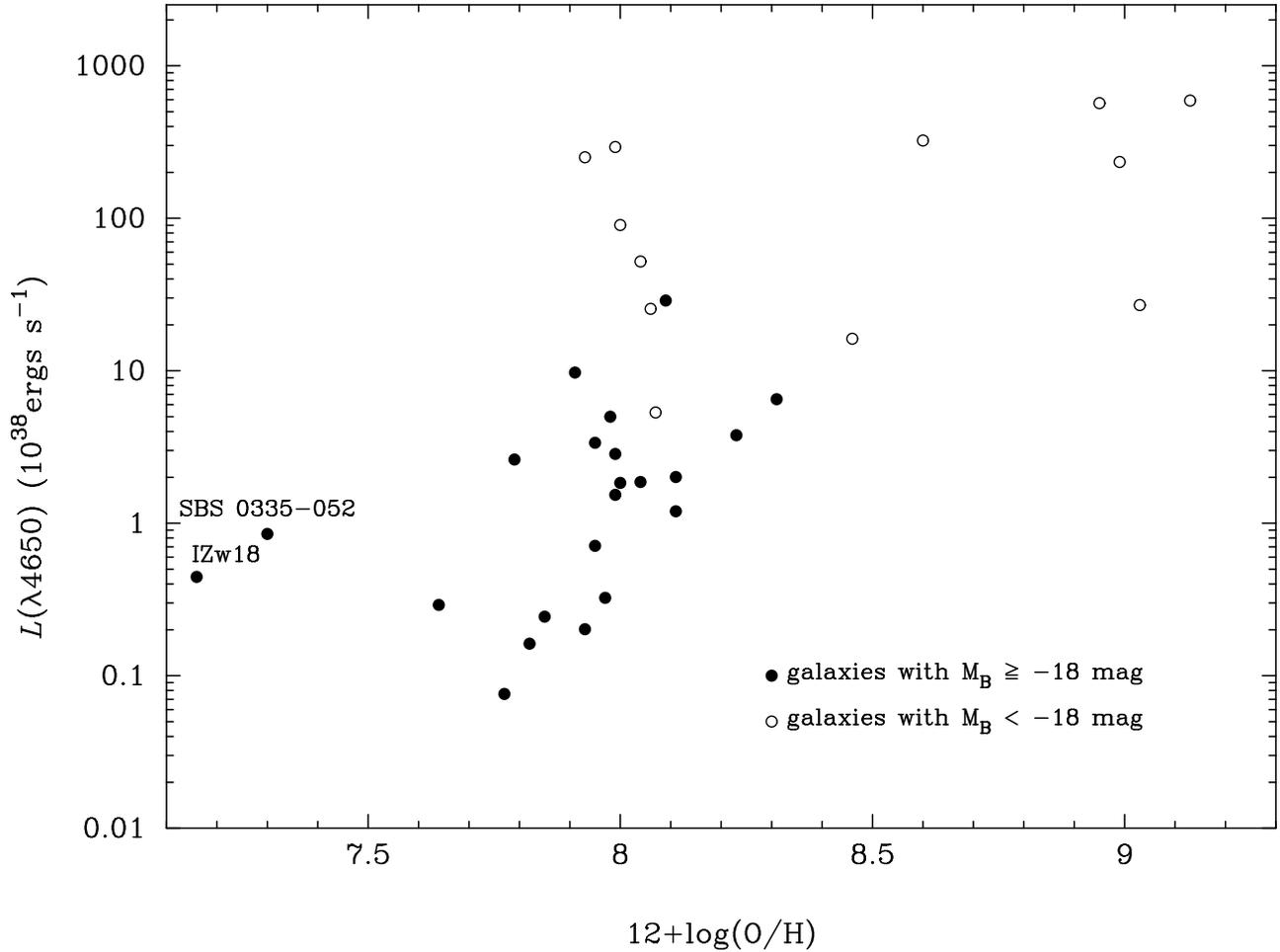}{0.cm}{270.}{80.}{80.}{-320.}{170.}
\vspace{10.cm}
\figcaption{The luminosity of the blue bump $L$($\lambda$4650) vs oxygen abundance
12 + log(O/H) for the WR galaxies. Filled circles denote 
galaxies with $M_B$ $\geq$ --18 and open circles those with $M_B$ $<$ --18. 
The data points belonging to the two most metal-deficient WR galaxies known,
I Zw 18 (Izotov et al. 1997, this paper) and SBS 0335--052 
(Izotov et al. 1999a), are marked.
}
\end{figure*}

\clearpage
%
%
\begin{figure*}
\figurenum{4}
\epsscale{2.0}
\plotfiddle{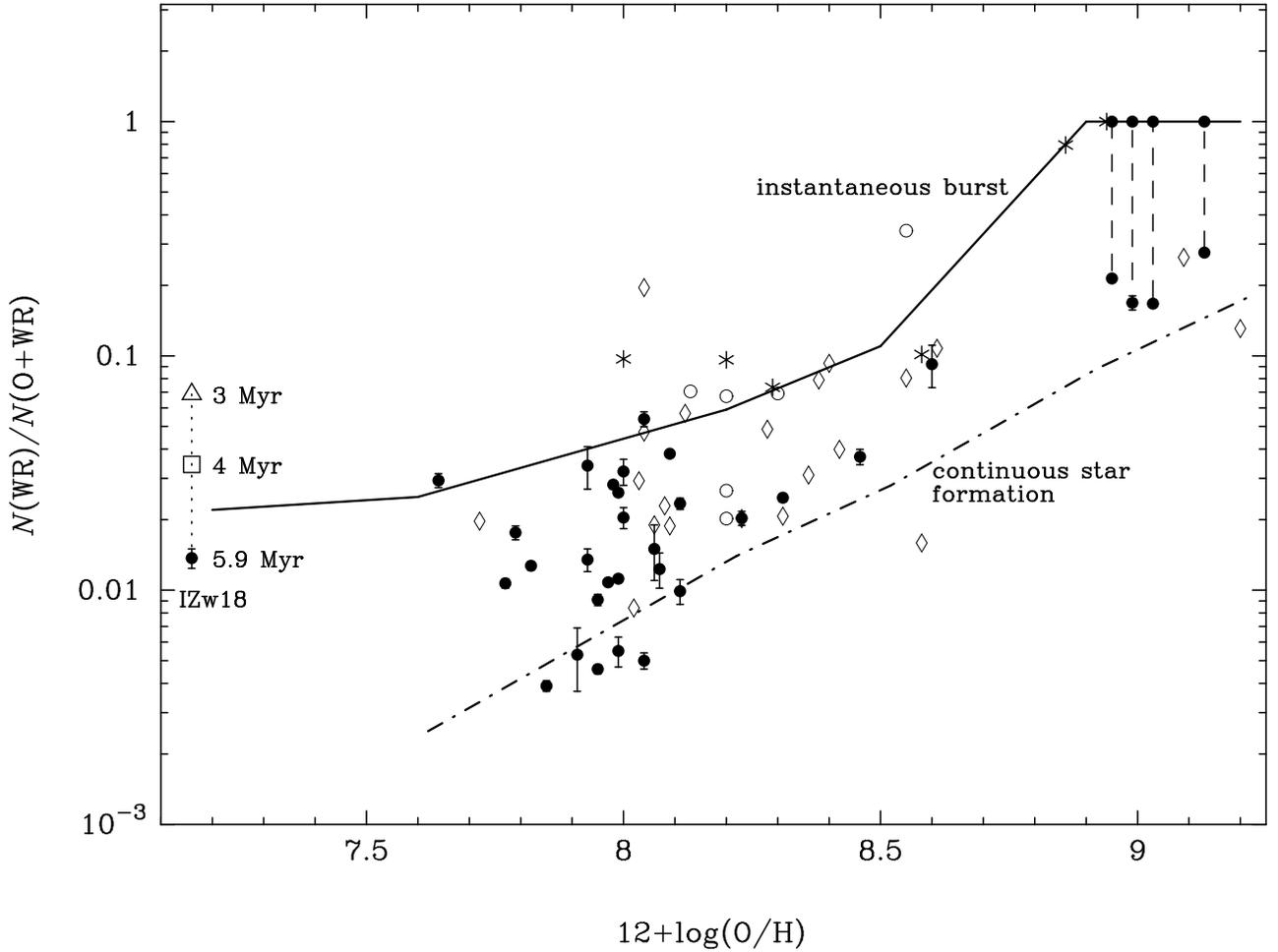}{0.cm}{270.}{80.}{80.}{-320.}{170.}
\vspace{10.cm}
\figcaption{The relative number of WR stars $N$(WR) / $N$(O+WR) vs oxygen abundance
12+log(O/H) for the galaxies in our sample (filled circles), and from Vacca \&
Conti (1992) (diamonds), Kunth \& Joubert (1985) (asterisks) and Schaerer
et al. (1999a) (open circles). For I Zw 18, three $N$(WR)/$N$(O+WR) values
are shown for three choices of the parameter $\eta_0$: 0.2 (filled circle),
0.5 (open rectangle) and 1 (open triangle) corresponding to burst ages
of 5.9, 4 and 3 Myr respectively. At the high metallicity end,
two values of $N$(WR)/$N$(O+WR) are shown for each galaxy connected by a 
dashed line: 1 and the value corresponding to $\eta_0$ = 1. The solid 
and dot-dashed lines 
show the model predictions of maximum values of 
$N$(WR)/$N$(O+WR) in the two limiting cases of an instantaneous burst
and continuous star formation, respectively  (SV98; Schaerer 1998, 
private communication).
}
\end{figure*}

\clearpage
%
%
\begin{figure*}
\figurenum{5}
\epsscale{2.0}
\plotfiddle{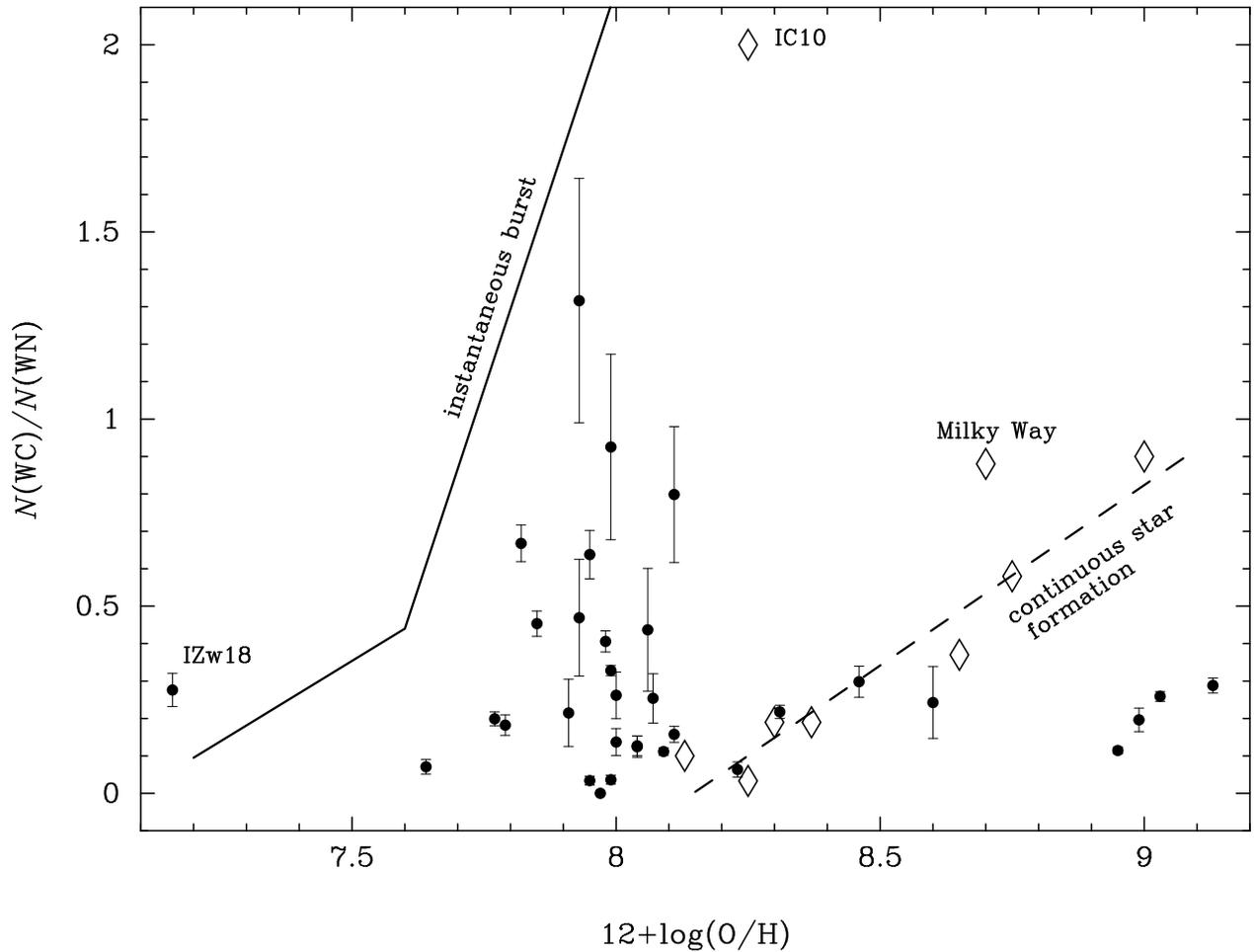}{0.cm}{270.}{80.}{80.}{-320.}{170.}
\vspace{10.cm}
\figcaption{The ratio of the WC to WN star numbers vs oxygen abundance 12 + log(O/H)
for the galaxies in our sample (filled circles). The solid line is the locus of
maximum values predicted by evolutionary synthesis models for an
instantaneous burst (SV98; Schaerer 1998, private communication). Diamonds are data from Massey \& Johnson (1998) for Local Group 
galaxies. The dashed line is those authors' fit to their data and represents
the case of continuous star formation.
}
\end{figure*}

\clearpage
%
%
\begin{figure*}
\figurenum{6}
\epsscale{2.0}
\plotfiddle{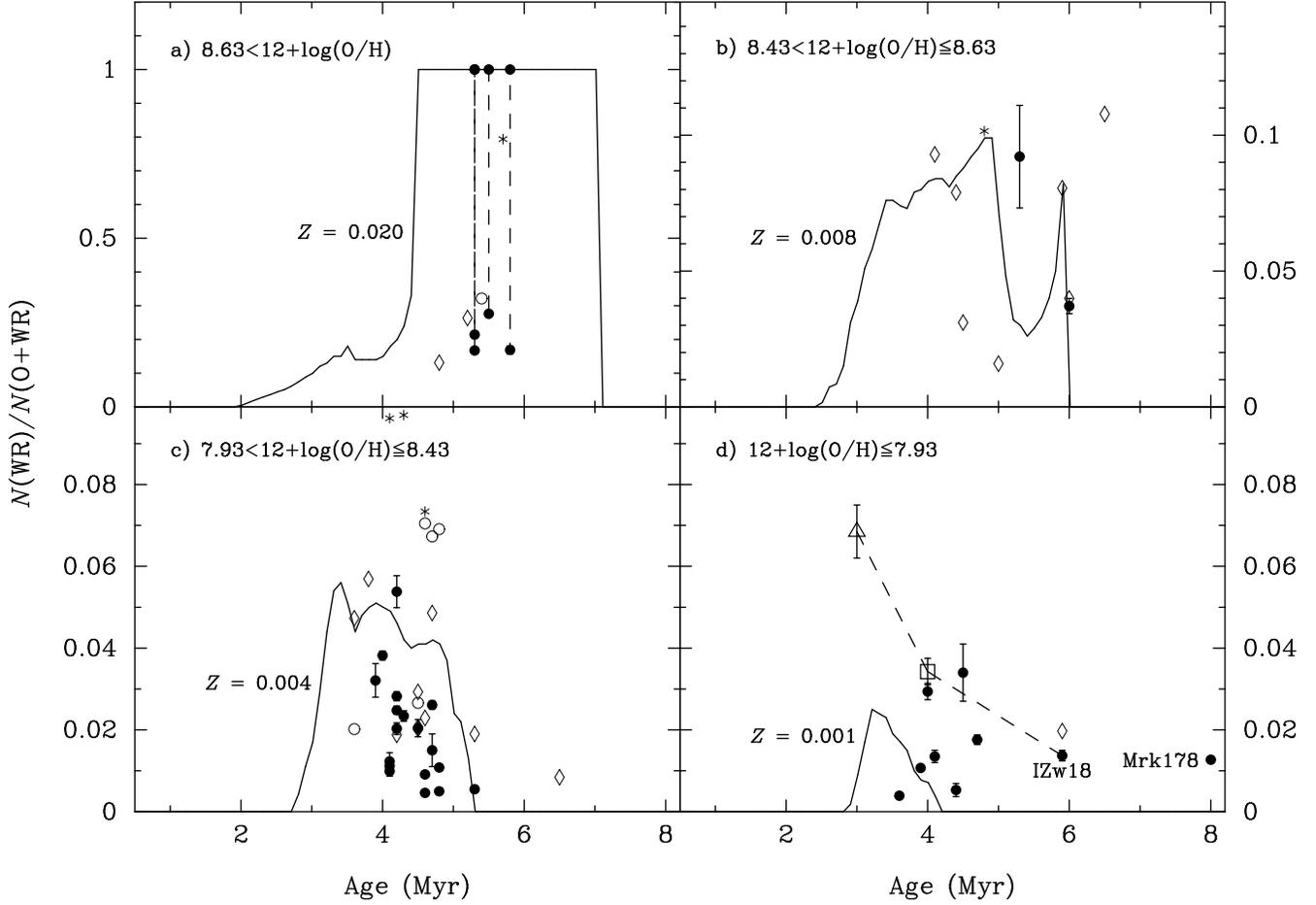}{0.cm}{270.}{70.}{70.}{-270.}{170.}
\vspace{10.cm}
\figcaption{$N$(WR)/$N$(O+WR) vs age of instantaneous burst diagrams for the galaxies
from our sample (filled circles), Vacca \& Conti (1992) (diamonds), Kunth \& 
Joubert (1985) (asterisks) and Schaerer et al. (1999a) (open circles). The data 
are divided into four groups according to oxygen abundances:
12 + log(O/H) $>$ 8.63 (Figure 6a), 8.43$<$ 12 + log(O/H) $\leq$ 8.63 (Figure 6b), 7.93$<$ 12 + log(O/H) $\leq$ 8.43 (Figure 6c) and 
12 + log(O/H) $\leq$ 7.93 (Figure 6d). In panel a) two values of $N$(WR)/$N$(O+WR) are shown for each high-metallicity galaxy,
connected by a dashed line (Figure 6a): 1 and the value corresponding to 
$\eta_0$ = 1. In panel d) three values of the $N$(WR)/$N$(O+WR) ratio 
connected by a dashed line are shown for I Zw 18. They are derived for three 
values of the parameter $\eta_0$: 0.2 (filled circle), 0.5 (open rectangle) and 
1 (open triangle) corresponding to burst ages of 5.9, 4 and 3 Myr respectively. 
In all four panels solid lines represent model predictions by SV98. 
They are labeled by the heavy element mass fraction of the model.
}
\end{figure*}

\clearpage
%
%
\begin{figure*}
\figurenum{7}
\epsscale{2.0}
\plotfiddle{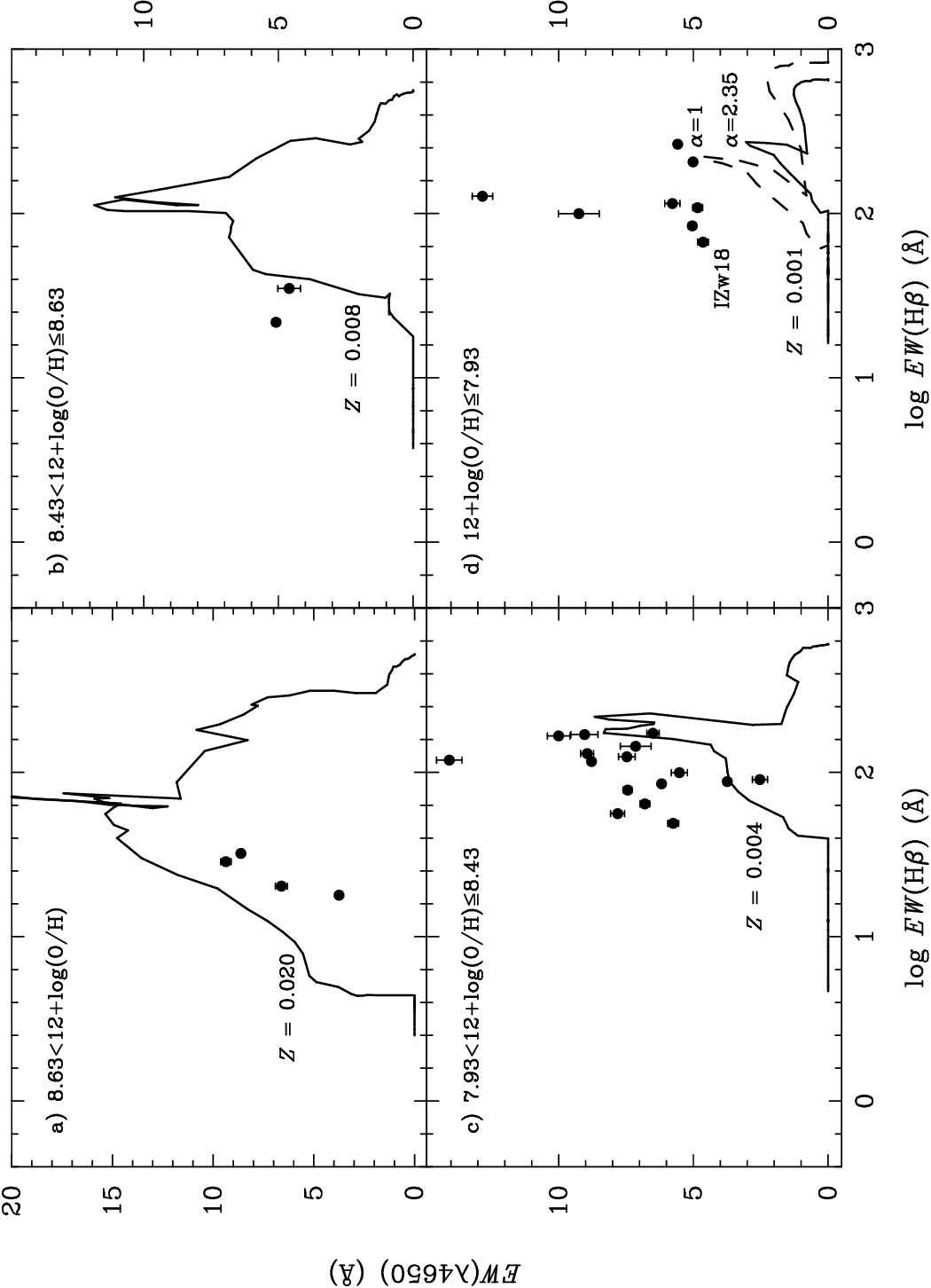}{0.cm}{270.}{70.}{70.}{-270.}{170.}
\vspace{10.cm}
\figcaption{The equivalent width of the blue bump $EW$($\lambda$4650) vs equivalent width
$EW$(H$\beta$) of the H$\beta$ emission line for our WR galaxy sample. 
The WR galaxies are divided into four groups with different ranges of
oxygen abundance as in Figure 6. The solid curves in all 4 panels show
theoretical predictions by SV98 for an instantaneous burst 
with  an initial mass function slope $\alpha$ = 2.35. In panel d), the 
predictions for an instantaneous burst with $\alpha$ = 1 are shown in addition.
Each model is labeled by its heavy element mass fraction.
}
\end{figure*}

\clearpage
%
%
\begin{figure*}
\figurenum{8}
\epsscale{2.0}
\plotfiddle{f8.ps}{0.cm}{270.}{70.}{70.}{-270.}{170.}
\vspace{10.cm}
\figcaption{Equivalent width of the red bump $EW$($\lambda$5808) vs equivalent width
$EW$(H$\beta$) of the H$\beta$ emission line for our WR galaxy sample. 
The WR galaxies are divided into four groups with different 
ranges of oxygen abundance as in Figure 6. The solid curves in all 4 panels show
theoretical predictions by SV98 for an instantaneous 
burst with an initial mass
function slope $\alpha$ = 2.35. In panel d), the predictions for an 
instantaneous burst with $\alpha$ = 1 are shown in addition. Each model is
labeled by its heavy element mass fraction.
}
\end{figure*}

\clearpage
%
%
\begin{figure*}
\figurenum{9}
\epsscale{2.0}
\plotfiddle{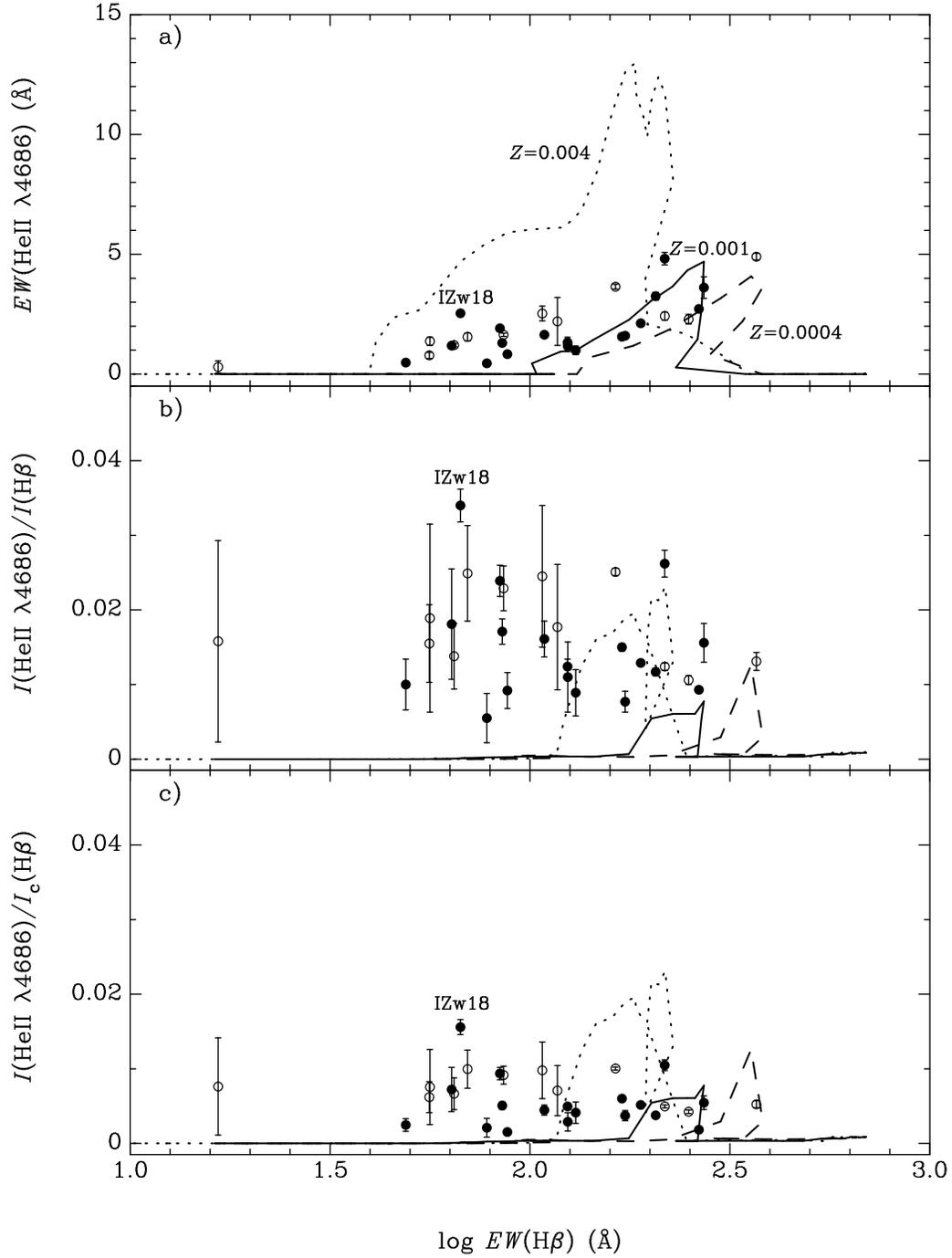}{0.cm}{0.}{70.}{70.}{-220.}{-530.}
\vspace{18.cm}
\figcaption{(a) Equivalent width of the He II $\lambda$4686 nebular emission line
vs equivalent width of the H$\beta$ emission line for a sample of
30 H II regions; 
(b) Intensity of the He II $\lambda$4686 nebular emission line
relative to H$\beta$, the latter not being
corrected for aperture, vs equivalent width of the H$\beta$ emission line;
(c) Intensity of the He II $\lambda$4686 nebular emission line
relative to aperture-corrected H$\beta$, 
vs equivalent width of the H$\beta$ emission line. 
In each panel, the lines show theoretical predictions for the heavy element 
mass fractions $Z$ = 0.004 (dotted), 0.001 (solid) and 0.0004 (dashed) 
respectively (SV98; Schaerer 1998, private
communication). H II regions with detected WR features are shown by filled 
circles while those with non-detected WR features are shown by open circles.
}
\end{figure*}

\end{document}